\pdfoutput=1
\documentclass[cernpreprint, atlasdraft=false, texlive=2023, UKenglish , texmf, orcidlogo]{atlasdoc}
 
\usepackage[backend=biber]{atlaspackage}

\usepackage[normalem]{ulem}

\usepackage{array}
\newcolumntype{C}[1]{>{\centering\let\newline\\\arraybackslash\hspace{0pt}}m{#1}}
 
\usepackage{atlasbiblatex}

\usepackage{atlasphysics}
 
\graphicspath{{logos/}{figures/}}

\usepackage{multirow}
 
\usepackage{ANA-EXOT-2019-26-PAPER-defs}

\AtlasTitle{Search for $t\bar tH/A \rightarrow t\bar tt\bar t$ production in the multilepton final state  in  proton--proton collisions at $\sqrt{s} = 13$~TeV with the ATLAS detector}
 
\AtlasAbstract{
A search for a new heavy scalar or pseudo-scalar Higgs boson ($H/A$)
produced in association with a pair of top quarks, with the Higgs boson decaying into a pair of top quarks ($H/A\rightarrow t\bar t$) is reported.
The search  targets a final state with exactly two leptons with same-sign electric charges or at least three leptons.
The analysed dataset corresponds to an integrated luminosity of 139~fb$^{-1}$~of  proton--proton collisions  collected at a centre-of-mass energy of 13~TeV with the ATLAS detector at the LHC. Two multivariate classifiers are used to separate the signal from the background.
No significant excess of events over the Standard Model expectation is observed.
The results are interpreted in the context of a type-II two-Higgs-doublet model. The observed (expected) upper limits at 95\%~confidence level on the $t\bar tH/A$ production cross-section
times the branching ratio of $H/A\rightarrow t\bar t$ range between 14~(10)~fb and 6~(5)~fb for a heavy Higgs boson with mass between 400~GeV and 1000~GeV, respectively.  Assuming  that only one  particle, either the scalar $H$ or the  pseudo-scalar $A$,  contributes to the $t\bar tt \bar t$ final state, values of $\tan\beta$ below 1.2 or 0.5 are excluded for a mass of 400~GeV or 1000~GeV, respectively.  These exclusion ranges increase to $\tan\beta$ below 1.6 or 0.6 when both particles are considered.
 
}
 
\author{The ATLAS Collaboration}

\PreprintIdNumber{CERN-EP-2022-170}
 
\AtlasDate{\today}

\AtlasJournal{\JHEP }
 
\AtlasJournalRef{\JHEP 07 (2023) 203}
\AtlasDOI{10.1007/JHEP07(2023)203}

\hypersetup{pdftitle={ATLAS document},pdfauthor={The ATLAS Collaboration}}
 
\begin{document}
 
\maketitle
 
\tableofcontents

\section{Introduction}

The discovery of the Higgs boson by the ATLAS and CMS Collaborations in 2012~\cite{Aad:2012tfa,Chatrchyan:2012ufa} and later
precision measurements~\cite{HIGG-2018-57,CMS-HIG-17-031} of Higgs boson production and decay properties
established the Standard Model (SM) of particle physics as an effective description of nature up to the TeV energy scale.
 
Within the SM, the Brout--Englert--Higgs mechanism~\cite{Englert:1964et,Higgs:1964ia,Higgs:1964pj,Guralnik:1964eu,Higgs:1966ev,Kibble:1967sv}
is responsible for generating the mass of the gauge bosons via electroweak symmetry breaking (EWSB). The Higgs boson emerges from the EWSB as the only physical spin-0 CP-even particle of the SM,
while the remaining components of the Higgs field are absorbed into the longitudinal components of the gauge bosons.
 
In many `beyond the Standard Model' (BSM) scenarios, the Higgs sector is extended to incorporate new degrees of freedom.
A well-motivated and minimal extension of the SM paradigm is provided by
two-Higgs-doublet models (2HDM)~\cite{Branco:2011iw} where the Higgs sector consists of two
complex doublets: a mixture of the two doublets plays the same role as the SM Higgs field and generates a Higgs boson ($h$); the other mixture produces three Higgs bosons -- one neutral CP-even ($H$), one neutral CP-odd ($A$), and one charged (\hp).
A generic CP-conserving 2HDM with natural flavour conservation, and a lighter CP-even Higgs boson playing the role of the SM Higgs boson, has five free parameters:
three Higgs boson masses ($m_H$, $m_A$, $m_{H^{\pm}}$), the mixing angle of the two CP-even Higgs fields ($\alpha$),
and the ratio of the vacuum expectation values of the two Higgs doublets  ($\tan\beta$).
A common realisation of the 2HDM is the Minimal Supersymmetric Standard Model (MSSM)~\cite{Fayet:1976et,Fayet:1977yc}. Similar Higgs sectors arise in axion models~\cite{Kim:1986ax}.
 
Existing constraints from direct searches for heavy neutral bosons by the ATLAS and CMS
Collaborations~\cite{HDBS-2018-13,ATLAS:2021ifb,HIGG-2018-27,HIGG-2018-09,HDBS-2018-46,HIGG-2016-32,CMS-HIG-17-033,CMS-HIG-18-012,CMS-HIG-18-023,CMS-HIG-18-010,CMS-HIG-18-005},
as well as precision measurements of the production cross-sections and decay rate of the SM Higgs boson,
restrict the available parameter-space to the so-called `alignment limit', $\sin(\beta-\alpha)\rightarrow 1$. In this limit the $h$ couplings are the same as for the SM Higgs boson.
 
For heavy neutral Higgs bosons with masses more than twice the top-quark mass, the dominant decay mode is \HAdec.
Inclusive searches for \HAdec\ are challenging because of destructive interference with the SM background, $gg\rightarrow \ttbar$, that largely dilutes a resonant peak in the \ttbar\
invariant mass spectrum~\cite{EXOT-2016-04,CMS-HIG-17-027}.
An alternative approach is to search for $\HA$ production in association with third-generation quarks~\cite{Craig:2016ygr}. In the low $\tan\beta$ region, the cross-section for $bbH/A$ is negligible, while is maximal for $t\bar{t}H/A$. Thus, the \ttHAnodec\ production mode provides a promising channel, characterised by large experimental acceptance and low SM background rate.
 
This paper presents a search for a new heavy scalar or pseudo-scalar Higgs boson, $\HA$,
produced in association with a pair of top quarks, with the Higgs boson decaying into a pair of top quarks, \ttHA. The Feynman diagram of this process is shown in Figure~\ref{fig:Feynmann}.  The mass of the heavy Higgs boson is assumed to be between 400~\GeV\ and 1000~\GeV, where a large $\HA \rightarrow t\overline{t}$ branching ratio and small $\HA$ total widths are expected.
The search  targets a final state with  exactly two leptons\footnote{In this paper, leptons refer to either electrons or muons, which can include those that come from $\tau$-lepton decays. } with same-sign electric charges or at least three leptons (SSML).
This particular signature is experimentally favoured due to the low level of background contamination, with the main contribution originating from the SM production of four top quarks and of $t\bar t$ in association with a $W$ boson ($t\bar tW$), $Z$ boson ($t\bar tZ$), or Higgs boson ($t\bar tH$).
Other significant sources of background are events where one of the leptons has a mis-assigned charge and events which contain a fake/non-prompt lepton.
Backgrounds from multiboson, $t\bar tWW$, single-top-quark or other rare top-quark processes are expected to be minor.
The analysed dataset corresponds to an integrated luminosity of 139~$\ifb$ of  proton--proton ($pp$) collisions  collected at a centre-of-mass energy of $\sqrt{s}=$13~\TeV\ with the ATLAS detector at the LHC.

\begin{figure}[htbp!]
\begin{center}
\includegraphics[width=0.5\linewidth, ]{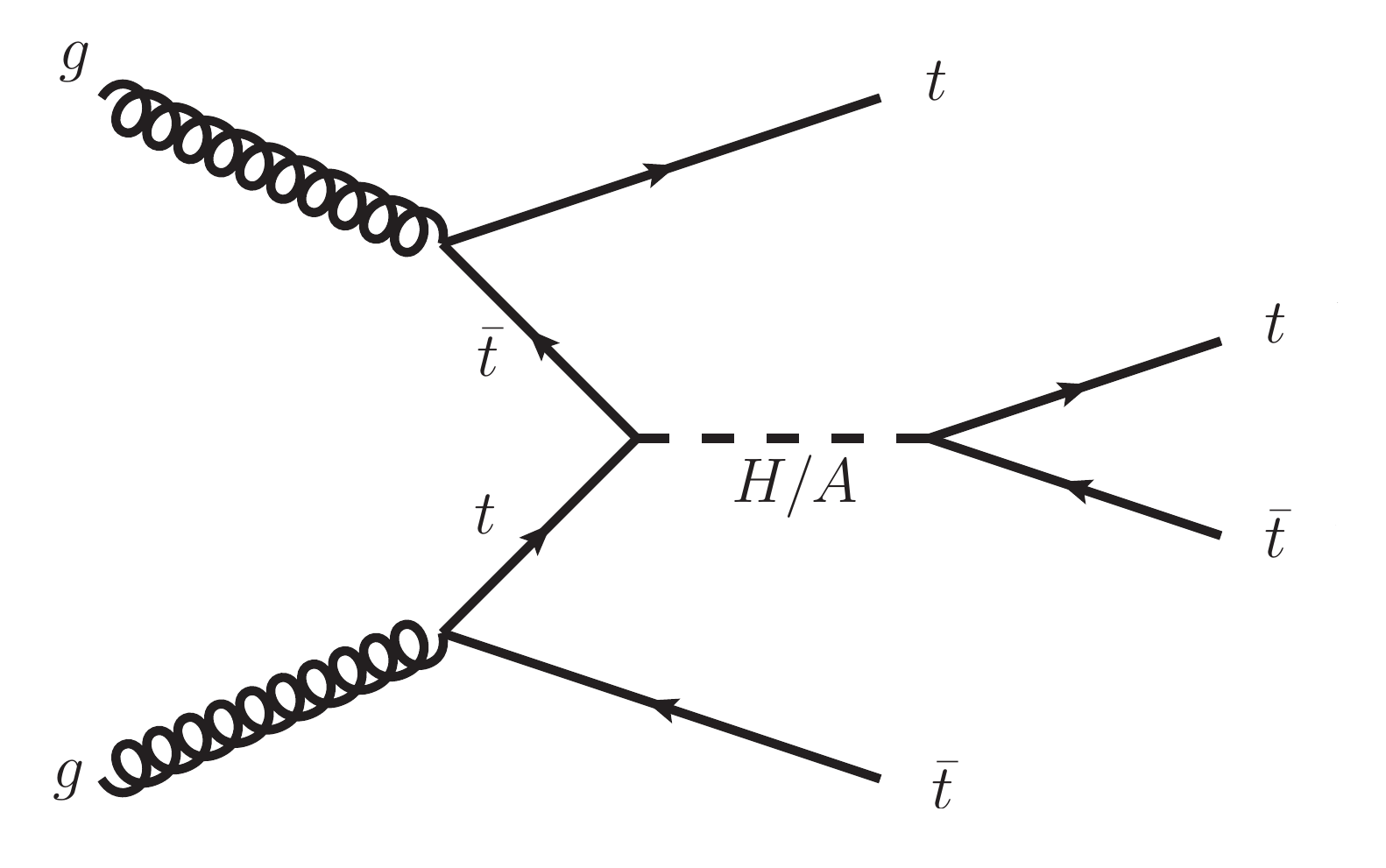}
\caption{Feynman diagram showing the production of a heavy scalar or pseudo-scalar Higgs boson, $H/A$,
produced in association with a pair of top quarks, with the Higgs boson decaying into a pair of top quarks.}
\label{fig:Feynmann}
\end{center}
\end{figure}

Previous searches for \ttHA\ in the SSML channel were performed by the ATLAS and CMS Co\-lla\-bo\-rations~\cite{EXOT-2016-16,CMS-TOP-18-003}.
A similar search  for BSM \tttt~events using an alternative experimental signature featuring exactly one charged lepton or two opposite-sign leptons was performed by the
CMS Collaboration~\cite{CMS-TOP-17-019}. 
Other related searches include those looking for SM production of four top quarks~\cite{TOPQ-2020-10,TOPQ-2018-05,CMS-TOP-18-003}.
The ATLAS and CMS measurements of SM four-top-quark production found the cross-section to be $24^{+7}_{-6}$~fb and $13^{+11}_{-9}$~fb, respectively, compared to a SM expectation
of $\sigma_{\tttt}=12.0\pm2.4$~fb~\cite{Frederix:2017wme}.

The results obtained in this search are interpreted in the context of a \typeII;
upper limits are placed on the  \ttHAnodec~production cross-section times the branching ratio of $\HA\rightarrow t\bar t$ as well as on $\tan\beta$ as a function of the heavy Higgs boson's mass.
Due to a significantly larger dataset and improved analysis techniques, including the use of multivariate classifiers based on
boosted decision trees (BDTs) to distinguish the signal from the SM background,
the expected sensitivity of the present \ttHA\ search exceeds that of the previous ATLAS \ttHA\ search with 36~\ifb~\cite{EXOT-2016-16} by about a factor of four.
 
This paper is structured as follows. The ATLAS detector is presented in Section~\ref{sec:detector}. The data and samples of simulated events are described in Section~\ref{sec:samples}. The object and event selections are detailed in Section~\ref{sec:eventSelection}. The analysis strategy is presented in Section~\ref{sec:strategy}. 
This is followed by a description of the systematic uncertainties in Section~\ref{sec:sistematic}.  The statistical analysis and  the results  are presented in Sections~\ref{sec:statistics} and~\ref{sec:result}, respectively. Finally, the conclusion is presented in Section~\ref{sec:conclusion}.


\section{ATLAS detector}
\label{sec:detector}

 
The ATLAS detector~\cite{PERF-2007-01} at the LHC covers nearly the entire solid angle around the collision point.\footnote{\AtlasCoordFootnote}
It consists of an inner tracking detector surrounded by a thin superconducting solenoid, electromagnetic and hadron calorimeters,
and a muon spectrometer incorporating three large superconducting air-core toroidal magnets with eight coils each.
 
The inner-detector system (ID) is immersed in a \SI{2}{\tesla} axial magnetic field
and provides charged-particle tracking in the range \(|\eta| < 2.5\).
The high-granularity silicon pixel detector covers the vertex region and typically provides four measurements per track,
the first hit normally being in the insertable B-layer (IBL) installed before Run~2~\cite{ATLAS-TDR-19,PIX-2018-001}.
It is surrounded by the silicon microstrip tracker (SCT), which usually provides eight measurements per track.
These silicon detectors are complemented by the transition radiation tracker (TRT),
which enables radially extended track reconstruction up to \(|\eta| = 2.0\).
The TRT also provides electron identification information
based on the fraction of hits (typically 30 in total) above a higher energy-deposit threshold corresponding to transition radiation.
 
The calorimeter system covers the pseudorapidity range \(|\eta| < 4.9\).
Within the region \(|\eta|< 3.2\), electromagnetic calorimetry is provided by barrel and
endcap high-granularity lead/liquid-argon (LAr) calorimeters,
with an additional thin LAr presampler covering \(|\eta| < 1.8\)
to correct for energy loss in material upstream of the calorimeters.
Hadron calorimetry is provided by the steel/scintillator-tile calorimeter,
segmented into three barrel structures within \(|\eta| < 1.7\), and two copper/LAr hadron endcap calorimeters.
The solid angle coverage is completed with forward copper/LAr and tungsten/LAr calorimeter modules
optimised for electromagnetic and hadronic energy measurements respectively.
 
The muon spectrometer (MS) comprises separate trigger and
high-precision tracking chambers measuring the deflection of muons in a magnetic field generated by the superconducting air-core toroidal magnets.
The field integral of the toroids ranges between \num{2.0} and \SI{6.0}{\tesla\metre}
across most of the detector.
A set of precision chambers covers the region \(|\eta| < 2.7\) with three layers of monitored drift tubes,
complemented by cathode-strip chambers in the forward region, where the background is highest.
The muon trigger system covers the range \(|\eta| < 2.4\) with resistive-plate chambers in the barrel, and thin-gap chambers in the endcap regions.
 
Interesting events are selected by the first-level trigger system implemented in custom hardware,
followed by selections made by algorithms implemented in software in the high-level trigger~\cite{TRIG-2016-01}.
The first-level trigger accepts events from the \SI{40}{\MHz} bunch crossings at a rate below \SI{100}{\kHz},
which the high-level trigger further reduces in order to record events to disk at about \SI{1}{\kHz}.
 
An extensive software suite~\cite{ATL-SOFT-PUB-2021-001} is used in data simulation, in the reconstruction and analysis of real
and simulated data, in detector operations, and in the trigger and data acquisition systems of the experiment.


\section{Data and samples of simulated events}
\label{sec:samples}

This analysis uses $pp$ collision data collected  between 2015 and 2018 by the ATLAS detector  at  $\sqrt{s}=13$~\TeV. The full dataset corresponds to an integrated luminosity of 139~$\ifb$. Only data recorded during stable beam conditions in which all detector subsystems  were operational~\cite{DAPR-2018-01} are included. Events were collected using single-lepton or dilepton triggers. The trigger $\pT$ threshold  depended on the lepton flavour and the data-taking period~\cite{TRIG-2018-05,TRIG-2018-01}. The lowest  $\pT$ threshold for the single-lepton trigger varied from 20 to 26~\GeV, while for the dilepton trigger it varied from 7 to 24~\GeV.
 
Monte Carlo (MC) simulation samples are used for the different signal and background processes. The generated events were processed through the simulation~\cite{SOFT-2010-01} of the ATLAS detector geometry and response based on \GEANT~\cite{Agostinelli:2002hh}, and through the same reconstruction software as data. Corrections are applied to the simulated events so that particle selection efficiencies, energy scales and energy resolutions match those determined from data. For the parton shower and hadronisation, unless otherwise stated, samples generated with  \MGNLO~\cite{Alwall:2014hca} and \POWHEGBOX[v2]~\cite{Frixione:2007nw,Nason:2004rx,Frixione:2007vw,Alioli:2010xd,Re:2010bp,Frederix:2012dh,Alioli:2009je} were interfaced to  \PYTHIA[8.230], 8.210 or 8.212~\cite{Sjostrand:2014zea}, using the A14 set of tuned parameters~\cite{ATL-PHYS-PUB-2014-021} and the \NNPDF[2.3lo]~\cite{Ball:2014uwa} parton distribution function (PDF) set. The decays of bottom and charm hadrons were simulated using the \EVTGEN[1.6.0]  or \EVTGEN[1.2.0] program~\cite{Lange:2001uf}.

Signal MC samples were generated using the \MGNLO[2.3.3] generator at leading order (LO) with the \NNPDF[3.1lo]~\cite{Ball:2014uwa} PDF set.
In order to account for the spin correlation of the particles in the final state, the full matrix-element for the $2 \rightarrow 12$ parton-scattering process was generated.
The signal MC events were generated assuming the production of \ttH\ in a \typeII.
A total of seven signal MC samples were generated with \mH ranging from
400~\gev\ to 1000~\gev\ with regular 100~\gev\ spacing. The width of the heavy Higgs boson is assumed to be small throughout the mass range, varying from 5~\gev\ to 30~\gev, consistent with the
expected width of a heavy Higgs boson in the \typeII\ for $\tan\beta \sim 1$.
Interference between the signal and SM four-top-quark production was not simulated. In the mass range of interest for this search, the interference leads to a change of up to 20$\%$
in the cross-section for a signal with $\mH = 400$~\gev\ and a width of 50~\gev, and the kinematic properties of the signal are found to be  consistent between $A$ and $H$. For $\tan\beta \sim 1$, the \ttH\ and \ttA\ production cross-sections are found to differ by $1\%$ at most.

The SM production of \tttt\ events was modelled in the following way. The nominal sample was produced with the \MGNLO[2.3.3]
generator, which provides matrix elements at next-to-leading order~(NLO) in the strong coupling constant \alphas
with the \NNPDF[3.1nlo] PDF set.
The functional form of the renormalisation and factorisation scales was set to $\muR=\muF=m_\textrm{T}/4$ where $m_\textrm{T}$ is defined as the scalar sum of the transverse masses $\sqrt{m^2+\pTsq}$  of the particles generated from the matrix element calculation, following Ref.~\cite{Frederix:2017wme}.
The decay of the top quark was simulated at LO using \MADSPIN~\cite{Frixione:2007zp,Artoisenet:2012st} to preserve all spin correlations.
To avoid the use of negative weights present in the nominal NLO sample in the training of the multivariate discriminant used to separate SM \tttt\ events from background (see Section~\ref{sec:signalDiscrimination}), a sample was produced with similar generator settings, but at LO.    An additional \tttt\ sample was produced with the \Pythia[8.230] parton shower of the nominal sample replaced by
\HERWIG[7.04]~\cite{Bahr:2008pv,Bellm:2015jjp} to evaluate the impact of using a different parton shower and hadronisation model. The H7UE set of
tuned parameters~\cite{Bellm:2015jjp} and the \MMHT[lo] PDF set~\cite{Harland-Lang:2014zoa} were used.
A sample of \tttt\ events modelled using the  \Sherpa[2.2.10]~\cite{Bothmann:2019yzt} generator with the \NNPDF[3.0nnlo]~\cite{Ball:2014uwa} PDF set is used to assess the uncertainty associated with the choice of event generator.  A sample including electroweak (EW) corrections~\cite{Gutschow:2018tuk} was generated with the same settings as in \Sherpa[2.2.10], but using \Sherpa[2.2.11], in order to evaluate the systematic uncertainty associated with the missing EW corrections in the nominal sample. The production of \tttt\  events is normalised to a cross-section of 12 fb computed at NLO in QCD including EW corrections~\cite{Frederix:2017wme}.

The  production of \ttW~events was modelled at NLO in QCD using the \Sherpa[2.2.10] generator with the  \NNPDF[3.0nnlo] PDF set, including up to one extra parton at NLO and up to two extra partons at LO. The additional partons were matched and merged with the \sherpa parton shower based on Catani--Seymour dipole factorisation~\cite{Schumann:2007mg} using the MEPS@NLO prescription~\cite{Hoeche:2011fd,Hoeche:2012yf,Catani:2001cc,Hoeche:2009rj} with a merging scale of 30~\GeV. The virtual QCD corrections for matrix elements at NLO accuracy were provided by the \OPENLOOPS[2]~\cite{Cascioli:2011va,Denner:2016kdg,Buccioni:2019sur}  library.  The renormalisation and factorisation scales were set to $\muR=\muF=m_\textrm{T}/2$.
Sub-leading EW corrections at order  $\mathcal{O}{(\alpha^{3}\alphas)}$~\cite{Frederix:2017wme}, were accounted for via the addition of an independent sample produced at LO in QCD for this final state using the same generator and PDF as in their QCD-only production.
The \ttW~cross-section predicted by these MC generator settings, including EW corrections, is 639~fb.  The impact of the systematic uncertainty associated with the choice of generator is evaluated with an alternative  \ttW~sample generated at NLO in QCD with no additional partons using the  \MGNLO[2.3.3] generator with the \NNPDF[3.0nlo] PDF set. 
The production of \ttW~events in  \MGNLO with only EW  corrections was modelled at LO in QCD with the  \NNPDF[2.3lo] PDF set. 
 
The production of $t\bar t(Z/\gamma^*) $ events was modelled using the \Sherpa[2.2.1] generator at NLO with the \NNPDF[3.0nnlo] PDF set. The invariant mass of the lepton pair was required to satisfy $m_{\ell^+\ell^-}>5$~\GeV. For the low mass range, i.e.\ $m_{\ell^+\ell^-}\in (1,5)$~\GeV,  events were modelled using the \MGNLO[2.3.3] generator at NLO with the \NNPDF[3.0nlo] PDF set.  These two samples were combined and together they form the `$t\bar t(Z/\gamma^*)$ (high  mass)' sample.   
In order to assess the uncertainty associated with the choice of generator, a \MGNLO sample generated with exactly the same settings as the nominal low-mass-range sample but with    $m_{\ell^+\ell^-}>5$~\GeV\ was used.

The production of $\ttbar H$ events was modelled using the
\POWHEGBOX[v2]
generator, which provided matrix elements at NLO in the strong coupling
constant \alphas in the five-flavour scheme with the \NNPDF[3.0nlo] PDF set.
The cross-section was calculated at NLO QCD and NLO EW accuracy using
\MGNLO as reported in Ref.~\cite{deFlorian:2016spz}.  An alternative sample generated with \MGNLO[2.3.3] is used to evaluate the impact of the uncertainty associated with the generator choice.
 
The production of \ttbar and single-top-quark events was modelled using the
\POWHEGBOX[v2]
generator at NLO in QCD with the \NNPDF[3.0nlo] PDF set. For the \ttbar sample, the \hdamp parameter\footnote{The
\hdamp parameter is a resummation damping factor and one of the
parameters that controls the matching of \POWHEG matrix elements to
the parton shower and thus effectively regulates the
high-\pt radiation against which the \ttbar system recoils.} was set
to 1.5\,\mtop~\cite{ATL-PHYS-PUB-2016-020}.  The diagram removal scheme~\cite{Frixione:2008yi} was used to
remove interference and overlap between \ttbar and $tW$ production.  For these two processes, the contribution from internal photon conversions ($\gamma^*\rightarrow \ell^+\ell^-$) with $m_{\ell^+\ell^-}< 1~\GeV$ was modelled by QED multiphoton radiation via the parton shower.   This  contribution is referred to in the following as the `Low $m_{\gamma^*}$' sample.

The production of $tWZ$, $tZ$, $\ttt$, $\ttWW$ and other rare top-quark processes, namely the $\ttZZ$, $\ttWZ$, $\ttHH$, and $\ttWH$ processes, was modelled using the \MGNLO generator.  The \tWZ\ sample was modelled at NLO, while the remaining processes were modelled at LO in QCD.  The contribution of $\ttWW$ is normalised to the NLO QCD theoretical cross-section~\cite{deFlorian:2016spz}.
 
The $WH$ and $ZH$ processes were modelled using the \PYTHIA[8.230] generator with the A14 tune  and  \NNPDF[2.3lo] PDF set and normalised to their theoretical cross-sections calculated at  NNLO QCD and NLO EW accuracies~\cite{Ciccolini:2003jy,Brein:2003wg,Brein:2011vx,Altenkamp:2012sx,Denner:2014cla,Brein:2012ne,Harlander:2014wda}. 
 
The production of $Z+$jets, $W$+jets, diboson ($VV$) and triboson ($VVV$) events  was modelled with the
\SHERPA[2.2.1] or 2.2.2 generator depending on the process.
The \NNPDF[3.0nnlo] set of PDFs was used, along with the
dedicated set of tuned parton-shower parameters developed by the
\SHERPA authors.

The effect of multiple interactions in the same and neighbouring bunch
crossings (\pileup) was modelled by overlaying the simulated hard-scattering event with
inelastic $pp$ events generated with \PYTHIA[8.186]~\cite{Sjostrand:2007gs}
using the \NNPDF[2.3lo] set of PDFs and the
A3 set of tuned parameters~\cite{ATL-PHYS-PUB-2016-017}.


\section{Object and event selections}
\label{sec:eventSelection}

Events are required to contain at least one primary vertex (PV) reconstructed from at least two ID tracks, each with $\pT>0.5~\GeV$.  If more than one PV  candidate satisfies these criteria, then the PV with the largest sum of $\pT^2$~ over all associated ID tracks is selected~\cite{ATL-PHYS-PUB-2015-026}.
 
Electron candidates are reconstructed from energy deposits in the electromagnetic calorimeter matched to a track in the ID~\cite{EGAM-2018-01}.  The identification working point for the nominal selection in this analysis is `TightLH'~\cite{EGAM-2018-01}. Only electron candidates with $\pT>28~\GeV$ and within $|\eta|<2.47$, excluding the calorimeter transition region $1.37<|\eta| < 1.52$, are selected. Electrons are required to be well isolated using criteria based on the properties of the topological clusters in the calorimeter and of ID tracks around the reconstructed electron.  Standard requirements on the longitudinal ($z_0$) and transverse ($d_0$) impact parameters are used to select electrons originating from the primary vertex. The requirements are   $|z_0\sin{\theta}|<0.5$~mm and $|d_0|/\sigma(d_0)<5$.
 
Muon candidates are reconstructed by combining tracks in the ID with tracks in the MS~\cite{MUON-2018-03}.  The identification working point used in this analysis is `Medium'~\cite{MUON-2018-03}.  Only muon candidates with $\pT>28~\GeV$ and within $|\eta|<2.5$ are selected.   Muons are required to satisfy isolation requirements based on the properties of ID tracks around the reconstructed muon.  Similarly to electrons, requirements on the longitudinal and transverse impact parameters, $|z_0\sin{\theta}|<0.5$~mm and $|d_0|/\sigma(d_0)<3$, are also applied.
 
An additional requirement is imposed on electrons in the $e^\pm e^\pm$ and $e^\pm \mu^\pm$ channels to reduce the background coming from electron charge misidentification. This requirement is based on a BDT discriminant  which uses the calorimeter and tracking variables to remove approximately 90\% of electrons with the wrong charge assignment while selecting 98\% of electrons with the correctly measured charge~\cite{EGAM-2018-01}.

The constituents for jet reconstruction are identified by combining measurements from both the ID and the calorimeter using a particle-flow algorithm~\cite{PERF-2015-09}. Jet candidates are reconstructed from these particle-flow objects using the anti-$k_t$ algorithm~\cite{Cacciari:2008gp,Fastjet} with a radius parameter of $R=0.4$.  They are calibrated using simulation with corrections obtained from data using in situ techniques~\cite{JETM-2018-05}. Only jet candidates with  $\pT> 25$~\GeV\ and within $|\eta|<2.5$ are selected.  To reduce the effect of pile-up,  each jet with $\pT<60$~\GeV\ and $|\eta|<2.4$ is required to satisfy the `Tight'  working point of the jet-vertex tagger (JVT)~\cite{ATLAS-CONF-2014-018} criteria used to identify the jets as originating from the selected primary vertex. A set of quality criteria are also applied to reject events containing at least one jet arising from non-collision sources or detector noise~\cite{ATLAS-CONF-2015-029}.
 
Jets containing $b$-hadrons  are identified ($b$-tagged) using the DL1r algorithm~\cite{ATL-PHYS-PUB-2017-013}. It consists of a deep-learning neural network based on the distinctive features of the $b$-hadrons in terms of the impact parameters of tracks and the displaced decay vertices reconstructed in the ID. The input to the DL1r network also includes discriminating variables constructed by a recurrent neural network~\cite{ATL-PHYS-PUB-2017-003}, which exploits the spatial and kinematic correlations between tracks originating from the same $b$-hadron. A jet is $b$-tagged if the DL1r score is above a certain threshold, referred to as an operating point (OP). Four OPs are defined with average  expected efficiencies for $b$-jets of 60\%, 70\%, 77\% and 85\%, as determined in simulated \ttbar\ events. A jet is considered $b$-tagged if it passes the OP corresponding to 77\% average efficiency for jets containing $b$-hadrons, with a misidentification rate of 1/130 (1/4.9) for light-flavour (charm) jets. Correction factors are applied to the simulated event samples to compensate for differences between data and simulation in the $b$-tagging efficiency for $b$-jets, $c$-jets and light-flavour jets. The correction for $b$-jets is derived from \ttbar\ events with final states containing two leptons, and the corrections are consistent with unity with uncertainties at the level of a few percent over most of the jet \pT\ range~\cite{FTAG-2018-01}.

An overlap removal procedure is applied to ensure that the same calorimeter energy deposit or the same track is not used in two different objects. First, any electron found to share a track with another electron with higher $\pT$ is removed. Next, electrons sharing their track with a muon candidate are removed. Then, jets are removed if they are within $\Delta R=0.2$ of an electron. Subsequently, in order to remove electrons arising from $b$- or $c$-decays, electrons are removed if they are within a $\Delta R=0.4$ of a jet.  Next, jets within $\Delta R=0.2$ of a muon are removed if they have less than three tracks. Finally, in order to remove muons arising from $b$- or $c$-decays, muons are removed if their tracks are within $\Delta R=0.4 + 10 \text{ \GeV} / p_{\text{T}}^{\mu}$ of any remaining jets.
 
The missing transverse momentum of the event, $\vec{\pT}^\text{miss}$, is defined as the negative vector sum of the $\pT$ of all selected and calibrated objects in the event. This sum includes a term to account for the transverse momenta of ID tracks matched to the  selected PV but which are not associated with any of the selected objects in the event~\cite{PERF-2016-07}. The magnitude of $\vec{\pT}^\text{miss}$ is denoted by \Etmiss.
 
Events are required to have exactly two leptons with the same electric charge or at least three leptons without any charge requirement. Each event must have at least one reconstructed lepton that matches a lepton that satisfied the trigger requirements. In order to reduce the  background coming from electron charge misidentification in the same-sign dilepton channel, and the contamination from the $Z$ boson decay in the trilepton channel, a $Z$-veto rejecting $m_{\ell\ell}\in [81,101]$~\GeV\ is applied, where $m_{\ell\ell}$ is checked for the two electrons defining the $e^\pm e^\pm$ channel, and for all opposite-sign same-flavour lepton pairs in  the trilepton channel.  In addition, events with an $e^\pm e^\pm$ final state are also required to satisfy $m_{ee} >15$~\GeV\ to reduce the background coming from low-mass resonances with electron charge misidentification.


\section{Analysis strategy}
\label{sec:strategy}

 
The analysis strategy follows the one used in the previous ATLAS $\tttt$ measurement~\cite{TOPQ-2018-05}, with dedicated control regions to constrain the dominant background processes, and  multivariate techniques to separate the signal from background. The background estimation is detailed in Section~\ref{sec:bkgEstimation}.
 
A signal-enriched region is defined by exploiting the high multiplicity of light-flavour and $b$-tagged jets as well as the high overall momentum of final-state jets and leptons in $\tttt$ events. Thus, events are required  to have at least six jets, among which at least two are $b$-tagged, and  $H_\text{T} > 500$~\GeV, where $H_\text{T}$ is defined as the scalar sum of the $\pT$ of all leptons and jets in the event. This signal region (SR) is referred to as the {\it baseline SR} in the following.
 
The signal is separated from the SM background by using two sequential BDT classifiers (see Section~\ref{sec:signalDiscrimination}). These BDTs are trained in
the {\it baseline SR}, inclusively in both lepton flavour and multiplicity. The first one, the `SM BDT', is used to separate SM $\tttt$ events from the remaining backgrounds. Due to the similar kinematics of the signal and SM $\tttt$ events, signal events are expected to be located in the same region of the BDT score distribution as  the SM $\tttt$ events. Therefore, the final signal region is a part of the {\it baseline SR}, and it is defined by the requirement SM~BDT~>~0.55.
This definition is referred to as the {\it BSM SR} in the following. The full selection for the {\it BSM SR} can be found in Table~\ref{tab:RegionDefinition}.
The second classifier, the `BSM pBDT', is a mass-parameterised BDT that discriminates signal events from all backgrounds.
To test for the presence of a signal, the distribution of the BSM pBDT score in the {\it BSM SR} is fitted jointly with the control regions defined in Section~\ref{sec:bkgEstimation}.  For the $t\bar tH(\rightarrow t\bar t)$ signal described in Section~\ref{sec:samples}, the acceptance times the efficiency to pass the {\it BSM SR} requirements ranges from 2\% to 3\% for the different signal hypotheses.


\subsection{Background estimation}
\label{sec:bkgEstimation}


Several SM processes can mimic a final state of two leptons with the same electric charge or three leptons. They can be  classified into two different categories depending on the origin of the  lepton:
 
\begin{description}
\item[Irreducible:] All selected leptons are prompt. They originate mainly from $W$ or $Z$ boson decays.  The main contribution in the signal region is given by $\tttt$, $\ttW$, $t\bar t(Z/\gamma^*)$ (high mass) and $t\bar{t}H$ production. Smaller contributions come from $VV$, $VVV$ and $VH$ production, and from rare processes ($\ttVV$, $tWZ$, $tZq$ and  $t\bar tt$).  This background is estimated  using the samples of simulated events described in Section~\ref{sec:samples} and normalised to the theory cross-sections.
The $\ttW$ contribution is split between its QCD and EW components.\footnote{The $\ttW$ QCD and EW samples are calculated at different orders: QCD is NLO in QCD, and EW is LO in QCD. Therefore, the QCD and EW components are decorrelated and treated as two independent samples with their own systematic uncertainties.} Because the normalisation of $\ttW$ has been found to be underestimated in the simulation in recent measurements~\cite{TOPQ-2016-11}, the normalisation of the QCD component is corrected using data in a dedicated control region as explained in Section~\ref{sub:fakesEstimation}. The Low $m_{\gamma^*}$ background is also included in this category. Since the normalisation of this background might not be correct in simulation, its normalisation is also estimated in a dedicated control region as explained in Section~\ref{sub:fakesEstimation}.
This search targets BSM production of four top quarks, so the background originating from SM four-top-quark production is normalised to its SM prediction with a Gaussian constraint given by the theoretical uncertainty.
 
\item[Reducible:]  At least one of the leptons is fake/non-prompt or it is a prompt lepton with its charge misidentified. It originates mainly from $\ttbar$+jets, $V$+jets and $tW$+jets production.  The fake/non-prompt lepton background is estimated together with the $\ttW$ QCD and Low $m_{\gamma^*}$ backgrounds by using the template fit method described in Section~\ref{sub:fakesEstimation}. The background coming from electron charge misidentification  (QmisID) is estimated using data-driven techniques as described in Section~\ref{sub:chargeMisID}.
 
\end{description}

\subsubsection{\label{sub:fakesEstimation}Fake/non-prompt lepton background,   Low $m_{\gamma^*}$  and $\ttW$ QCD normalisations}
 
Fake/non-prompt lepton background arises from different sources:
 
\begin{itemize}
\item Events with one lepton from a detector-material (or beam pipe) photon conversion (Mat.\ Conv.).
\item Events with one  electron (muon) from a heavy-flavour meson decay (HF $e$/$\mu$).
\item Events with one  lepton from a light-meson decay or a jet misidentified as a lepton  (LF).
\item Events with one fake/non-prompt lepton not arising from any of the above (other fake).
\end{itemize}
 
The true generator-level information from the $\ttbar$+jets, $V$+jets and single-top samples described in Section~\ref{sec:samples} is used to separate the different sources of fake/non-prompt leptons. The last two components are very small and are fully estimated with samples of simulated events. The contributions  from detector-material photon conversions and HF $e$/$\mu$ are estimated with the template fit method.  This method relies on the simulation to model the kinematic distributions of fake/non-prompt leptons from different sources  and uses dedicated control regions to determine their normalisations.
 
Several control regions were defined, with each region designed to maximise the background component which is mainly fitted in that region. The variable to be fitted in each region was chosen so as to provide the most discrimination for the targeted background.  The normalisation of the Low $m_{\gamma^*}$ irreducible background is mainly estimated in the control region enriched in background from detector-material photon conversions. A dedicated control region was defined for  $\ttW$~QCD production. An extra control region was defined so as to be as close as possible to the {\it BSM SR}. In total, five control regions are used in the analysis, closely following the definitions in Ref.~\cite{TOPQ-2018-05}. They are described in Table~\ref{tab:RegionDefinition} and summarised below.

\begin{itemize}
\item `CR Conv':  It is enriched in detector-material photon conversions and  Low $m_{\gamma^*}$ background. Events are required to have an $e^\pm e^\pm$ or $e^\pm \mu^\pm$ pair. For each electron in the event, the invariant mass of the system formed by the track associated with the electron and the closest track at the conversion (primary) vertex $m_{ee}^\text{CV}$ ($m_{ee}^\text{PV}$) is computed.  The conversion vertex is defined at the point where the track from the electron and its closest track in  $\Delta R$ have the same $\phi$. Virtual photons lead to a lepton pair originating from the primary vertex, having a low $m_{ee}^\text{PV}\sim ~m_{\gamma^*}$. Detector-material photon conversions have a large conversion radius, and the track extrapolation induces a large apparent $m_{ee}^\text{PV}$.  The control region is then obtained by selecting events with low $m_{ee}^\text{CV}$ and fitting the $m_{ee}^\text{PV}$ distribution to separate the detector-material photon conversions and Low $m_{\gamma^*}$ background from each other.  Events are also required to have four or five jets, at least one $b$-tagged jet, and low $H_\text{T}$.
 
\item `CR HF $e$' ( `CR HF $\mu$'):  It is enriched in events with one electron (muon) coming from heavy-flavour decay. The selection targets $\ttbar$ dilepton decays with an extra non-prompt lepton. This region is then defined by selecting events with  three leptons, namely $eee$ and $ee\mu$ ($e\mu\mu$ and $\mu\mu\mu$) for CR HF $e$ (CR HF $\mu$), low $H_\text{T}$ and exactly one $b$-jet.  The number of events in this region is used as the fitted variable.
 
\item `CR $\ttW$': It is enriched in $\ttW$ events.  Events are required to have an $e^\pm \mu^\pm$  or $\mu^\pm \mu^\pm$  pair, at least four jets and at least two $b$-jets.  In order to reduce the  background coming from electron charge misidentification, events containing electrons with $|\eta|>1.5$ are removed. This region is also required to be orthogonal to the  `CR Conv' region and to the {\it baseline SR}. The  fitted variable is the scalar sum of the lepton $\pT$.
 
\item `CR lowBDT': It is not enriched in any particular background, but rather used as a control region which is very close to the {\it BSM SR}.  Events are required to be in the {\it baseline SR}, but with SM~BDT~<~0.55. The fitted variable is the SM BDT output score.
 
\end{itemize}

These control regions are fitted simultaneously  with the {\it BSM SR} to determine both the strength of any BSM $\tttt$ signal and the five normalisation factors: $\lambda_{\ttW \text{QCD}}$ for $\ttW$~QCD production,  $\lambda_{\text{Mat.\ Conv.}}$ for the background from detector-material photon conversions,  $\lambda_{\text{Low }m_{\gamma^*}}$ for the contribution from virtual photons leading to $e^+e^-$ pairs, and $\lambda_{\text{HF }e}$ ($\lambda_{\text{HF }\mu }$) for the non-prompt electron (muon) background from heavy-flavour decays.

\begin{table}[htbp!]
\centering
\caption{Definition of the signal region and control regions used in the analysis. The first column shows the region name as used in the text. The event selection requirements are defined in the middle columns. The last column shows the fitted variable in each  region.  The variable  $m_{ee}^\text{CV}$ ($m_{ee}^\text{PV}$)  is defined as the invariant mass of the system formed by the track associated with the electron and the closest track at the conversion (primary) vertex. $N_{\text{j}}$ ($N_\text{b}$) indicates the jet ($b$-tagged jet) multiplicity. $H_\text{T}$ is defined as the scalar sum of the transverse momenta of the isolated leptons and jets.  The {\it baseline SR} is equal to the {\it BSM SR} + CR lowBDT.}

\begin{tabular}{c | c | c | c | c | c  }
\toprule
Region & Channel & $\nj $ & $\nb$ & Other selection requirements & Fitted variable\\
\toprule
\multirow{2}{*}{CR Conv} & \multirow{2}{*}{$e^\pm e^\pm$~$||$~$e^\pm \mu^\pm$} & \multirow{2}{*}{$4 \le $\nj $ < 6$} & \multirow{2}{*}{$\ge 1$} & $ \MeeConvV \in[0, 0.1]$~\GeV  &  \multirow{2}{*}{$\MeePV$} \\
& & & & $200 < \HT < 500$~\GeV & \\
\hline
CR HF $e$ & $eee$~$||$~$ee\mu$ & & $= 1$ & $100 < \HT < 250$~\GeV & Yield \\
\hline
CR HF $\mu$ & $e\mu\mu$~$||$~$\mu\mu\mu$ & & $= 1$ & $100 < \HT < 250$~\GeV  & Yield \\
\hline
\multirow{3}{*}{CR \ttW} & \multirow{3}{*}{$e^\pm\mu^\pm$~$||$~$\mu^\pm \mu^\pm$} & \multirow{3}{*}{$\ge 4$}  & \multirow{3}{*}{$\ge 2$} & $\MeeConvV \notin [0,0.1]$~\GeV , $\etae < 1.5$ & \multirow{3}{*}{$\sumptlep$} \\
& & & & for $\nb=2$, $\HT < 500$~\GeV~ or $\nj < 6$;  & \\
& & & &  for $\nb\ge 3$, $\HT < 500$~\GeV   & \\
\hline
CR lowBDT & SS+3L & $\ge 6$ & $\ge 2$ & $\HT>500$~\GeV , SM BDT < 0.55 & SM BDT\\
\toprule{\it BSM SR} & SS+3L & $\ge 6$ & $\ge 2$ & $\HT>500$~\GeV , SM BDT $\ge$ 0.55 & BSM pBDT \\
\toprule
\end{tabular}


\label{tab:RegionDefinition}
\end{table}

\subsubsection{\label{sub:chargeMisID}Electron charge misidentification background}
 
Background from electron charge misidentification is relevant only in the same-sign dilepton channel. It arises when the sign of the electric charge of one of the two  leptons in the selected same-sign pair is misreconstructed either because of  bremsstrahlung photon emission followed by its conversion ($e^\pm \rightarrow e^\pm\gamma \rightarrow e^\pm e^+e^-$) or due to mismeasured track curvature.  In the signal region, it mainly comes from $\ttbar$+jets production.   Due to the low probability of bremsstrahlung for muons and because of the large lever arm of the MS, the muon charge misidentification rate is very low.  Thus, this background is only relevant for the $e^\pm e^\pm$ and  $e^\pm \mu^\pm$ channels.

The probability for an electron to have its charge incorrectly reconstructed is measured in a data sample of dielectron events with invariant mass $m_{ee}$ within 10~\GeV\ of the $Z$ boson mass. The sideband method is used to subtract the background contamination. The electron charge misidentification probability  is calculated in bins of electron $|\eta|$ and $\pT$. For the conversion region defined in  Section~\ref{sub:fakesEstimation}, a parametrisation in bins of $m_{ee}^\text{PV}$ is also added.  A likelihood fit that adjusts these binned probabilities is used to find the best agreement with the observed numbers of same-charge and opposite-charge electron pairs. The  electron charge misidentification rates vary from 0.002\% for  low-$\pT$ electrons ($\pT\le 60$~\GeV) at $|\eta|\le 0.6$, to ${\sim}10\%$ for high-$\pT$ electrons ($\pT\ge 200$~\GeV) at $|\eta| \in [2.3, 2.5]$.
 
To estimate the event yields in the $e^\pm e^\pm$ and  $e^\pm \mu^\pm$  channels, $ee$ and $e\mu$ events are selected using all the criteria applied in the analysis, except that the leptons are required to have opposite charges. Then, the final background yield is obtained by weighting these opposite-sign dilepton events by the probability of one electron charge being misreconstructed.


\subsection{Signal discrimination}
\label{sec:signalDiscrimination}

 
Multivariate techniques are used to separate the signal from the SM backgrounds. This is done through two sequential BDT classifiers: the first one, the background rejection BDT, namely SM BDT, separates  SM $\fourtops$ events from other SM backgrounds. Then, the second one, the BSM mass-parameterised BDT (BSM pBDT) discriminates between BSM $\fourtops$ events and all background.  The BSM pBDT is parameterised as a function of the mass of the heavy Higgs boson  by introducing the mass as a labelled input in the training~\cite{Baldi:2016fzo}.   In the parametrisation, all signal and  background samples are further reweighted to have a uniform truth mass distribution in the training. Both the SM BDT and BSM pBDT are trained in the {\it baseline SR} with the XGBoost (Extreme Gradient Boosting) algorithm~\cite{2016XGBoost}. In both cases, the input variables are optimised to maximise the integral under the receiver operating characteristic (ROC) curve of each BDT.
 
The input variables for the SM BDT include those from the SM $\fourtops$ search~\cite{TOPQ-2018-05}, except that the highest-ranked variable, the sum of the pseudo-continuous $b$-tagging score,\footnote{The pseudo-continuous $b$-tagging score is an integer from 1 to 5 assigned to a jet, based on the operating point of the $b$-tagging algorithm it passes, with a value of 5 assigned to the jet most similar to a $b$-jet~\cite{FTAG-2018-01}.} is  not over all jets but instead only over the four jets with the highest score.  The jet multiplicity is also a new input variable for the discriminant, and it is introduced to further distinguish SM $\fourtops$ events from the other SM backgrounds. The distributions of these two variables are shown in Figure~\ref{fig:inputSMBDT}. Other input variables for the SM BDT are the minimum distance $\Delta R$ between two leptons among all possible pairs, the leading lepton $\pT$, $\Etmiss$, the $\pT$ of the leading and sub-leading jets, the $\pT$ of the sixth jet, the $\pT$ of the leading $b$-jet, the scalar sum of the transverse momenta over all leptons and jets except the leading $\pT$ jet, the sum of distances $\Delta R$ between two leptons for all possible pairs, and the minimum distance $\Delta R$ between a jet and a $b$-jet among all possible pairs.   The SM BDT score is used to split the {\it baseline SR}:  the high SM BDT region defines the  {\it BSM SR} (SM~BDT~$\ge 0.55$), while the low part defines the CR lowBDT (SM~BDT~<~0.55).   The SM BDT is used as an input variable to build the BSM pBDT and as the fitted variable in the CR lowBDT.

\begin{figure}[htbp!]
\centering
\subfloat[]{
\includegraphics[width=0.49\linewidth]{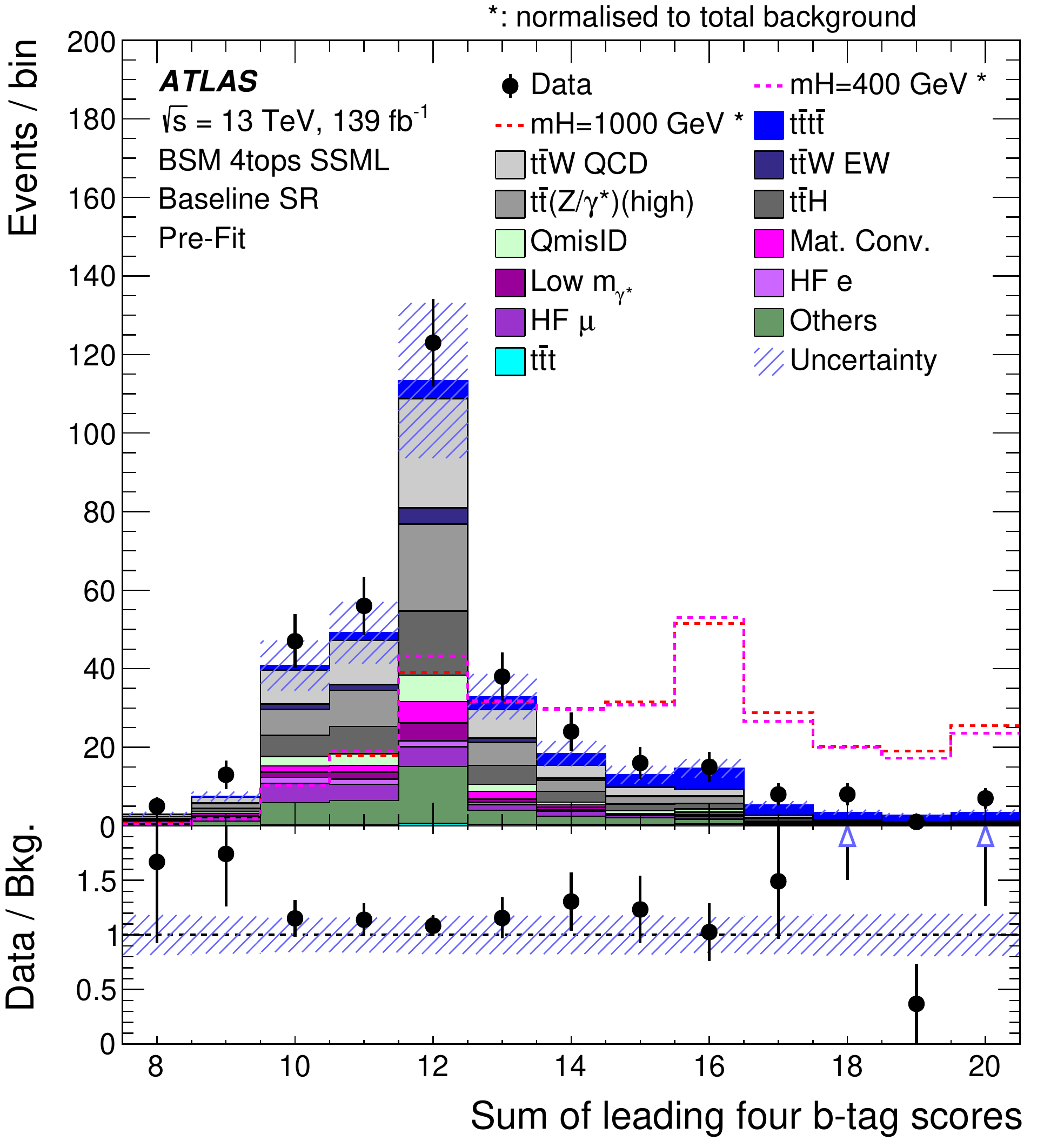}
}
\subfloat[]{
\includegraphics[width=0.49\linewidth]{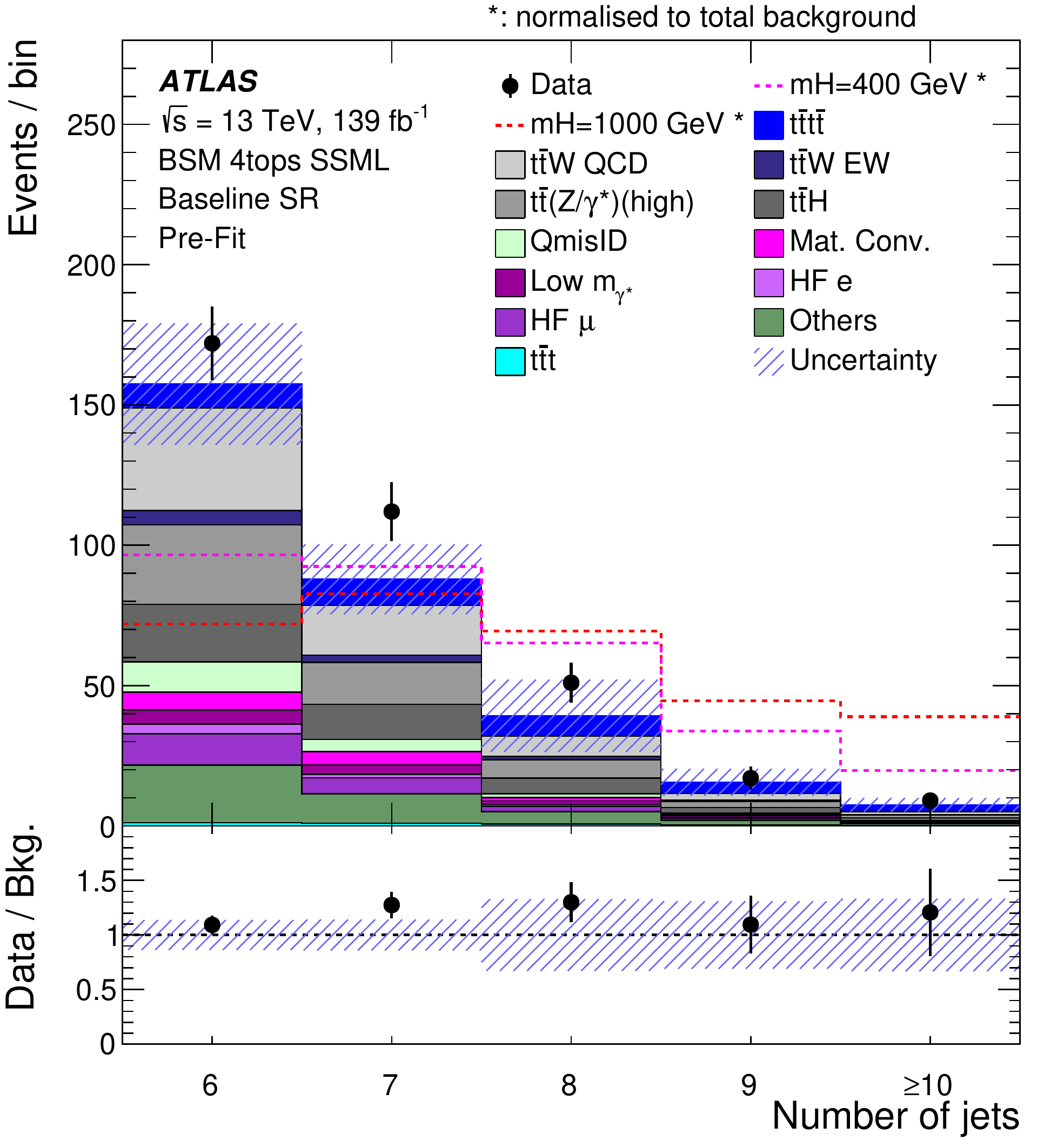}
}
\caption{Pre-fit comparison between data and background in the {\it baseline SR} for two of the  variables used as input for the SM BDT:  (a)~the sum of the leading four jets' $b$-tagging scores and (b)~the number of jets. The signal samples corresponding to $m_{H}=400$~\GeV\ and $1000$~\GeV\ are also shown.  The signal distributions are normalised to the total background to compare the shapes. The dashed band includes the total background uncertainty. The first and last bins contain underflow and overflow events, respectively. Upward arrows indicate that the value is out of the plotted range of the $y$-axis.}
\label{fig:inputSMBDT}
\end{figure}

The BSM pBDT is used in the {\it BSM SR}, but because there are too few events it is trained in the {\it baseline SR}. Since the {\it BSM SR} is $\fourtops$ enriched, the training is done after reweighting the SM $\fourtops$ events to mimic their fraction of the total background in the {\it BSM SR}.  The other backgrounds are reweighted so that they preserve the total background yield.  The SM BDT score is the most discriminating input variable for the BSM pBDT. The second most discriminating is $H_\text{T}$. The distributions of these two variables in the {\it baseline SR} are shown in Figure~\ref{fig:inputpBDT}.  Other variables used as input are the event shape variable associated with hadronic activity (sphericity)~\cite{PhysRevD.1.1416},  the sphericity in the transverse plane, the minimum distance $\Delta R$ between two leptons among all possible pairs,  the sum of distances $\Delta R$ between two leptons for all possible pairs, and the $\Etmiss$ of the event.
The BSM pBDT output score is used as the fitted variable in the {\it BSM SR.}

\begin{figure}[htbp!]
\centering
\subfloat[]{
\includegraphics[width=0.49\linewidth]{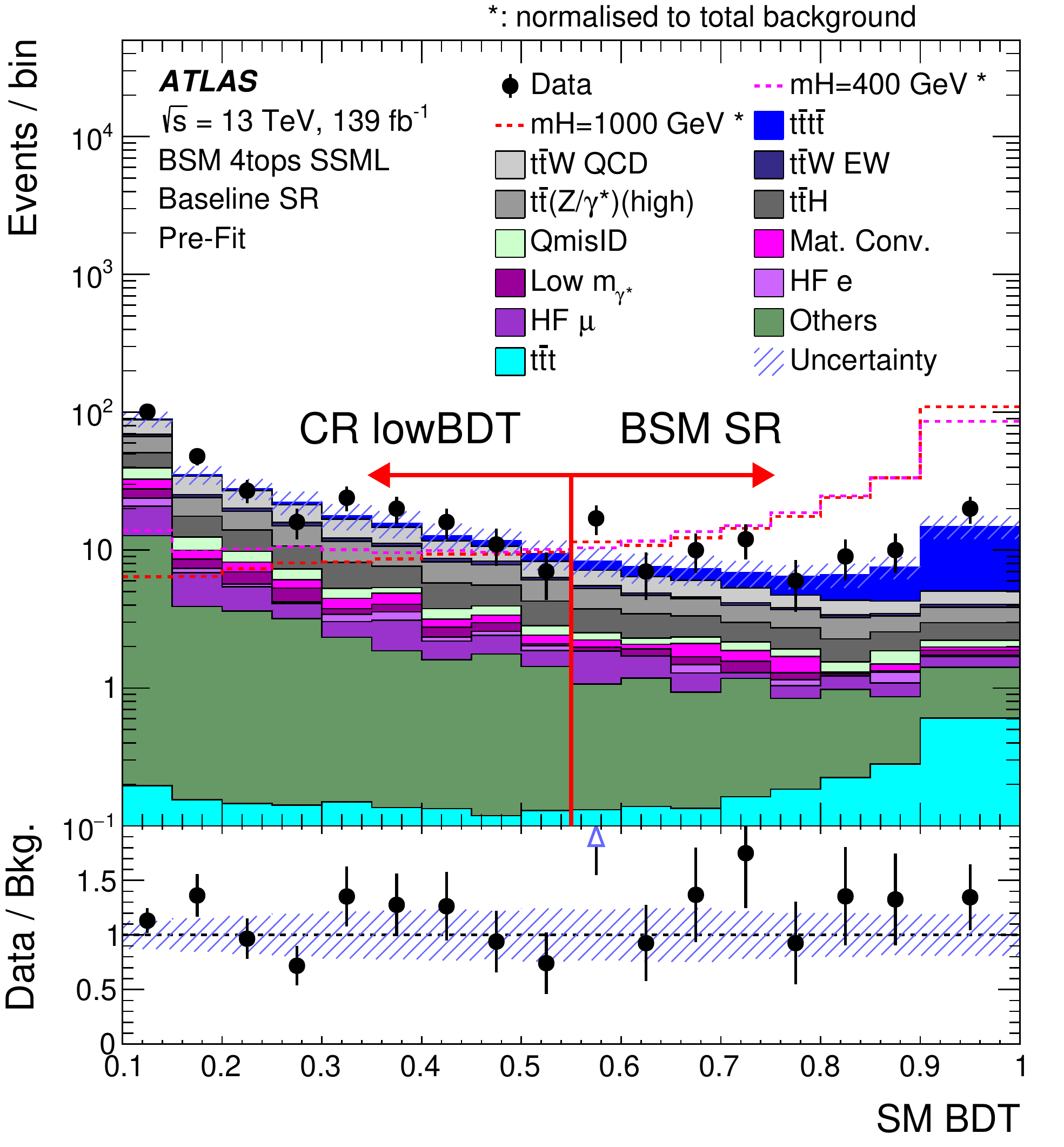}
}
\subfloat[]{
\includegraphics[width=0.49\linewidth]{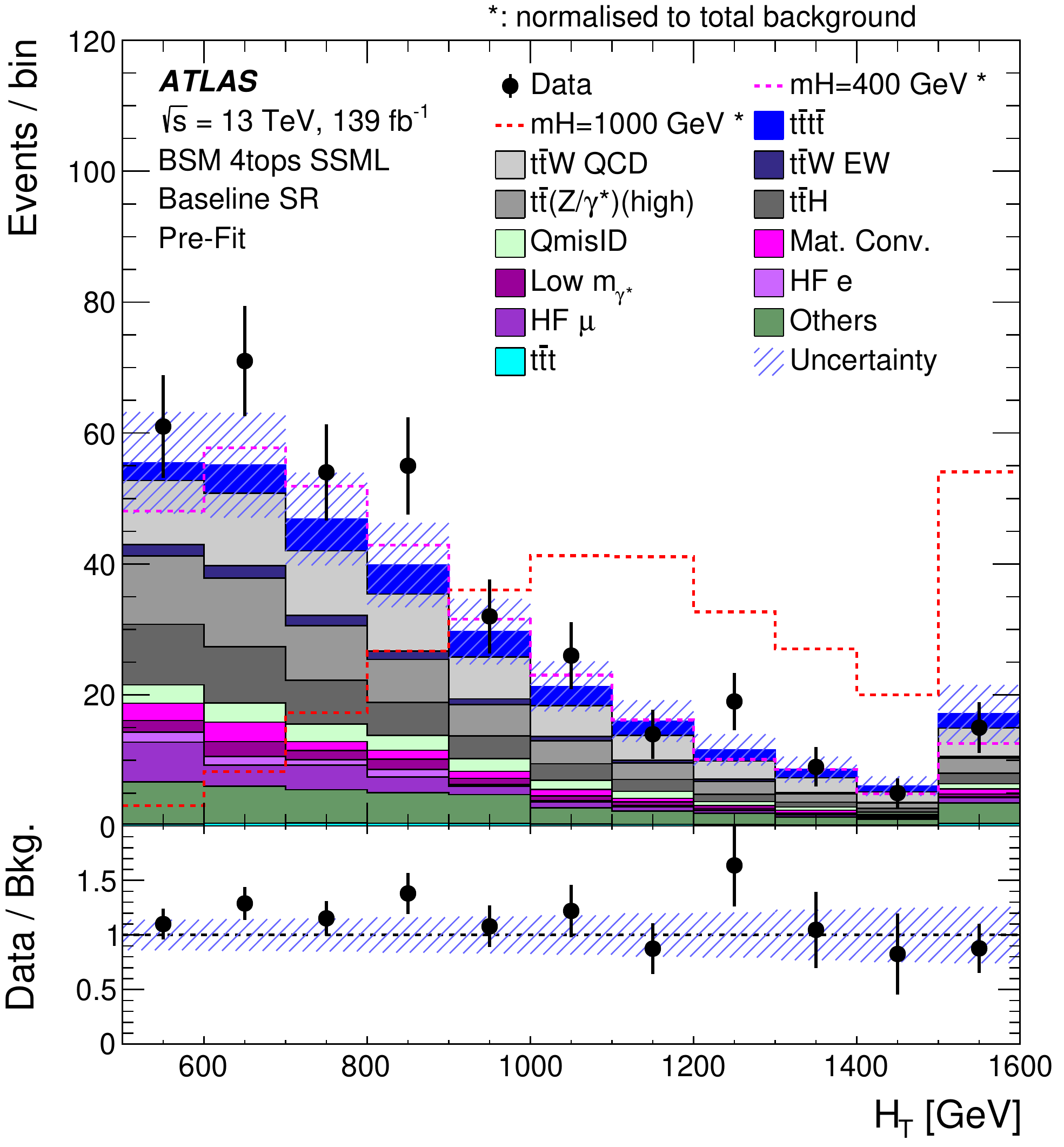}
}
\caption{Pre-fit comparison between data and background in the {\it baseline SR} for two of the variables used as input for the BSM pBDT:  (a)~the SM BDT score and (b) $H_\text{T}$.  The signal samples corresponding to $m_{H}=400$~\GeV\ and $1000$~\GeV\ are also shown. The signal distributions are normalised to the total background to compare the shapes. The dashed band includes the total background uncertainty. The first and last bins contain underflow and overflow events, respectively. Upward arrows indicate that the value is out of the plotted range of the $y$-axis.}
\label{fig:inputpBDT}
\end{figure}


\section{Systematic uncertainties}
\label{sec:sistematic}

Systematic uncertainties arise from the reconstruction of the various physics objects and
from theoretical and/or modelling uncertainties affecting the predictions for both the background and signal processes. They can be classified as either experimental uncertainties or modelling uncertainties of the signal,  irreducible, and reducible backgrounds.  These uncertainties manifest themselves as uncertainties in both the overall yield and
shape of the final observable, and they are treated as fully correlated across all fit regions.
 
\subsection{Experimental uncertainties}
 
The uncertainty in the integrated luminosity of the combined 2015--2018 dataset is 1.7\%~\cite{ATLAS-CONF-2019-021}, obtained using the LUCID-2 detector~\cite{LUCID2} for the primary luminosity measurements. The uncertainties related to the reweighting factors that correct the pile-up profile in simulations to match the one in data are also included.
 
Uncertainties in the modelling of leptons arise from their  momentum resolution and scale, as well as the trigger, reconstruction, identification and isolation efficiencies  used to correct for the difference between the simulation and data~\cite{TRIG-2018-05,TRIG-2018-01,EGAM-2018-01,MUON-2018-03}.
 
Uncertainties in the modelling of jets come from their energy scale  and resolution, containing  the effect of jet flavour composition, single-particle response, and pile-up~\cite{JETM-2018-05}. The uncertainty in the efficiency to pass the JVT requirement for pile-up suppression is also applied~\cite{PERF-2014-03}.  Uncertainties from the calibration of the $b$-tagging efficiencies, including the efficiencies of tagging $b$-jets as well as the rates of mis-tagging $c$-jets and light-flavour jets are also considered, and decomposed respectively into 45, 20 and 20 independent eigenvectors~\cite{ATL-PHYS-PUB-2017-013}.
 
The   uncertainty in \MET arising from a possible mis-calibration of its soft-track component is also included~\cite{PERF-2016-07}.

\subsection{Uncertainties in modelling the signal and irreducible background}
 
Uncertainties in modelling the signal arise from the PDF and from missing higher-order QCD corrections. The former is evaluated following the PDF4LHC recommendations~\cite{Butterworth:2015oua}, while the latter is estimated by varying the renormalisation and factorisation scales simultaneously by factors of 2.0 or 0.5 relative to the central value.

Several sources of uncertainty are considered for the background coming from  \tttt\  production. The uncertainties related to missing higher-order QCD corrections and  to the PDF are evaluated in the same way as for the signal. The uncertainty associated with the choice of parton shower and hadronisation model  is estimated by comparing the nominal prediction  with the alternative sample generated with \MGNLO matched to \Herwig instead of \Pythia. The uncertainty related to the choice of generator is obtained by comparing the nominal sample with the one generated with \Sherpa[2.2.10]. The uncertainty associated with the lack of EW corrections in the nominal sample affects only the shape of the distributions, and it is evaluated by comparing the  \Sherpa[2.2.11] (QCD-only) and  \Sherpa[2.2.11] (QCD+EW)\footnote{EW corrections are included in the \Sherpa[2.2.11] sample as additional weights. Thus, when using the QCD-only version the EW weights are not used.} simulated samples.  A separate 20\% uncertainty computed at NLO in QCD including EW corrections~\cite{Frederix:2017wme} is applied to the total cross-section.

Several sources of uncertainty affect the modelling of $\ttW$, $\ttZ$, and $t\bar tH$ production. Uncertainties associated with the generator are estimated by comparing the predictions from the nominal samples with those from alternative samples generated with \MGNLO as described in Section~\ref{sec:samples}.  The uncertainties due to missing higher-order QCD corrections are evaluated in the same way as for the signal.  An uncertainty of~12\%~(10\%)  is applied to the $\ttZ$~($t\bar tH$) total cross-section~\cite{deFlorian:2016spz}.  A  1\% uncertainty from the PDF, estimated following the PDF4LHC prescription~\cite{Butterworth:2015oua},  is applied to both the $\ttZ$ and $t\bar tH$ processes. No cross-section uncertainty is assigned to $\ttW$  QCD  production because its normalisation is estimated in the fit (see Section~\ref{sec:bkgEstimation}).  An uncertainty of~20\%~\cite{PhysRevD.103.094014} is applied to the  $\ttW$ EW cross-section. In order to cover the difference between data and prediction in the $\ttW$ validation region defined in  Section~\ref{sec:result}, where a data excess is observed for high jet multiplicities, an additional 133\% (208\%) uncertainty is assigned to $\ttW$ QCD production with seven (eight or more) jets\footnote{No uncertainty is assigned for events with less than seven jets, as good agreement between data and prediction is observed.}. Events arising from $\ttW$, $\ttZ$, and $t\bar tH$ production can have additional heavy-flavour jets. As these processes are difficult to model in the simulation, an uncertainty of 50\% is assigned to the events with an additional true $b$-jet and a separate 50\% uncertainty is assigned to the events with two or more additional true $b$-jets. Those estimates are taken from Ref.~\cite{TOPQ-2018-05}, and treated as uncorrelated between the three different processes because of the different set-ups used to simulate the events.

The uncertainty in the $\ttt$ production  cross-section is set to 100\% as done in Ref.~\cite{TOPQ-2018-05}. Following the same motivation given earlier about the uncertainties for processes with additional heavy-flavour jets for $t\bar tV$ samples, an extra uncertainty of 50\% is applied to $\ttt$  events with at least one additional true $b$-jet.

An uncertainty of 30\% is applied to the cross-sections of the $tZ$  and $tWZ$ single-top-quark processes~\cite{TOPQ-2016-14,Demartin:2016axk}. The uncertainty in the diboson cross-section is set to 40\%  following studies of $WZ{+}b$ production. For the rare top-quark processes, i.e.\ \ttWW, \ttZZ, \ttWZ, \ttHH, and \ttWH, an uncertainty of 50\% is applied to the cross-section~\cite{EXOT-2016-16}.   For the remaining small backgrounds, an uncertainty of 50\% is assigned to the cross-section.  An additional uncertainty of 50\% is applied to all processes described in this paragraph. This uncertainty is applied to events with one additional true $b$-jet, and  separately to events with at least two additional true $b$-jets.
 
\subsection{Uncertainties in modelling the reducible background}
 
The uncertainties in the background due to charge misidentification arise from the uncertainties in the measurement of the electron charge misidentification rates. The following contributions are considered:  the statistical uncertainty from the likelihood fit to data, the difference between the rates extracted in simulated events with the likelihood method and the true rates, and  the changes in the measured rates when the size of the sideband regions used to subtract the background from the $Z$-peak is varied from 8~\GeV\ to 12~\GeV.
 
The uncertainties in the detector-material photon conversion and Low $m_{\gamma^*}$ backgrounds only come from the shape of the distributions used in the template fit. No uncertainty is assigned to the normalisation of these processes since the overall normalisations of these backgrounds are estimated in the fit.  The shape uncertainty is estimated in a region enriched in $Z(\rightarrow \mu\mu){+}\gamma$ events by comparing data with the \POWPY[8] simulation of $Z(\rightarrow \mu\mu){+}\gamma/$jets production.  The uncertainty is supposed to cover the extrapolation from the `CR Conv', i.e.\ events with $\MeeConvV \in [0,0.1]$~\GeV, to the regions with events with larger \MeeConvV.  The uncertainty was estimated to be 50\% (100\%) for the detector-material photon conversion (Low $m_{\gamma^*}$) background, and it is applied to events with  $\MeeConvV> 0.1$~\GeV\ in all control and signal regions.
 
Similarly to the detector-material photon conversion and Low $m_{\gamma^*}$ backgrounds, the only uncertainty in the heavy-flavour non-prompt lepton background  comes from the shape of the distributions used in the fit. This uncertainty is estimated by a bin-by-bin comparison of the data and the post-fit background prediction in every region used in the analysis, after enhancing the contribution of non-prompt leptons. This is done by using  a loose lepton selection with the isolation requirements dropped and the identification criteria relaxed. The shape uncertainty is treated as correlated among all regions. It is derived separately for electrons and muons.
 
The uncertainty assigned to the normalisation of the background coming from light-flavour non-prompt leptons is 100\%.  This value was derived from the level of agreement between data and simulation in a region enriched in light-flavour non-prompt leptons. For the remaining sources of fake/non-prompt leptons, an uncertainty of 30\% is applied as done in Ref.~\cite{TOPQ-2018-05}.
 
Other uncertainties affecting the shape of the distributions of the background coming from fake/non-prompt leptons arise from the modelling of the heavy-flavour content of the main source of reducible background, $\ttbar$+jets production.  Based on the measurement of $\ttbar$ production with additional heavy-flavour jets~\cite{TOPQ-2017-12}, an uncertainty of 30\% is assigned to events  with three true $b$-jets, and a separate  30\% uncertainty is assigned to events with at least four true $b$-jets.


\section{Statistical analysis}
\label{sec:statistics}

The presence of a signal is tested for by fitting the BSM pBDT score in the {\it BSM SR} jointly with the control regions as described in Section~\ref{sec:strategy}.
The statistical analysis uses a binned likelihood function ${\cal L}(\mu,\boldsymbol{\theta},\boldsymbol{\lambda})$ constructed as
a product of Poisson probability terms over all bins considered in the search. This function depends on the following:
the signal-strength parameter $\mu$, defined as a factor multiplying the expected yield of \ttHA\ signal events; $\boldsymbol{\theta}$, a set of nuisance parameters that encode the effect of systematic uncertainties on the signal and background expectations; and $\boldsymbol{\lambda}$, a set of multiplicative factors that normalise the background expectations.
All nuisance parameters are subject to Gaussian constraints in the likelihood.
Therefore, the expected total number of events in a given bin depends on $\mu$, $\boldsymbol{\theta}$ and $\boldsymbol{\lambda}$.
 
For a given value of $\mu$, the nuisance parameters $\boldsymbol{\theta}$ and the normalisation factors $\boldsymbol{\lambda}$ allow variations of the expectations for signal and background
according to the corresponding systematic uncertainties, and their fitted values result in the deviations from
the nominal expectations that globally provide the best fit to the data.
This procedure allows a reduction of the impact of systematic uncertainties on
the search sensitivity by taking advantage of the highly populated background-dominated bins included in the likelihood fit.
Simulation statistical uncertainties in each bin are taken into account by dedicated parameters in the fit, which are modelled with Poisson constraints.
The best-fit branching fraction is obtained by performing a binned likelihood fit to the data under the signal-plus-background
hypothesis, maximising the likelihood function ${\cal L}(\mu,\boldsymbol{\theta}, \boldsymbol{\lambda})$ over $\mu$, $\boldsymbol{\theta}$ and $\boldsymbol{\lambda}$.

The test statistic $q_\mu$ is defined as the profile likelihood ratio,
$q_\mu = -2\ln({\cal L}(\mu, {\hat{\boldsymbol{\theta}}}_\mu, {\hat{\boldsymbol{\lambda}}}_\mu)/{\cal L}(\hat{\mu},\hat{\boldsymbol{\theta}}_{\hat{\mu}}, \hat{\boldsymbol{\lambda}}_{\hat{\mu}}) )$,
where $\hat{\mu}$, $\hat{\boldsymbol{\theta}}_{\hat{\mu}}$ and $\hat{\boldsymbol{\lambda}}_{\hat{\mu}}$ are the values of the parameters that
maximise the likelihood function (subject to the constraint $0\leq \hat{\mu} \leq \mu$); ${\hat{\boldsymbol{\theta}}}_\mu$ and ${\hat{\boldsymbol{\lambda}}}_\mu$ are respectively the values of the
nuisance parameters and normalisation factors that maximise the likelihood function for a given value of $\mu$.
The test statistic $q_\mu$ is evaluated with the RooFit package~\cite{Verkerke:2003ir,RooFitManual}.
A related test statistic is used to determine whether the observed data is compatible with the background-only hypothesis 
by setting $\mu=0$ in the profile likelihood ratio and leaving $\hat{\mu}$ unconstrained:
$q_0 = -2\ln({\cal L}(0,{\hat{\boldsymbol{\theta}}}_0, {\hat{\boldsymbol{\lambda}}}_0 )/{\cal L}(\hat{\mu},\hat{\boldsymbol{\theta}}, \hat{\boldsymbol{\lambda}}))$.
The $p$-value 
representing the level of agreement between the data and the background-only hypothesis is estimated by integrating
the distribution of $q_0$, based on the asymptotic formulae in Ref.~\cite{Cowan:2010js}, above the $q_0$ value observed in the data.
Upper limits on $\mu$ are derived by using $q_\mu$ in the CL$_{\textrm{s}}$ method~\cite{Junk:1999kv,Read:2002hq}.
For a given signal scenario, values of $\mu$ yielding CL$_{\textrm{s}} < 0.05$,
where CL$_{\textrm{s}}$ is computed using the asymptotic approximation~\cite{Cowan:2010js}, are excluded at ${\geq}95\%$ confidence level (CL).


\section{Results and interpretation}
\label{sec:result}

The signal strength for every signal hypothesis and the normalisation factors for the background processes described in Section~\ref{sub:fakesEstimation} are determined via a binned likelihood fit performed simultaneously in all signal and control regions defined in Table~\ref{tab:RegionDefinition}. 
 
For the {\it BSM SR},  the binning of the BSM pBDT distribution is optimised for every signal hypothesis to provide the best discrimination between the tested signal  and the background, avoiding the presence of bins with no contribution from the major backgrounds.  The results obtained for the background-only fit shown in this section are the ones obtained when the fitted distribution in the signal region is the BSM pBDT score assigned to the background events when the signal hypothesis is $\mH=400$~\GeV. Similar  results are obtained for other signal hypotheses.

The fit was initially validated through extensive studies using background-only fits to real data where bins of the SM BDT score with a signal contamination above 5\% were excluded (referred to as blinding requirements). The robustness of the model for systematic uncertainties is established by verifying the stability of the fitted background when varying assumptions about some of the leading sources of uncertainty. After this, the blinding requirements on the SM BDT are removed in the data and a fit under the signal-plus-background hypothesis is performed to all analysis regions.  Further checks involve the comparison of the fitted nuisance parameters before and after removal of the blinding requirements, and their values are found to be consistent confirming that the signal-rich analysis bins do not contribute to the background estimation. The modelling of a number of kinematic variables in all considered analysis regions is also studied before and after the fit, confirming that the fit consistently improves the modelling of  all of them.

The normalisation factors from the background-only fit to data for the different background processes are shown in Table~\ref{tab:NFs}.  The data and  post-fit background comparison for the distributions of the discriminating variables fitted in the control regions is shown in Figure~\ref{fig:BSM_dis_CR_mH400_0}. Good agreement between data and post-fit background is observed.  Pre-fit and post-fit  background yields in the {\it BSM SR} are shown in Table~\ref{tab:postFitYields_400}. The signal yields corresponding to the pre-fit estimate in a 2HDM with $\tan\beta = 1$ are shown for two signal hypotheses. The total post-fit background yield is less than one standard deviation from the SM prediction. For the $\ttW$ QCD background, the large difference between the pre-fit and post-fit yields arises not only from the change in its normalisation factor $\lambda_{\ttW \text{QCD}}$, but also from several nuisance parameters associated with this process,
which are found to be  pulled from their nominal values in the fit.
These are the  ones related  to the generator, the scale variations, and the systematic uncertainty assigned to events with seven, eight or more jets. The largest single impact is provided by the latter, with a contribution of 32\% to the $\ttW$ QCD yield.
The measured normalisation factor $\lambda_{\ttW \text{QCD}}$ is found to agree with that in Ref.~\cite{TOPQ-2016-11}.
No other nuisance parameters were found to be significantly pulled or constrained in the fit.

\begin{table}[htp!]
 
\begin{center}
\caption{ Normalisation factors for the different background processes obtained from the background-only fit. The fit is  done simultaneously in all signal and control regions. The fitted distribution in the {\it BSM SR} is the BSM pBDT score assigned to the background events when the signal hypothesis is $m_{H}=400$~\GeV.  The uncertainties include both statistical and systematic uncertainties.}
 
\begin{tabular}{c|ccccc}
\toprule
Parameter & $\lambda_{\ttW \text{QCD}}$  &$\lambda_{\text{Mat.\ Conv.}}$  & $\lambda_{\text{Low }m_{\gamma^*}}$ & $\lambda_{\text{HF }e}$ & $\lambda_{\text{HF }\mu }$ \\
\hline
Value        & $1.3\pm 0.3$                   & $1.5\pm 0.5$               &  $0.6\pm 0.5$     & $0.9\pm 0.4$              & $1.0\pm 0.2$  \\
\toprule
\end{tabular}
\label{tab:NFs}
\end{center}
\end{table}

\begin{figure}[htbp!]
\centering
 
\subfloat[CR HF $e$]{\includegraphics[width=0.34\linewidth]{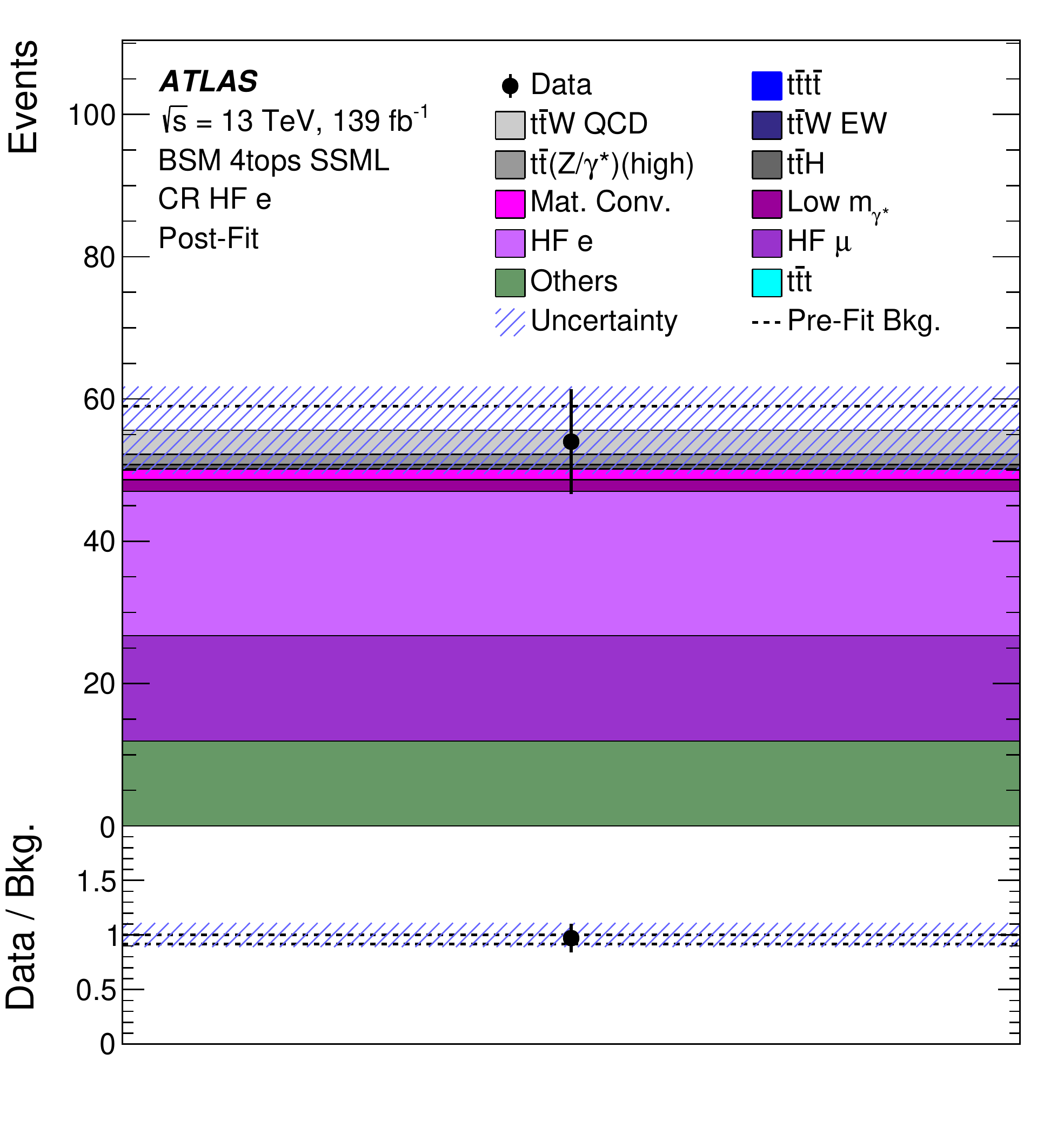}}
\subfloat[CR HF $\mu$]{\includegraphics[width=0.34\linewidth]{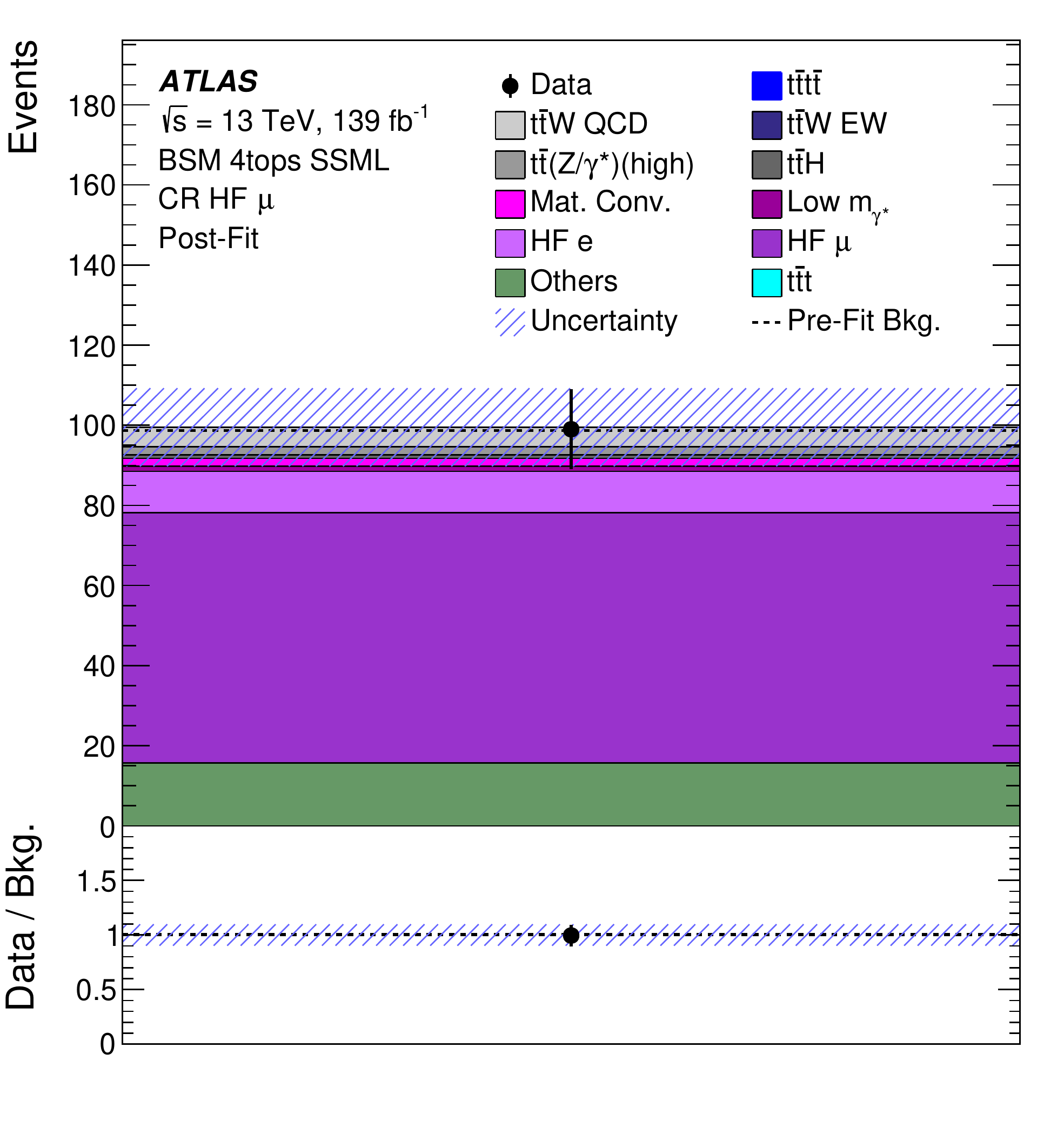}}
 
\subfloat[CR Conv]{\includegraphics[width=0.34\linewidth]{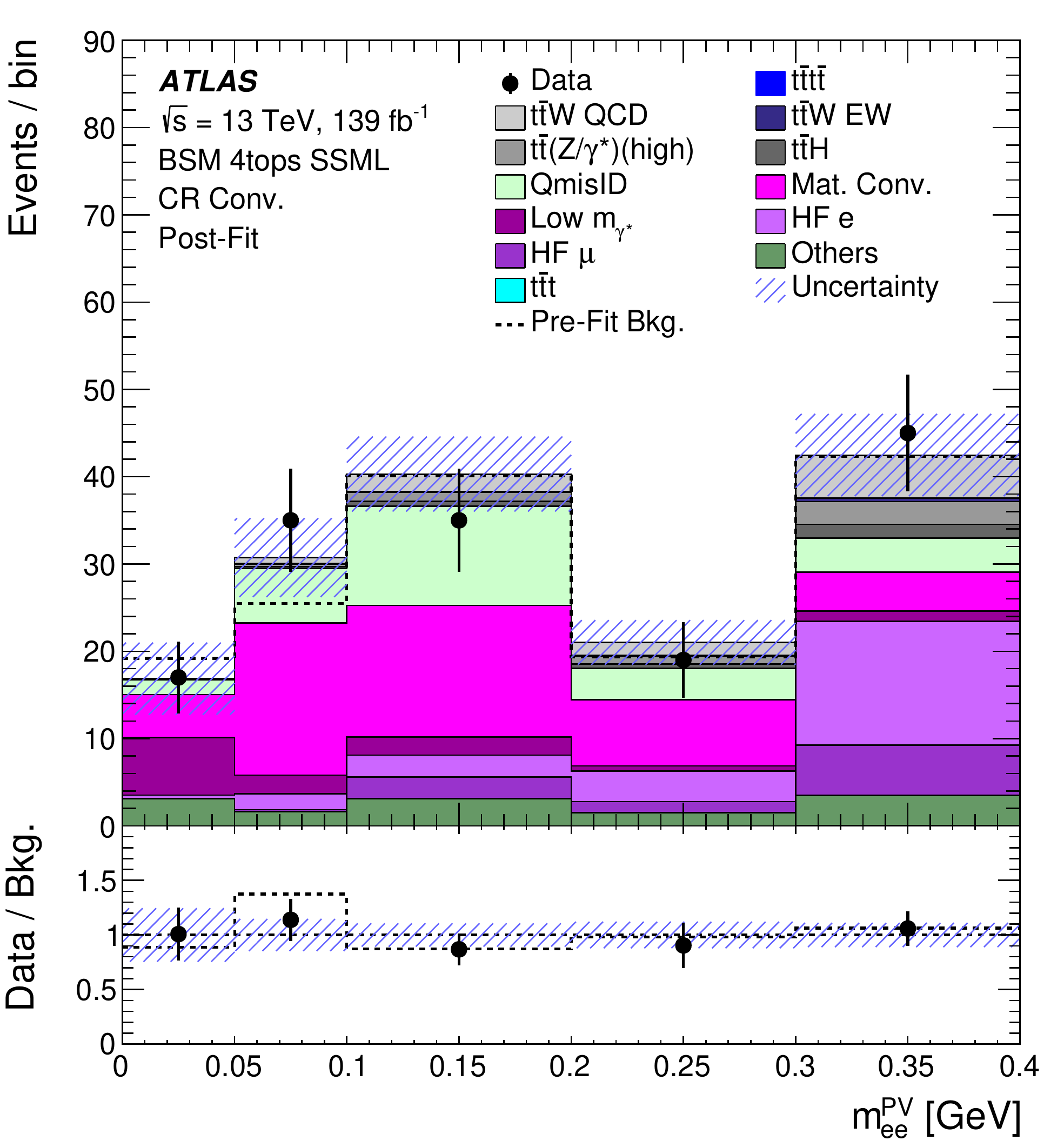}}
\subfloat[CR $t\bar{t}W$]{\includegraphics[width=0.34\linewidth]{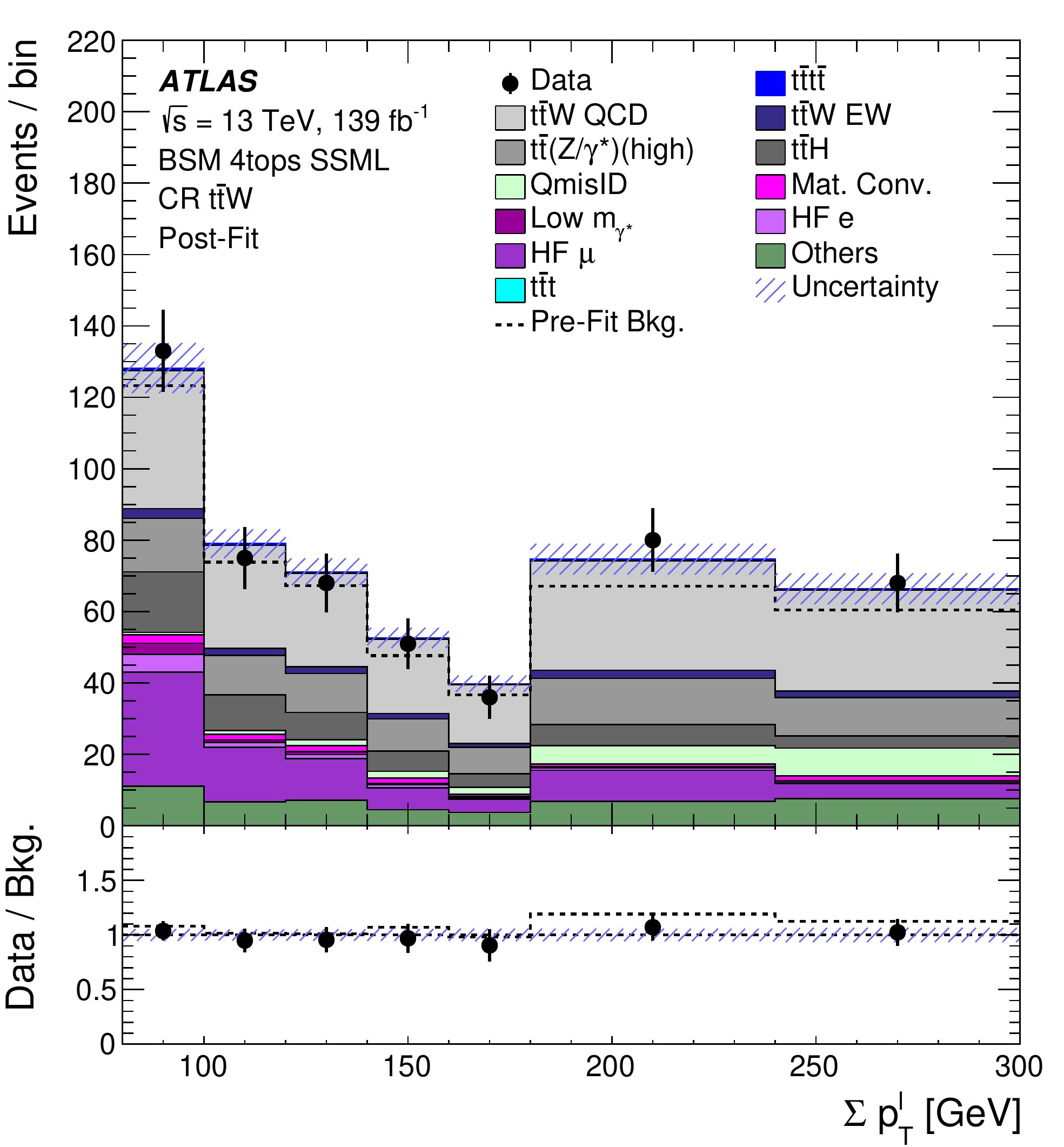}	}

\subfloat[CR lowBDT]{\includegraphics[width=0.34\linewidth]{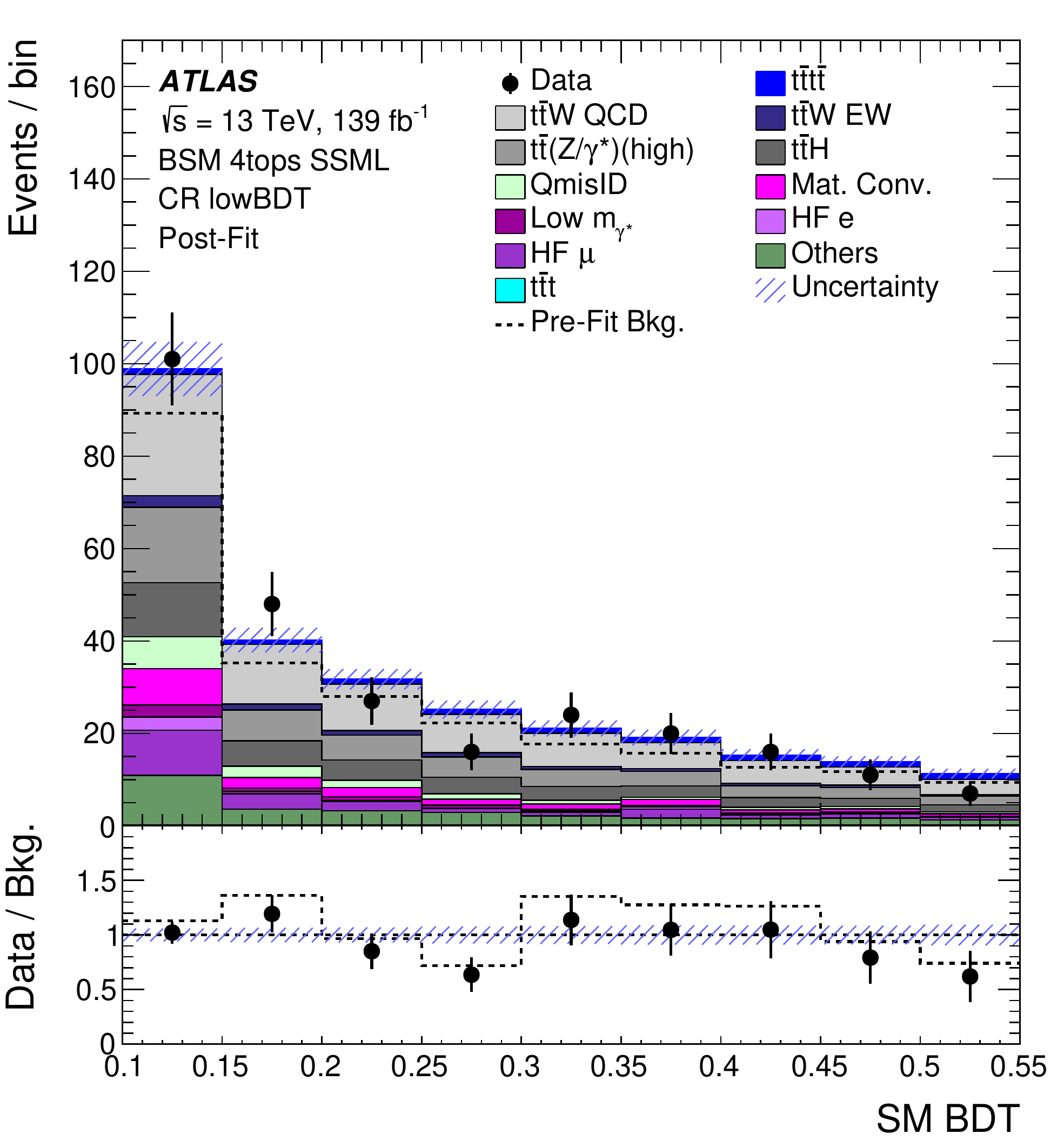}}
\caption{Data and  post-fit background comparison for the distributions of the discriminating variables fitted in the control regions obtained with the background-only fit. The fit is done simultaneously in all signal and control regions. The fitted distribution in the  {\it BSM SR} is the BSM pBDT score assigned to the background events when the signal hypothesis is  $m_{H}=400$~\GeV.  The band includes the total uncertainty of the post-fit estimate. The total background prediction before the likelihood fit to data (`Pre-Fit Bkg.') is shown as a dashed black line. The ratio of the data to the background prediction is shown in the lower panel, separately for post-fit background  (black points) and pre-fit background (dashed black line). The first and last bins of (c), (d) and (e) contain underflow and overflow events, respectively. }
\label{fig:BSM_dis_CR_mH400_0}
\end{figure}

\begin{table}[htp!]
\caption{Pre-fit and post-fit background yields in the {\it BSM SR}  obtained with the  background-only fit. The fit is done simultaneously in all signal and control regions. The fitted distribution in the  {\it BSM SR} is the BSM pBDT score assigned to the background events when the signal hypothesis is $m_{H}=400$~\GeV.  The signal yields correspond to the pre-fit estimate in the type-II 2HDM for $\tan\beta = 1$ and  for  $m_{H}=400$~\GeV\ and 1000~\GeV, assuming only the production of a new scalar $H$ boson.  The number of data events is also shown. The total systematic uncertainty differs from the sum in quadrature of the different uncertainties because of correlations and anticorrelations.}
\begin{center}
 
\begin{tabular}{l|C{3cm}|C{3cm}}
 
\toprule
Process & {Pre-fit}  & {Post-fit}
\tabularnewline
\toprule
\ensuremath{  t\bar{t}t\bar{t}   } &   \hspace{-0.07in}\ensuremath{ 22.3 \pm 5.3 } & \hspace{-0.07in}\ensuremath{ 25.8 \pm 5.3 } \tabularnewline
$t\bar{t}W$ QCD                                                  &  \ensuremath{ 9.4 \pm 9.6 }   & \hspace{-0.07in}\ensuremath{ 17.2 \pm 6.9 } \tabularnewline
$t\bar{t}W$ EW  						   &  \ensuremath{ 1.3 \pm 0.5 }  &\ensuremath{ 1.4 \pm 0.6 } \tabularnewline
$t\bar{t}WW$     						 &  \ensuremath{ 1.8 \pm 1.0 }    &\ensuremath{ 1.9 \pm 1.0 } \tabularnewline
$t\bar{t}(Z/\gamma^*)$ (high mass) 		 &  \ensuremath{ 8.5 \pm 2.1 }  &\ensuremath{ 9.2 \pm 2.2 } \tabularnewline
$t\bar{t}H$   							&  \ensuremath{ 7.2 \pm 1.7 } &\ensuremath{ 7.8 \pm 1.7 } \tabularnewline
QmisID      							   &  \ensuremath{ 2.1 \pm 0.2 }  &\ensuremath{ 2.1 \pm 0.2 } \tabularnewline
Mat.\ Conv.  							  &    \ensuremath{ 1.8 \pm 0.6 } &\ensuremath{ 3.0 \pm 1.2 } \tabularnewline
$\text{Low }m_{\gamma^*}$   				&  \ensuremath{ 1.2 \pm 0.6 }   &\ensuremath{ 0.8 \pm 0.8 } \tabularnewline
HF $e$             						&    \ensuremath{ 0.6 \pm 0.7 } &\ensuremath{ 0.6 \pm 0.7 } \tabularnewline
HF $\mu$      							   &    \ensuremath{ 2.7 \pm 1.1 } &\ensuremath{ 2.9 \pm 1.2 } \tabularnewline
LF                   							 &  \ensuremath{ 1.1 \pm 1.2 }  &\ensuremath{ 0.4 \pm 1.0 } \tabularnewline
Other fake  							  &  \ensuremath{ 1.1 \pm 0.7 }  &\ensuremath{ 1.3 \pm 0.7 } \tabularnewline
$tZ$, $tWZ$  							&  \ensuremath{ 0.9 \pm 0.3 }  &\ensuremath{ 0.9 \pm 0.3 } \tabularnewline
$VV$, $VH$, $VVV$  					&  \ensuremath{ 0.3\pm 0.1 }   &\ensuremath{ 0.3 \pm 0.1 } \tabularnewline
$t\bar tt$                   				      &\ensuremath{ 1.9 \pm 1.9 }   &\ensuremath{ 2.3 \pm 2.1 } \tabularnewline
$t\bar t WZ$, $t\bar t ZZ$, $t\bar t WH$, $t\bar t HH$   &\ensuremath{ 1.3 \pm 0.7 }   &\ensuremath{ 1.4 \pm 0.8 } \tabularnewline
\toprule
Total background   						&\ensuremath{ 65.6 \pm 13.4 } & \hspace{-0.07in}\ensuremath{ 79.5 \pm 6.7 } \tabularnewline
\toprule
$\ttbar H(\rightarrow \ttbar )$,  $\mH=400$~\GeV 	& \hspace{-0.08in}\ensuremath{ 38.6 \pm 2.4 } & -- \tabularnewline
$\ttbar H(\rightarrow \ttbar )$,  $\mH=1000$~\GeV &   \hspace{-0.01in}\ensuremath{ 4.4 \pm 0.2 } &  --  \tabularnewline
\toprule
Data    &   \multicolumn{2}{c}{91}
 
\tabularnewline
\toprule
\end{tabular}
\label{tab:postFitYields_400}
\end{center}
\end{table}

The modelling of the  $\ttW$ QCD background was validated by using the charge asymmetry of $\ttW$ production.
The $\ttW$  validation region is defined to be a looser version than the {\it baseline SR}, with no requirement on $H_\text{T}$, and with the cut on the number of jets relaxed to four.  The charges of selected leptons are examined for this selection, and the  difference between the number of events with a positive sum
of lepton charges and the number of events with a negative sum is calculated. The jet multiplicity  and SM BDT distributions of this difference are shown in Figure~\ref{fig:VRttW}.  A residual disagreement is observed  between the data and the post-fit background at high jet multiplicities due to the constrained nuisance parameters mentioned above, and to the BSM pBDT not being sensitive to the difference of lepton charges as a function of numbers of jets. However, a good agreement is observed in the SM BDT distribution.

\begin{figure}[htbp!]
\centering
\subfloat[]{\includegraphics[width=0.5\linewidth]{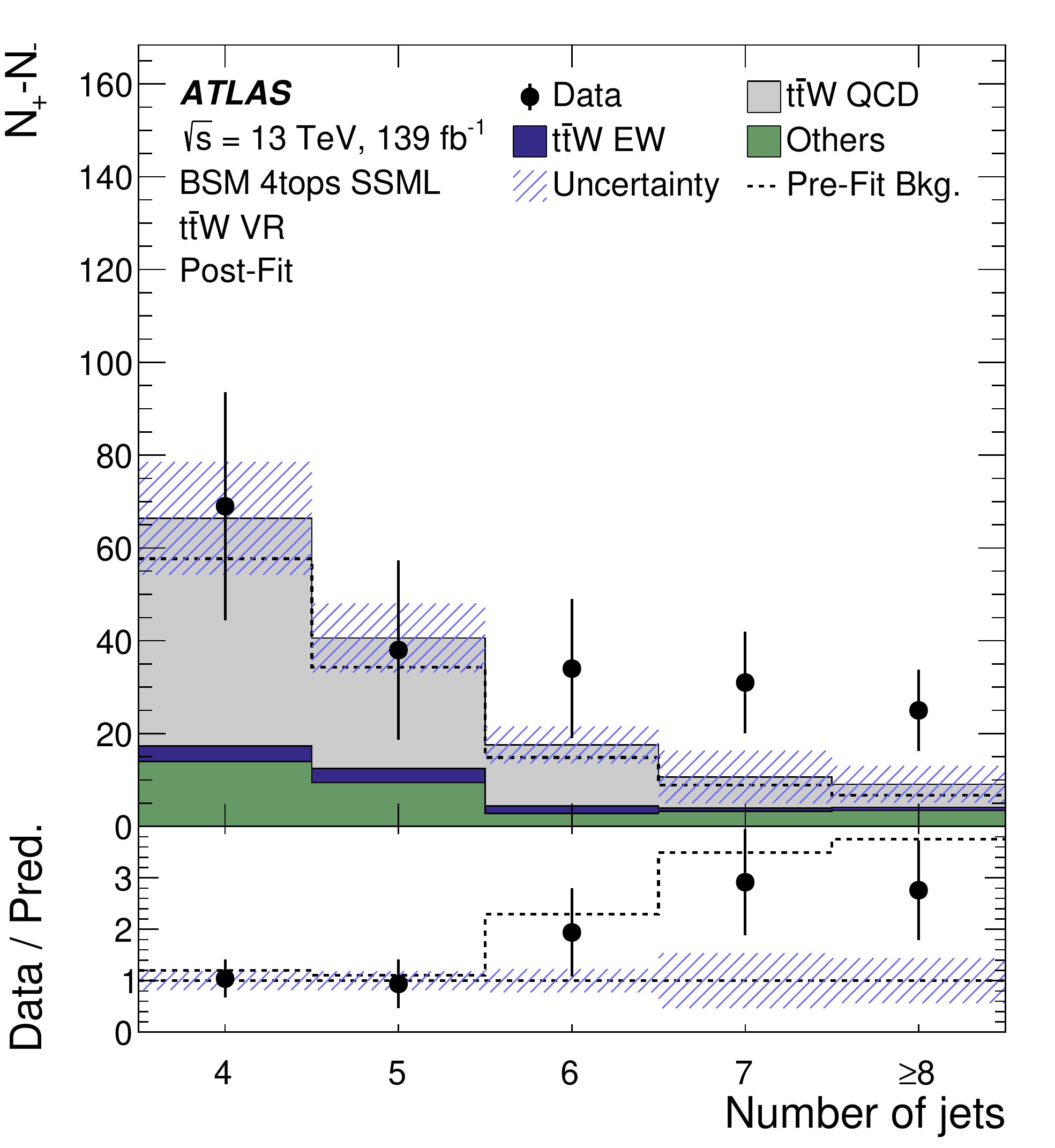}}
\subfloat[]{\includegraphics[width=0.5\linewidth]{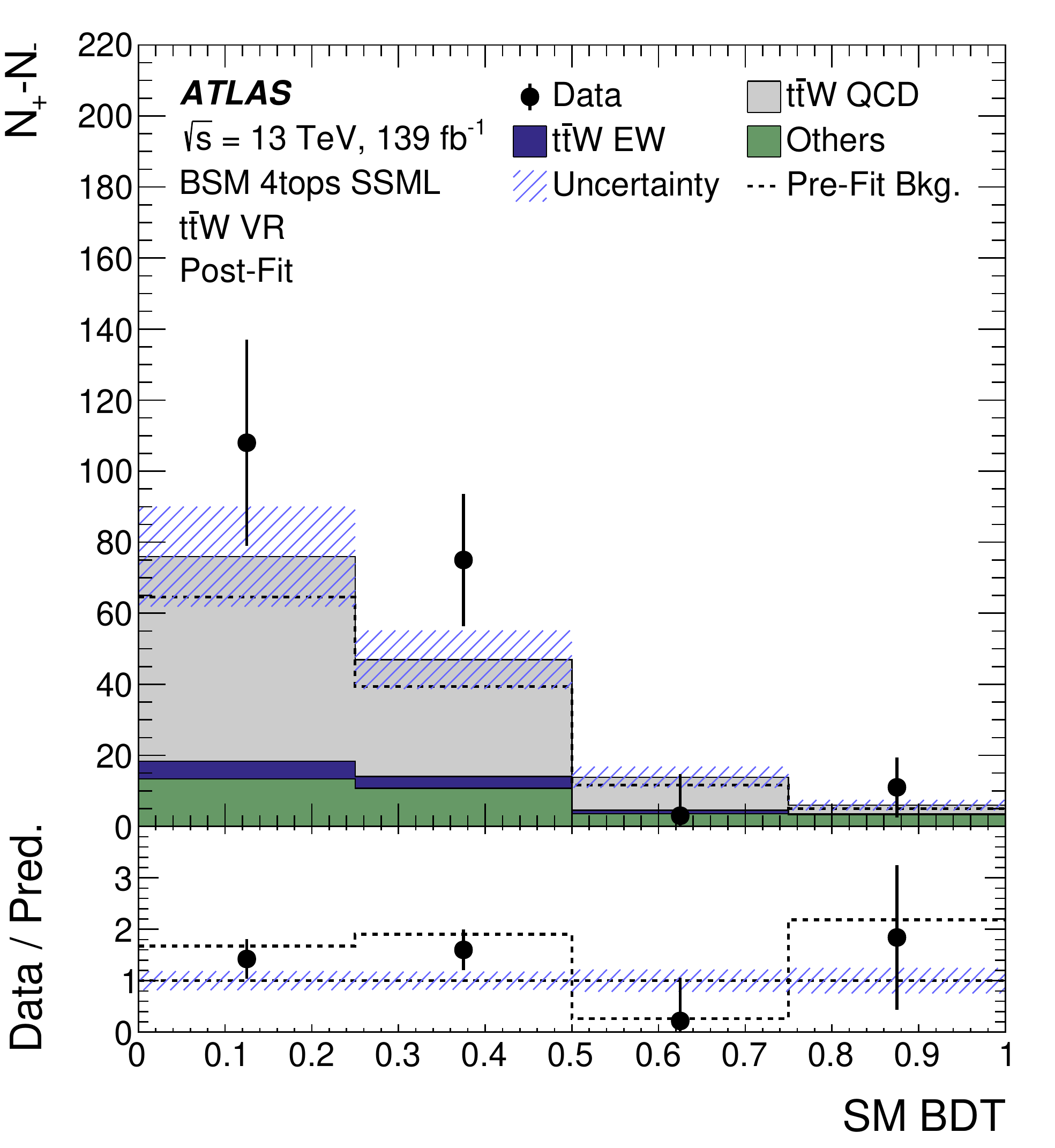}}
\caption{Data and  post-fit background comparison obtained with the background-only fit in the validation region defined for $t\bar{t}W$ events.  The fit is done simultaneously in all signal and control regions. The $y$-axis shows the difference between the number of events with a positive sum of the charges of the selected leptons and the number of events with a negative sum.  The band includes the total uncertainty of the post-fit estimate. The total background prediction before the likelihood fit to data (`Pre-Fit Bkg.') is shown as a dashed black line. The ratio of the data to the background prediction is shown in the lower panel, separately for post-fit background  (black points) and pre-fit background (dashed black line).}
\label{fig:VRttW}
\end{figure}

The different post-fit BSM pBDT distributions corresponding to the signal benchmarks $\mH=400$ and 1000~\GeV\ in the {\it BSM SR} for the background-only fit are shown in Figure~\ref{fig:BSMdata_dis_mH5001000}.
Under the signal-plus-background hypothesis the best-fit signal cross-section ranges between $4^{+6}_{-5}$~fb and $2^{+2}_{-2}$~fb for a heavy Higgs boson with mass between 400~\GeV\ and 1000~\GeV, respectively.
No significant excess of events above the SM prediction is observed.
The results are then interpreted in the context of a \typeII.
The observed and expected upper limits on the $\ttbar H/A$~cross-section times branching fraction  of $H/A \rightarrow \ttbar$ at 95\% CL as a function of $m_{H/A}$  are shown in Figure~\ref{fig:BSMdata_limit}. The observed upper limits  range between 14~fb and 6~fb  for the studied mass range. 
These upper limits on the cross-section can be translated into limits in the $\tan\beta$ vs $m_{H/A}$ plane. Two different scenarios are considered:  one where the scalar $H$ and pseudo-scalar $A$ bosons  have equal masses and both contribute to BSM $\tttt$ production, and the other where only the  scalar $H$ boson contributes. The observed and expected 95\% CL exclusion regions  for these scenarios are shown in Figure~\ref{fig:BSMdata_tanb}.   If both particles contribute to BSM $\tttt$ production, the values of  $\tan\beta$ below 1.6 and 0.6 are excluded  for $m_{H/A}$ between 400~\GeV\ and 1000~\GeV, respectively.  When considering only the scalar $H$ boson, the values of $\tan\beta$ below 1.2 and 0.5 are excluded for the same mass range. In the parameter space studied, the limits for the pseudo-scalar $A$ boson alone are similar to those for the scalar $H$ boson.

The robustness of the results was checked by reproducing the fit results using data split by data-taking period, or by splitting the {\it BSM SR} into events in the same-sign dilepton channel and events in the trilepton channel.  An additional test was performed by using same-sign dilepton events that are only positively charged or only negatively charged. The tests show that the response of the fit is stable with regard to the fitted nuisance parameters and signal strength.

The impact of the different sources of systematic uncertainty on the signal strength is summarised in Table~\ref{tab:groupedSyst}. The systematic uncertainties having the highest impact on the best-fit $\mu$ are the ones associated with the dominant backgrounds in the {\it BSM SR}.  Thus, the highest-ranked systematic uncertainties are the ones related to the modelling of SM $\tttt$ and  $\ttW$.

All of the results described above were obtained with the $\tttt$ contribution constrained to the SM prediction. The binned likelihood fit was also performed with the goal of measuring the normalisation factor for $\tttt$~events assuming no contribution from the BSM $\tttt$ signal, and by fitting the SM BDT distribution in the {\it baseline SR}.   The best-fit value is found to be compatible with the observation from  the previous ATLAS $\tttt$~cross-section measurement~\cite{TOPQ-2018-05}.

\begin{figure}[htbp!]
\centering
\subfloat[]{\includegraphics[width=0.49\linewidth]{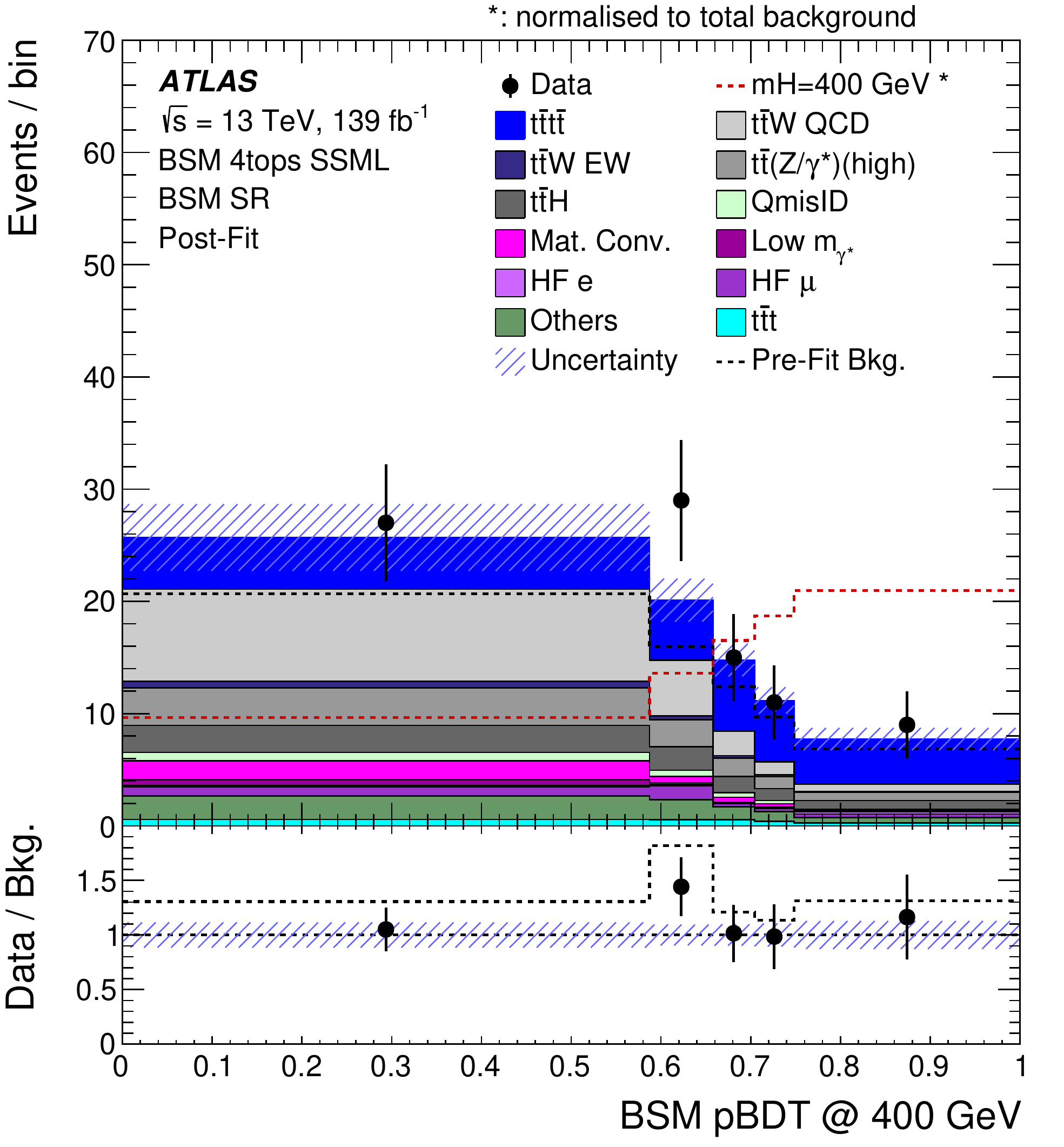}}
\subfloat[]{\includegraphics[width=0.49\linewidth]{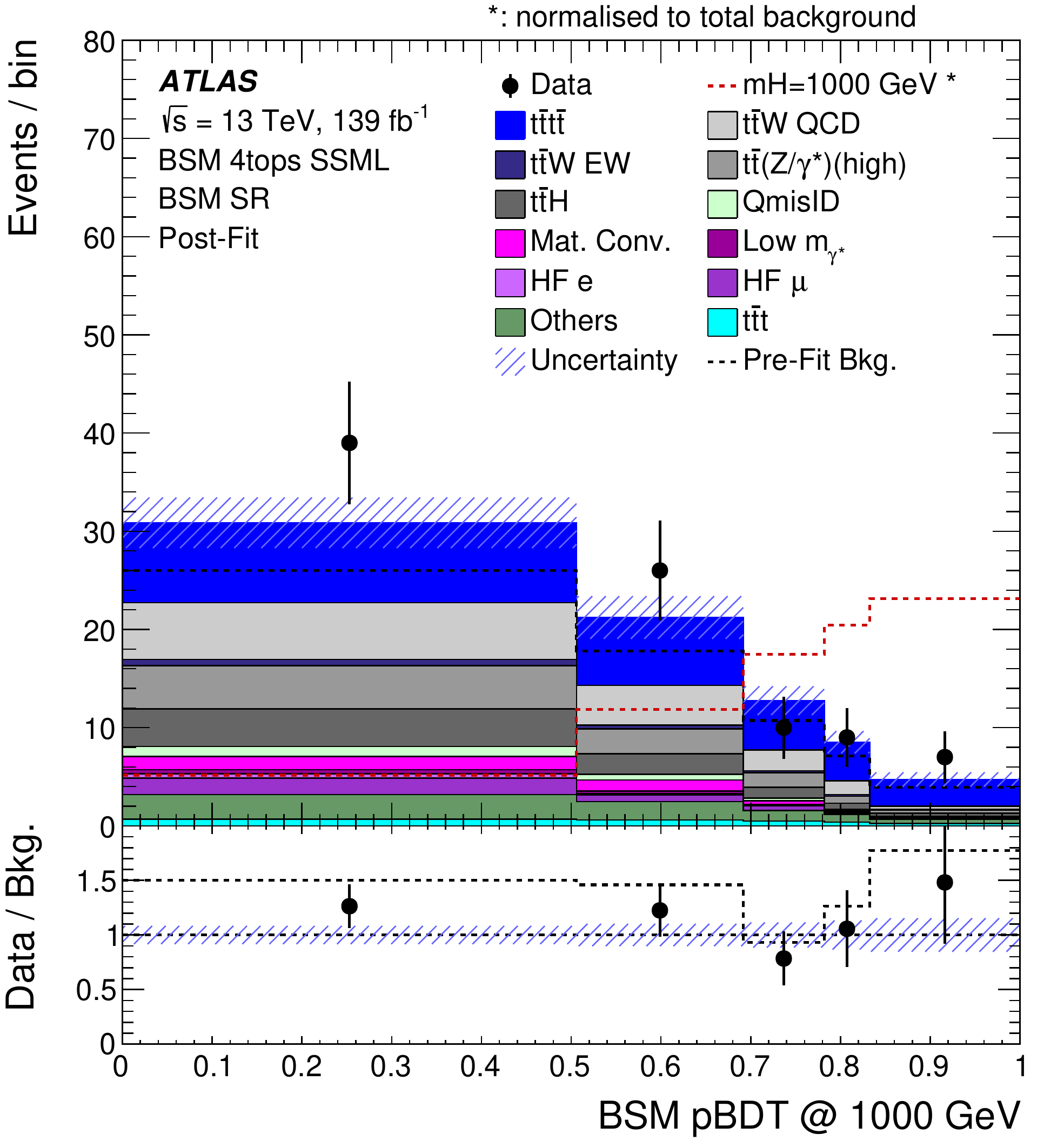}}
 
\caption{Data and  post-fit background comparison obtained with the background-only fit to the {\it BSM SR} for the BSM pBDT distribution used for (a)~$m_{H}=400$~\GeV\ and (b)~$m_{H}=1000$~\GeV.  The fit is done simultaneously in all signal and control regions.  The band includes the total uncertainty of the post-fit estimate. The respective signal hypothesis is also shown. The signal is normalised to the total background for better visibility. The total background prediction before the likelihood fit to data (`Pre-Fit Bkg.') is shown as a dashed black line. The ratio of the data to the background prediction is shown in the lower panel, separately for post-fit background  (black points) and pre-fit background (dashed black line). The binning of the BSM pBDT is optimised for every signal hypothesis to provide the best discrimination between the tested signal and the background, avoiding the presence of bins with no contribution from the major backgrounds.
}
\label{fig:BSMdata_dis_mH5001000}
\end{figure}

\begin{figure}[htbp!]
\centering
\includegraphics[width=0.7\linewidth]{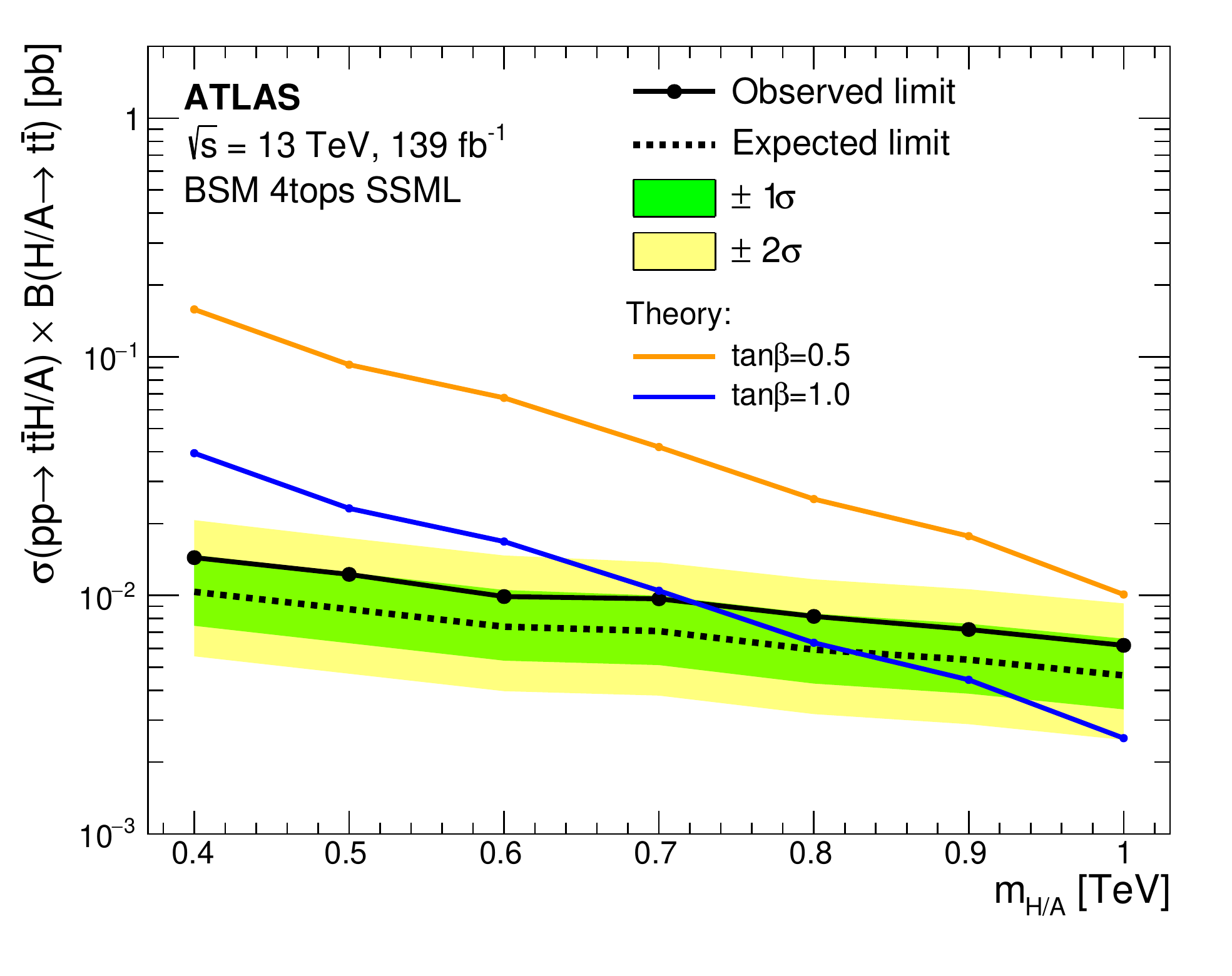}
\caption{Observed (black solid line) and expected (black dashed line) $95\%$ CL upper limits on the $t\bar{t}H/A$~cross-section times branching fraction of  $H/A \rightarrow t\bar{t}$ as a function of $m_{H/A}$.  The limits are estimated assuming that both a heavy scalar $H$ boson and a pseudo-scalar $A$ boson contribute to the $t\bar{t}t\bar{t}$ final state and have the same mass  $m_{H}=m_{A}$. The green (yellow) band shows the $\pm 1\sigma$ ($\pm 2\sigma$) variation of the expected limits. Theoretical predictions for two values of $\tan\beta$ are shown.}
\label{fig:BSMdata_limit}
\end{figure}

\begin{figure}[htbp!]
\centering
\subfloat[]{\includegraphics[width=0.49\linewidth]{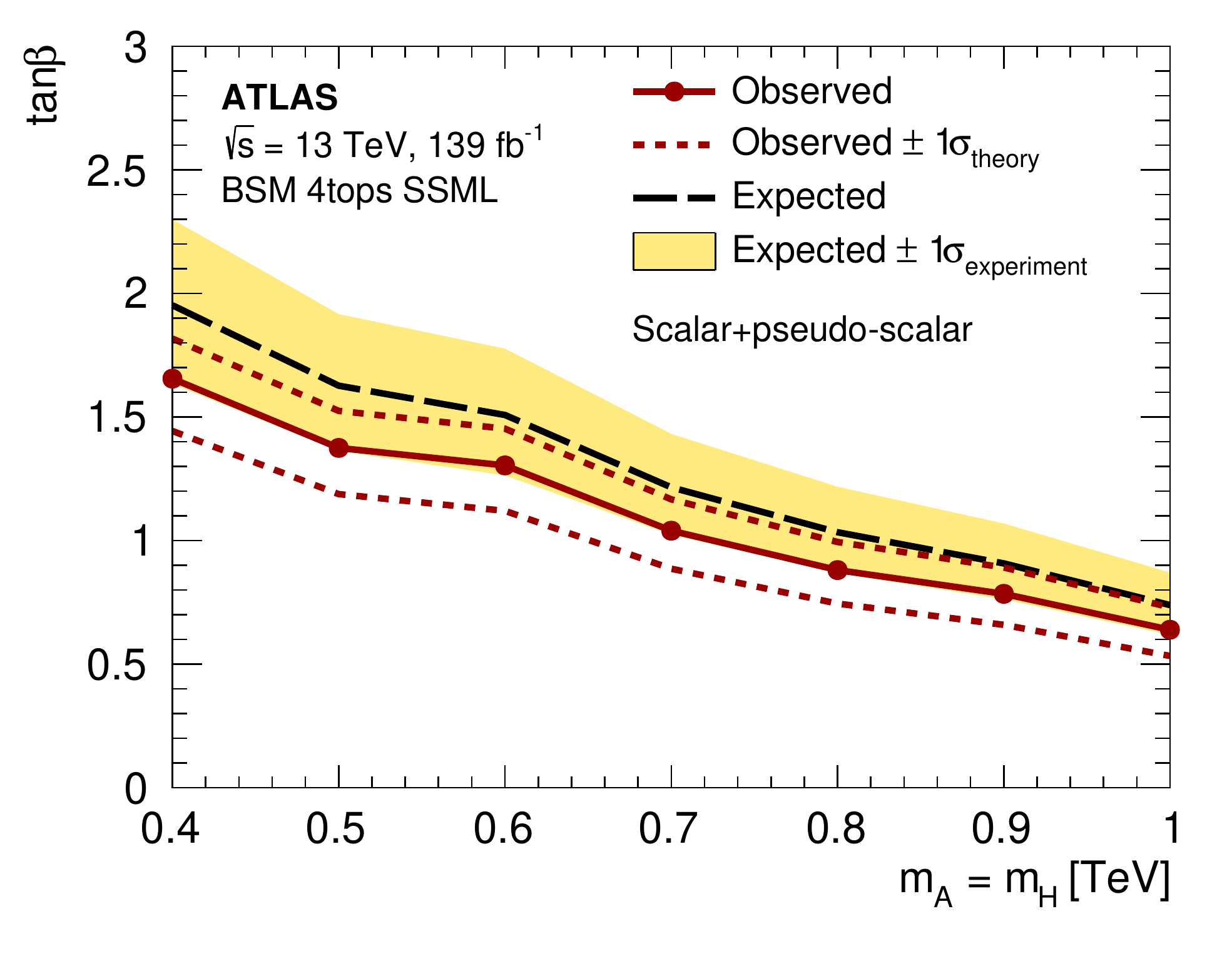}}
\subfloat[]{\includegraphics[width=0.49\linewidth]{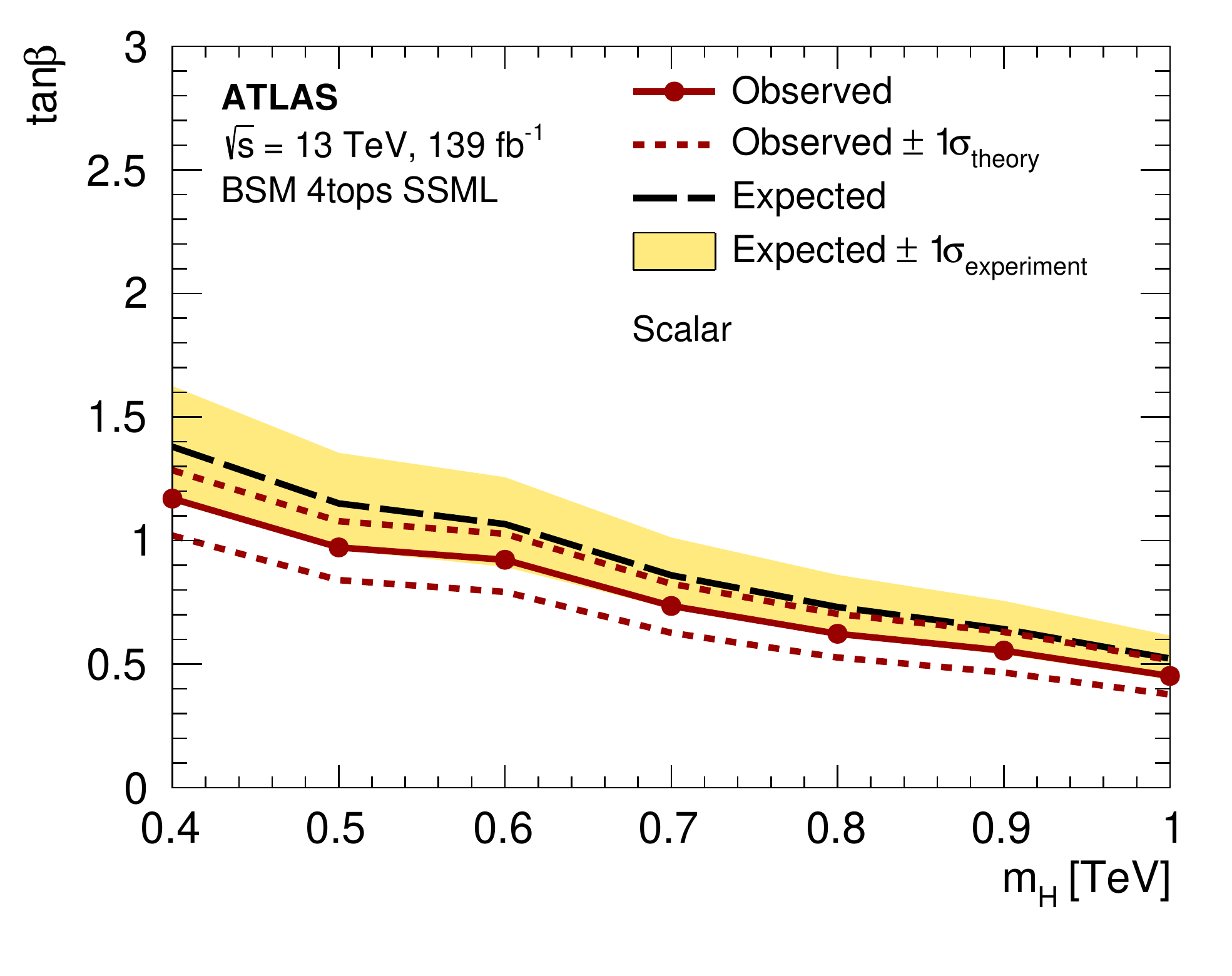}}
 
\caption{Observed (red line) and expected (black dashed line) exclusion regions at $95\%$ CL in the $\tan\beta$ vs $m_{H/A}$ plane (a) assuming that both a heavy scalar $H$ boson and heavy pseudo-scalar $A$ boson contribute to the $t\bar{t}t\bar{t}$ final state and have the same mass  $m_{H}=m_{A}$ or (b) assuming that only the scalar $H$ boson contributes. The limits for the pseudo-scalar $A$ boson alone are similar to those in (b). The yellow band shows the $\pm 1\sigma$ variation of the expected exclusion limit. The region below the red solid line is excluded at $95\%$ CL. The exclusion regions are derived in the context of a type-II 2HDM. Variations of the theory cross-section are also shown as dashed red lines.}
\label{fig:BSMdata_tanb}
\end{figure}

\begin{table}[htp!]
\caption{ Post-fit impact of the different systematic uncertainties on the signal strength $\mu$ assuming $m_{H}=400$~\GeV, grouped in categories.  For each uncertainty source, the fit is repeated with the corresponding group of nuisance parameters fixed to their best-fit values.
The square of each group's contribution is evaluated as the difference of the squares of the full-fit uncertainty and the uncertainty obtained in the repeated fit.
The contributions from individual groups are compared with the total systematic uncertainty and the statistical uncertainty.  The total uncertainty is different from the sum in quadrature of the components due to correlations among nuisance parameters in the fit.}
 
\begin{center}
\begin{tabular}{l  c  c  }
\hline
Uncertainty source & \multicolumn{2}{c}{$\Delta\mu$}  \\
\hline
\textbf{Signal modelling} & \multicolumn{2}{c}{}  \\
$t\bar tH (\rightarrow t\bar t)$ &  $+0.01$ & $-0.00$ \\
\hline
\textbf{Background modelling} & \multicolumn{2}{c}{}  \\
 
$t\bar t t\bar t$  & $+0.17$ & $-0.17$  \\
$t\bar tW$  & $+0.07$ & $-0.07$  \\
$t\bar tt$  & $+0.06$ & $-0.05$ \\
Non-prompt leptons  & $+0.05$ & $-0.05$ \\
$t\bar tZ$    & $+0.05$ & $-0.05$ \\
$t\bar tH$ &  $+0.03$ & $-0.03$ \\
Other background  &  $+0.03$ & $-0.02$ \\
\hline
\textbf{Instrumental} & \multicolumn{2}{c}{}  \\
 
Jet uncertainties &  $+0.12$ & $-0.09$ \\
Jet flavour tagging ($b$-jets) &   $+0.05$ & $-0.04$ \\
 
Jet flavour tagging (light-flavour jets) &   $+0.04$ & $-0.03$ \\
Luminosity &  $+0.03$ & $-0.02$ \\
Jet flavour tagging ($c$-jets) &   $+0.02$ & $-0.02$ \\
Other experimental uncertainties &  $+0.02$ & $-0.02$ \\
\hline
\textbf{MC statistical uncertainty} & \multicolumn{2}{c}{}  \\
Simulation sample size &  $+0.04$ & $-0.04$ \\
\hline
Total systematic uncertainty & $+0.31$ & $-0.28$ \\
\hline
\textbf{Statistical} & & \\
 
HF, Mat. Conv., and Low $m_{\gamma*}$ normalisation & $+0.05$ & $-0.04$ \\
$t\bar tW$ QCD normalisation & $+0.05$ & $-0.04$ \\
\hline
Total statistical uncertainty & $+0.35$ & $-0.32$ \\
\hline
\hline
 
Total uncertainty  &  $+0.46$ & $-0.41$ \\
\hline
 
\end{tabular}
\label{tab:groupedSyst}
\end{center}
\end{table}


\clearpage
\section{Conclusion}
\label{sec:conclusion}

 
A search for a new heavy Higgs boson in the process $t\bar tH/A\rightarrow \tttt$  in the same-sign dilepton channel and the multilepton channel is presented. The search makes use of 139~fb$^{-1}$  of proton--proton collision data  collected by the ATLAS detector at the LHC at a centre-of-mass energy of 13~\TeV.

No significant excess of events over the Standard Model expectation is observed.  The results are then interpreted in the context of the \typeII, assuming no interference with the Standard Model production of four-top-quarks.  The expected sensitivity of the present \ttHA\ search exceeds that of the previous ATLAS \ttHA\ search with 36~\ifb~\cite{EXOT-2016-16} by about a factor of four.  The observed (expected) upper limits at 95\%~CL on the $\ttbar H/A$ cross-section times branching ratio of  $H/A \rightarrow \ttbar$   range between 14~(10)~fb and 6~(5)~fb  for a heavy Higgs boson with mass  between 400~\GeV\ and 1000~\GeV, respectively.   	These upper limits on the cross-section are translated into limits in the $\tan\beta$ vs  $m_{H/A}$ plane. Assuming  that only one  particle, either the scalar $H$ boson or the  pseudo-scalar $A$ boson,  contributes to the \tttt\ final state,  the values of $\tan\beta$ below 1.2 or 0.5 are excluded for a mass of 400~\GeV\ or 1000~\GeV, respectively.  These exclusion ranges increase to $\tan\beta$ below 1.6 or 0.6 when both particles are considered.


\section*{Acknowledgements}


We thank CERN for the very successful operation of the LHC, as well as the
support staff from our institutions without whom ATLAS could not be
operated efficiently.
 
We acknowledge the support of
ANPCyT, Argentina;
YerPhI, Armenia;
ARC, Australia;
BMWFW and FWF, Austria;
ANAS, Azerbaijan;
SSTC, Belarus;
CNPq and FAPESP, Brazil;
NSERC, NRC and CFI, Canada;
CERN;
ANID, Chile;
CAS, MOST and NSFC, China;
Minciencias, Colombia;
MEYS CR, Czech Republic;
DNRF and DNSRC, Denmark;
IN2P3-CNRS and CEA-DRF/IRFU, France;
SRNSFG, Georgia;
BMBF, HGF and MPG, Germany;
GSRI, Greece;
RGC and Hong Kong SAR, China;
ISF and Benoziyo Center, Israel;
INFN, Italy;
MEXT and JSPS, Japan;
CNRST, Morocco;
NWO, Netherlands;
RCN, Norway;
MEiN, Poland;
FCT, Portugal;
MNE/IFA, Romania;
JINR;
MES of Russia and NRC KI, Russian Federation;
MESTD, Serbia;
MSSR, Slovakia;
ARRS and MIZ\v{S}, Slovenia;
DSI/NRF, South Africa;
MICINN, Spain;
SRC and Wallenberg Foundation, Sweden;
SERI, SNSF and Cantons of Bern and Geneva, Switzerland;
MOST, Taiwan;
TENMAK, T\"urkiye;
STFC, United Kingdom;
DOE and NSF, United States of America.
In addition, individual groups and members have received support from
BCKDF, CANARIE, Compute Canada and CRC, Canada;
COST, ERC, ERDF, Horizon 2020 and Marie Sk{\l}odowska-Curie Actions, European Union;
Investissements d'Avenir Labex, Investissements d'Avenir Idex and ANR, France;
DFG and AvH Foundation, Germany;
Herakleitos, Thales and Aristeia programmes co-financed by EU-ESF and the Greek NSRF, Greece;
BSF-NSF and GIF, Israel;
Norwegian Financial Mechanism 2014-2021, Norway;
NCN and NAWA, Poland;
La Caixa Banking Foundation, CERCA Programme Generalitat de Catalunya and PROMETEO and GenT Programmes Generalitat Valenciana, Spain;
G\"{o}ran Gustafssons Stiftelse, Sweden;
The Royal Society and Leverhulme Trust, United Kingdom.
 
The crucial computing support from all WLCG partners is acknowledged gratefully, in particular from CERN, the ATLAS Tier-1 facilities at TRIUMF (Canada), NDGF (Denmark, Norway, Sweden), CC-IN2P3 (France), KIT/GridKA (Germany), INFN-CNAF (Italy), NL-T1 (Netherlands), PIC (Spain), ASGC (Taiwan), RAL (UK) and BNL (USA), the Tier-2 facilities worldwide and large non-WLCG resource providers. Major contributors of computing resources are listed in Ref.~\cite{ATL-SOFT-PUB-2021-003}.


\clearpage
\appendix

\printbibliography

@article{PhysRevD.1.1416,
  title = {Statistical Model for Electron-Positron Annihilation into Hadrons},
  author = {Bjorken, James D. and Brodsky, Stanley J.},
  journal = {Phys. Rev. D},
  volume = {1},
  pages = {1416--1420},
  numpages = {0},
  year = {1970},
  month = {Mar},
  publisher = {American Physical Society},
  doi = {10.1103/PhysRevD.1.1416},
  url = {https://link.aps.org/doi/10.1103/PhysRevD.1.1416}
}

@article{Gutschow:2018tuk,
    author = {G\"utschow, Christian and Lindert, Jonas M. and Sch\"onherr, Marek},
    title = "{Multi-jet merged top-pair production including electroweak corrections}",
    eprint = "1803.00950",
    archivePrefix = "arXiv",
    primaryClass = "hep-ph",
    reportNumber = "CERN-TH-18-039, CERN-TH-2018-039, IPPP/18/13, MCnet-18-05, IPPP-18-13, MCNET-18-05",
    doi = "10.1140/epjc/s10052-018-5804-2",
    journal = "Eur. Phys. J. C",
    volume = "78",
    number = "4",
    pages = "317",
    year = "2018"
}

@article{2016XGBoost,
   title={XGBoost: A Scalable Tree Boosting System},
   url={http://dx.doi.org/10.1145/2939672.2939785},
   DOI={10.1145/2939672.2939785},
   journal={Proceedings of the 22nd ACM SIGKDD International Conference on Knowledge Discovery and Data Mining},
   publisher={ACM},
   author={Chen, Tianqi and Guestrin, Carlos},
   year={2016},
   month={Aug}
}

@article{Baldi:2016fzo,
      author         = "Baldi, Pierre and Cranmer, Kyle and Faucett, Taylor and
                        Sadowski, Peter and Whiteson, Daniel",
      title          = "{Parameterized neural networks for high-energy physics}",
      journal        = "Eur. Phys. J. C",
      volume         = "76",
      year           = "2016",
      number         = "5",
      pages          = "235",
      doi            = "10.1140/epjc/s10052-016-4099-4",
      eprint         = "1601.07913",
      archivePrefix  = "arXiv",
      primaryClass   = "hep-ex",
      SLACcitation   = "%%CITATION = ARXIV:1601.07913;%%"
}

@article{Aad:2012tfa,
      xauthor         = "G. Add et al",
      title          = "{Observation of a new particle in the search for the Standard Model Higgs boson with the ATLAS detector at the LHC}",
      author  = "{ATLAS Collaboration}",
      journal        = "Phys. Lett. B",
      volume         = "716",
      pages          = "1",
      doi            = "10.1016/j.physletb.2012.08.020",
      year           = "2012",
      eprint         = "1207.7214",
      archivePrefix  = "arXiv",
      primaryClass   = "hep-ex",
      reportNumber   = "CERN-PH-EP-2012-218",
      SLACcitation   = "%%CITATION = ARXIV:1207.7214;%%",                                                                                                                      
}

@article{Chatrchyan:2012ufa,
      xauthor         = "Chatrchyan, Serguei and others",
      title          = "{Observation of a new boson at a mass of 125 GeV with the CMS experiment at the LHC}",
      author  = "{CMS Collaboration}",
      journal        = "Phys. Lett. B",
      volume         = "716",
      pages          = "30",
      doi            = "10.1016/j.physletb.2012.08.021",
      year           = "2012",
      eprint         = "1207.7235",
      archivePrefix  = "arXiv",
      primaryClass   = "hep-ex",
      reportNumber   = "CMS-HIG-12-028, CERN-PH-EP-2012-220",
      SLACcitation   = "%%CITATION = ARXIV:1207.7235;%%",                                                                                                                       
}

@article{Englert:1964et,
    author = "Englert, F. and Brout, R.",
    editor = "Taylor, J. C.",
    title = "{Broken Symmetry and the Mass of Gauge Vector Mesons}",
    doi = "10.1103/PhysRevLett.13.321",
    journal = "Phys. Rev. Lett.",
    volume = "13",
    pages = "321--323",
    year = "1964"
}

@article{Higgs:1964ia,
    author = "Higgs, Peter W.",
    title = "{Broken symmetries, massless particles and gauge fields}",
    doi = "10.1016/0031-9163(64)91136-9",
    journal = "Phys. Lett.",
    volume = "12",
    pages = "132--133",
    year = "1964"
}

@article{Higgs:1964pj,
    author = "Higgs, Peter W.",
    editor = "Taylor, J. C.",
    title = "{Broken Symmetries and the Masses of Gauge Bosons}",
    doi = "10.1103/PhysRevLett.13.508",
    journal = "Phys. Rev. Lett.",
    volume = "13",
    pages = "508--509",
    year = "1964"
}

@article{Guralnik:1964eu,
    author = "Guralnik, G. S. and Hagen, C. R. and Kibble, T. W. B.",
    editor = "Taylor, J. C.",
    title = "{Global Conservation Laws and Massless Particles}",
    doi = "10.1103/PhysRevLett.13.585",
    journal = "Phys. Rev. Lett.",
    volume = "13",
    pages = "585--587",
    year = "1964"
}

@article{Higgs:1966ev,
    author = "Higgs, Peter W.",
    title = "{Spontaneous Symmetry Breakdown without Massless Bosons}",
    doi = "10.1103/PhysRev.145.1156",
    journal = "Phys. Rev.",
    volume = "145",
    pages = "1156--1163",
    year = "1966"
}

@article{Kibble:1967sv,
    author = "Kibble, T. W. B.",
    editor = "Taylor, J. C.",
    title = "{Symmetry Breaking in Non-Abelian Gauge Theories}",
    doi = "10.1103/PhysRev.155.1554",
    journal = "Phys. Rev.",
    volume = "155",
    pages = "1554--1561",
    year = "1967"
}

@article{Branco:2011iw,
    author = "Branco, G. C. and Ferreira, P. M. and Lavoura, L. and Rebelo, M. N. and Sher, Marc and Silva, Joao P.",
    title = "{Theory and phenomenology of two-Higgs-doublet models}",
    eprint = "1106.0034",
    archivePrefix = "arXiv",
    primaryClass = "hep-ph",
    doi = "10.1016/j.physrep.2012.02.002",
    journal = "Phys. Rept.",
    volume = "516",
    pages = "1--102",
    year = "2012"
}

@article{Fayet:1976et,
    author = "Fayet, Pierre",
    title = "{Supersymmetry and Weak, Electromagnetic and Strong Interactions}",
    reportNumber = "PTENS 76/14",
    doi = "10.1016/0370-2693(76)90319-1",
    journal = "Phys. Lett. B",
    volume = "64",
    pages = "159",
    year = "1976"
}

@article{Fayet:1977yc,
    author = "Fayet, Pierre",
    title = "{Spontaneously Broken Supersymmetric Theories of Weak, Electromagnetic and Strong Interactions}",
    reportNumber = "LPTENS 77/11",
    doi = "10.1016/0370-2693(77)90852-8",
    journal = "Phys. Lett. B",
    volume = "69",
    pages = "489",
    year = "1977"
}

@article{Kim:1986ax,
    author = "Kim, Jihn E.",
    title = "{Light Pseudoscalars, Particle Physics and Cosmology}",
    reportNumber = "SNUHE-86-09",
    doi = "10.1016/0370-1573(87)90017-2",
    journal = "Phys. Rept.",
    volume = "150",
    pages = "1--177",
    year = "1987"
}

@article{Craig:2016ygr,
    author = "Craig, Nathaniel and Hajer, Jan and Li, Ying-Ying and Liu, Tao and Zhang, Hao",
    title = "{Heavy Higgs bosons at low $\tan \beta$: from the LHC to 100 TeV}",
    eprint = "1605.08744",
    archivePrefix = "arXiv",
    primaryClass = "hep-ph",
    doi = "10.1007/JHEP01(2017)018",
    journal = "JHEP",
    volume = "01",
    pages = "018",
    year = "2017"
}

@article{ATLAS:2021ifb,
    author = "{ATLAS Collaboration}",
    title = "{Search for Higgs boson pair production in the two bottom quarks plus two photons final state in $pp$ collisions at $\sqrt{s}=13$ TeV with the ATLAS detector}",
    eprint = "2112.11876",
    archivePrefix = "arXiv",
    primaryClass = "hep-ex",
    reportNumber = "CERN-EP-2021-180",
    month = "12",
    year = "2021"
}

@article{Junk:1999kv,
      author         = "Junk, Thomas",
      title          = "{Confidence level computation for combining searches with
                        small statistics}",
      journal        = "Nucl. Instrum. Meth. A",
      volume         = "434",
      pages          = "435",
      doi            = "10.1016/S0168-9002(99)00498-2",
      year           = "1999",
      eprint         = "hep-ex/9902006",
      archivePrefix  = "arXiv",
      reportNumber   = "CARLETON-OPAL-PHYS-99-01, CERN-EP-99-041",
      SLACcitation   = "%%CITATION = HEP-EX/9902006;%%",
}

@article{Verkerke:2003ir,
      author         = "Verkerke, Wouter and Kirkby, David P.",
      title          = "{The RooFit toolkit for data modeling}",
      journal        = "eConf",
      volume         = "C0303241",
      pages          = "MOLT007",
      year           = "2003",
      eprint         = "physics/0306116",
      archivePrefix  = "arXiv",
      primaryClass   = "physics.data-an",
      reportNumber   = "CHEP-2003-MOLT007",
      SLACcitation   = "%%CITATION = PHYSICS/0306116;%%",
}

@booklet{RooFitManual,
  author        = "Wouter Verkerke and David Kirkby",
  title         = "RooFit Users Manual",
  url = "http://roofit.sourceforge.net",
}

@article{PhysRevD.103.094014,
  title = {Top-quark pair production in association with a ${W}^{\ifmmode\pm\else\textpm\fi{}}$ gauge boson in the POWHEG-BOX},
  author = {Cordero, F. Febres and Kraus, M. and Reina, L.},
  journal = {Phys. Rev. D},
  volume = {103},
  pages = {094014},
  numpages = {27},
  year = {2021},
  month = {May},
  publisher = {American Physical Society},
  doi = {10.1103/PhysRevD.103.094014},
  url = {https://link.aps.org/doi/10.1103/PhysRevD.103.094014}
}

@article{LUCID2,
  author={G. Avoni and others},
  title={The new LUCID-2 detector for luminosity measurement and monitoring in ATLAS},
  journal={JINST},
  volume={13},
  number={07},
  pages={P07017},
  doi="10.1088/1748-0221/13/07/P07017",
  year={2018},
}

@Article{Cacciari:2008gp,
     author    = "Cacciari, Matteo and Salam, Gavin P. and Soyez, Gregory",
     title     = "{The anti-\(k_{t}\) jet clustering algorithm}",
     journal   = "JHEP",
     volume    = "04",
     year      = "2008",
     pages     = "063",
     eprint    = "0802.1189",
     archivePrefix = "arXiv",
     primaryClass  =  "hep-ph",
     doi       = "10.1088/1126-6708/2008/04/063",
     SLACcitation  = "%%CITATION = 0802.1189;%%"
}

@Article{Fastjet,
      author         = "Cacciari, Matteo and Salam, Gavin P. and Soyez, Gregory",
      title          = "{FastJet user manual}",
      journal        = "Eur. Phys. J. C",
      volume         = "72",
      year           = "2012",
      pages          = "1896",
      doi            = "10.1140/epjc/s10052-012-1896-2",
      eprint         = "1111.6097",
      archivePrefix  = "arXiv",
      primaryClass   = "hep-ph",
      reportNumber   = "CERN-PH-TH-2011-297",
      SLACcitation   = "%%CITATION = ARXIV:1111.6097;%%"
}

@Article{Butterworth:2015oua,
      author         = "Butterworth, Jon and others",
      title          = "{PDF4LHC recommendations for LHC Run II}",
      journal        = "J. Phys. G",
      volume         = "43",
      year           = "2016",
      pages          = "023001",
      doi            = "10.1088/0954-3899/43/2/023001",
      eprint         = "1510.03865",
      archivePrefix  = "arXiv",
      primaryClass   = "hep-ph",
      reportNumber   = "OUTP-15-17P, SMU-HEP-15-12, TIF-UNIMI-2015-14,
                        LCTS-2015-27, CERN-PH-TH-2015-249",
      SLACcitation   = "%%CITATION = ARXIV:1510.03865;%%"
}

@article{Harland-Lang:2014zoa,
      author         = "{L. A. Harland-Lang, A. D. Martin, P. Motylinski
                        and R. S. Thorne}",
      title          = "{Parton distributions in the LHC era: MMHT 2014 PDFs}",
      journal        = "Eur. Phys. J. C",
      number         = "5",
      volume         = "75",
      pages          = "204",
      doi            = "10.1140/epjc/s10052-015-3397-6",
      year           = "2015",
      eprint         = "1412.3989",
      archivePrefix  = "arXiv",
      primaryClass   = "hep-ph",
      reportNumber   = "LCTS-2014-47, IPPP-14-97, DCPT-14-194",
      SLACcitation   = "%%CITATION = ARXIV:1412.3989;%%",
}

@article{Ball:2014uwa,
      author         = "Ball, Richard D. and others",
      title          = "{Parton distributions for the LHC run II}",
      collaboration  = "NNPDF",
      journal        = "JHEP",
      volume         = "04",
      year           = "2015",
      pages          = "040",
      doi            = "10.1007/JHEP04(2015)040",
      eprint         = "1410.8849",
      archivePrefix  = "arXiv",
      primaryClass   = "hep-ph",
      reportNumber   = "EDINBURGH-2014-15, IFUM-1034-FT, CERN-PH-TH-2013-253,
                        OUTP-14-11P, CAVENDISH-HEP-14-11",
      SLACcitation   = "%%CITATION = ARXIV:1410.8849;%%"
}

@Article{Sjostrand:2007gs,
	author    = "Sj{\"o}strand, T. and Mrenna, S. and Skands, P.",
	title     = "{A brief introduction to PYTHIA 8.1}",
	journal   = "Comput. Phys. Commun.",
	volume    = "178",
	year      = "2008",
	pages     = "852-867",
	eprint    = "0710.3820",
	archivePrefix = "arXiv",
	primaryClass  =  "hep-ph",
	doi       = "10.1016/j.cpc.2008.01.036",
	SLACcitation  = "%%CITATION = 0710.3820;%%"
}

@article{Sjostrand:2014zea,
      author         = "Sj{\"o}strand, Torbj{\"o}rn and Ask, Stefan and Christiansen,
                        Jesper R. and Corke, Richard and Desai, Nishita and Ilten,
                        Philip and Mrenna, Stephen and Prestel, Stefan and
                        Rasmussen, Christine O. and Skands, Peter Z.",
      title          = "{An introduction to PYTHIA 8.2}",
      journal        = "Comput. Phys. Commun.",
      volume         = "191",
      year           = "2015",
      pages          = "159",
      doi            = "10.1016/j.cpc.2015.01.024",
      eprint         = "1410.3012",
      archivePrefix  = "arXiv",
      primaryClass   = "hep-ph",
      reportNumber   = "LU-TP-14-36, MCNET-14-22, CERN-PH-TH-2014-190,
                        FERMILAB-PUB-14-316-CD, DESY-14-178, SLAC-PUB-16122,
                        --FERMILAB-PUB-14-316-CD",
      SLACcitation   = "%%CITATION = ARXIV:1410.3012;%%"
}

@Article{Lange:2001uf,
      author         = "Lange, D. J.",
      title          = "{The EvtGen particle decay simulation package}",
      booktitle      = "{Proceedings, 7th International Conference on B physics
                        at hadron machines (BEAUTY 2000)}",
      journal        = "Nucl. Instrum. Meth. A",
      volume         = "462",
      year           = "2001",
      pages          = "152",
      doi            = "10.1016/S0168-9002(01)00089-4",
      SLACcitation   = "%%CITATION = NUIMA,A462,152;%%"
}

@Article{Frixione:2008yi,
     author    = "Frixione, Stefano and Laenen, Eric and Motylinski, Patrick and White, Chris and Webber, Bryan R.",
     title     = "{Single-top hadroproduction in association with a \(W\) boson}",
     journal   = "JHEP",
     volume    = "07",
     year      = "2008",
     pages     = "029",
     eprint    = "0805.3067",
     archivePrefix = "arXiv",
     primaryClass  =  "hep-ph",
     doi       = "10.1088/1126-6708/2008/07/029"
}

@Article{Alwall:2014hca,
      author         = "Alwall, J. and Frederix, R. and Frixione, S. and Hirschi,
                        V. and Maltoni, F. and Mattelaer, O. and Shao, H. -S. and
                        Stelzer, T. and Torrielli, P. and Zaro, M.",
      title          = "{The automated computation of tree-level and
                        next-to-leading order differential cross sections, and
                        their matching to parton shower simulations}",
      journal        = "JHEP",
      volume         = "07",
      year           = "2014",
      pages          = "079",
      doi            = "10.1007/JHEP07(2014)079",
      eprint         = "1405.0301",
      archivePrefix  = "arXiv",
      primaryClass   = "hep-ph",
      reportNumber   = "CERN-PH-TH-2014-064, CP3-14-18, LPN14-066, MCNET-14-09,
                        ZU-TH-14-14",
      SLACcitation   = "%%CITATION = ARXIV:1405.0301;%%"
}

@article{Artoisenet:2012st,
      author         = "Artoisenet, Pierre and Frederix, Rikkert and Mattelaer,
                        Olivier and Rietkerk, Robbert",
      title          = "{Automatic spin-entangled decays of heavy resonances in
                        Monte Carlo simulations}",
      journal        = "JHEP",
      volume         = "03",
      pages          = "015",
      doi            = "10.1007/JHEP03(2013)015",
      year           = "2013",
      eprint         = "1212.3460",
      archivePrefix  = "arXiv",
      primaryClass   = "hep-ph",
      reportNumber   = "NIKHEF-2012-021, CERN-PH-TH-2012-329",
      SLACcitation   = "%%CITATION = ARXIV:1212.3460;%%",
}

@article{Frixione:2007zp,
      author         = "Frixione, Stefano and Laenen, Eric and Motylinski,
                        Patrick and Webber, Bryan R.",
      title          = "{Angular correlations of lepton pairs from vector boson
                        and top quark decays in Monte Carlo simulations}",
      journal        = "JHEP",
      volume         = "04",
      year           = "2007",
      pages          = "081",
      doi            = "10.1088/1126-6708/2007/04/081",
      eprint         = "hep-ph/0702198",
      archivePrefix  = "arXiv",
      OPTprimaryClass   = "HEP-PH",
      reportNumber   = "CAVENDISH-HEP-07-01, GEF-TH-09-2007, ITP-UU-07-10,
                        NIKHEF-2007-004",
      SLACcitation   = "%%CITATION = HEP-PH/0702198;%%"
}

@Article{Bahr:2008pv,
      author         = "B{\"a}hr, M. and others",
      title          = "{Herwig++ physics and manual}",
      journal        = "Eur. Phys. J. C",
      volume         = "58",
      year           = "2008",
      pages          = "639",
      doi            = "10.1140/epjc/s10052-008-0798-9",
      eprint         = "0803.0883",
      archivePrefix  = "arXiv",
      primaryClass   = "hep-ph",
      reportNumber   = "CERN-PH-TH-2008-038, CAVENDISH-HEP-08-03, KA-TP-05-2008,
                        DCPT-08-22, IPPP-08-11, CP3-08-05",
      SLACcitation   = "%%CITATION = ARXIV:0803.0883;%%"
}

@Article{Bellm:2015jjp,
      author         = "Bellm, Johannes and others",
      title          = "{Herwig 7.0/Herwig++ 3.0 release note}",
      journal        = "Eur. Phys. J. C",
      volume         = "76",
      year           = "2016",
      number         = "4",
      pages          = "196",
      doi            = "10.1140/epjc/s10052-016-4018-8",
      eprint         = "1512.01178",
      archivePrefix  = "arXiv",
      primaryClass   = "hep-ph",
      reportNumber   = "CERN-PH-TH-2015-289, MAN-HEP-2015-15, IFJPAN-IV-2015-13,
                        HERWIG-2015-01, KA-TP-18-2015, DCPT-15-142, MCNET-15-28,
                        IPPP-15-71, --HERWIG-2015-01",
      SLACcitation   = "%%CITATION = ARXIV:1512.01178;%%"
}

@Article{Nason:2004rx,
      author         = "Nason, Paolo",
      title          = "{A new method for combining NLO QCD with shower Monte Carlo algorithms}",
      journal        = "JHEP",
      volume         = "11",
      pages          = "040",
      doi            = "10.1088/1126-6708/2004/11/040",
      year           = "2004",
      eprint         = "hep-ph/0409146",
      archivePrefix  = "arXiv",
}

@Article{Frixione:2007vw,
      author         = "Frixione, Stefano and Nason, Paolo and Oleari, Carlo",
      title          = "{Matching NLO QCD computations with parton shower
                        simulations: the POWHEG method}",
      journal        = "JHEP",
      volume         = "11",
      pages          = "070",
      doi            = "10.1088/1126-6708/2007/11/070",
      year           = "2007",
      eprint         = "0709.2092",
      archivePrefix  = "arXiv",
      primaryClass   = "hep-ph",
}

@Article{Alioli:2010xd,
      author         = "Alioli, Simone and Nason, Paolo and Oleari, Carlo and Re,
                        Emanuele",
      title          = "{A general framework for implementing NLO calculations in
                        shower Monte Carlo programs: the POWHEG BOX}",
      journal        = "JHEP",
      volume         = "06",
      pages          = "043",
      doi            = "10.1007/JHEP06(2010)043",
      year           = "2010",
      eprint         = "1002.2581",
      archivePrefix  = "arXiv",
      primaryClass   = "hep-ph",
}

@article{Alioli:2009je,
      author         = "Alioli, Simone and Nason, Paolo and Oleari, Carlo and Re,
                        Emanuele",
      title          = "{NLO single-top production matched with shower in POWHEG:
                        \(s\)- and \(t\)-channel contributions}",
      journal        = "JHEP",
      volume         = "09",
      year           = "2009",
      pages          = "111",
      doi            = "10.1088/1126-6708/2009/09/111",
      eprint         = "0907.4076",
      archivePrefix  = "arXiv",
      primaryClass   = "hep-ph",
      SLACcitation   = "%%CITATION = ARXIV:0907.4076;%%",
      related        = "Alioli:2009je-err",
      relatedstring  = "Erratum:",
}

@article{Frederix:2012dh,
      author         = "Frederix, Rikkert and Re, Emanuele and Torrielli, Paolo",
      title          = "{Single-top \(t\)-channel hadroproduction in the four-flavour
                        scheme with POWHEG and aMC@NLO}",
      journal        = "JHEP",
      volume         = "09",
      year           = "2012",
      pages          = "130",
      doi            = "10.1007/JHEP09(2012)130",
      eprint         = "1207.5391",
      archivePrefix  = "arXiv",
      primaryClass   = "hep-ph",
      reportNumber   = "ZU-TH-14-12, IPPP-12-54, DCPT-12-108,
                        CERN-PH-TH-2012-206, MCNET-12-10, LPN12-084",
      SLACcitation   = "%%CITATION = ARXIV:1207.5391;%%"
}

@article{Re:2010bp,
      author         = "Re, Emanuele",
      title          = "{Single-top \(Wt\)-channel production matched with parton
                        showers using the POWHEG method}",
      journal        = "Eur. Phys. J. C",
      volume         = "71",
      year           = "2011",
      pages          = "1547",
      doi            = "10.1140/epjc/s10052-011-1547-z",
      eprint         = "1009.2450",
      archivePrefix  = "arXiv",
      primaryClass   = "hep-ph",
      reportNumber   = "IPPP-10-74, DCPT-10-148",
      SLACcitation   = "%%CITATION = ARXIV:1009.2450;%%"
}

@article{Frixione:2007nw,
      author         = "Frixione, Stefano and Nason, Paolo and Ridolfi, Giovanni",
      title          = "{A positive-weight next-to-leading-order Monte Carlo for
                        heavy flavour hadroproduction}",
      journal        = "JHEP",
      volume         = "09",
      pages          = "126",
      doi            = "10.1088/1126-6708/2007/09/126",
      year           = "2007",
      eprint         = "0707.3088",
      archivePrefix  = "arXiv",
      primaryClass   = "hep-ph",
      SLACcitation   = "%%CITATION = ARXIV:0707.3088;%%",
}

@article{Frederix:2017wme,
      author         = "Frederix, Rikkert and Pagani, Davide and Zaro, Marco",
      title          = "{Large NLO corrections in \(t\bar{t}W^{\pm}\) and
                        \(t\bar{t}t\bar{t}\) hadroproduction from supposedly
                        subleading EW contributions}",
      journal        = "JHEP",
      volume         = "02",
      year           = "2018",
      pages          = "031",
      doi            = "10.1007/JHEP02(2018)031",
      eprint         = "1711.02116",
      archivePrefix  = "arXiv",
      primaryClass   = "hep-ph",
      SLACcitation   = "%%CITATION = ARXIV:1711.02116;%%"
}

@article{deFlorian:2016spz,
      author         = "de Florian, D. and others",
      title          = "{Handbook of LHC Higgs Cross Sections: 4. Deciphering the
                        Nature of the Higgs Sector}",
      collaboration  = "LHC Higgs Cross Section Working Group",
      doi            = "10.23731/CYRM-2017-002",
      year           = "2016",
      eprint         = "1610.07922",
      archivePrefix  = "arXiv",
      primaryClass   = "hep-ph",
      reportNumber   = "FERMILAB-FN-1025-T, CERN-2017-002-M",
      SLACcitation   = "%%CITATION = ARXIV:1610.07922;%%"
}

@Article{Hoeche:2009rj,
      author         = "H{\"o}che, Stefan and Krauss, Frank and Schumann, Steffen and Siegert, Frank",
      title          = "{QCD matrix elements and truncated showers}",
      journal        = "JHEP",
      volume         = "05",
      pages          = "053",
      doi            = "10.1088/1126-6708/2009/05/053",
      year           = "2009",
      eprint         = "0903.1219",
      archivePrefix  = "arXiv",
      primaryClass   = "hep-ph",
}

@Article{Schumann:2007mg,
      author         = "Schumann, Steffen and Krauss, Frank",
      title          = "{A parton shower algorithm based on Catani--Seymour dipole factorisation}",
      journal        = "JHEP",
      volume         = "03",
      pages          = "038",
      doi            = "10.1088/1126-6708/2008/03/038",
      year           = "2008",
      eprint         = "0709.1027",
      archivePrefix  = "arXiv",
      primaryClass   = "hep-ph",
}

@article{Hoeche:2012yf,
      author         = "H{\"o}che, Stefan and Krauss, Frank and Sch{\"o}nherr, Marek and
                        Siegert, Frank",
      title          = "{QCD matrix elements + parton showers. The NLO case}",
      journal        = "JHEP",
      volume         = "04",
      year           = "2013",
      pages          = "027",
      doi            = "10.1007/JHEP04(2013)027",
      eprint         = "1207.5030",
      archivePrefix  = "arXiv",
      primaryClass   = "hep-ph",
      reportNumber   = "SLAC-PUB-15191, IPPP-12-52, DCPT-12-104, LPN12-081,
                        FR-PHENO-2012-017, MCNET-12-09, --FR-PHENO-2012-017",
      SLACcitation   = "%%CITATION = ARXIV:1207.5030;%%"
}

@article{Bothmann:2019yzt,
      author = "Bothmann, Enrico and others",
      title = "{Event generation with Sherpa 2.2}",
      journal = "SciPost Phys.",
      volume = "7",
      year = "2019",
      number = "3",
      pages = "034",
      doi = "10.21468/SciPostPhys.7.3.034",
      reportNumber = "FERMILAB-PUB-19-218-T, SLAC-PUB-17433, IPPP/19/42, MCNET-19-11",
      eprint = "1905.09127",
      archivePrefix = "arXiv",
      primaryClass = "hep-ph",
}

@article{Hoeche:2011fd,
      author         = "H{\"o}che, Stefan and Krauss, Frank and Sch{\"o}nherr, Marek and
                        Siegert, Frank",
      title          = "{A critical appraisal of NLO+PS matching methods}",
      journal        = "JHEP",
      volume         = "09",
      year           = "2012",
      pages          = "049",
      doi            = "10.1007/JHEP09(2012)049",
      eprint         = "1111.1220",
      archivePrefix  = "arXiv",
      primaryClass   = "hep-ph",
      reportNumber   = "SLAC-PUB-14661, IPPP-11-67, DCPT-11-134, LPN11-58,
                        FR-PHENO-2011-019, MCNET-11-24",
      SLACcitation   = "%%CITATION = ARXIV:1111.1220;%%"
}

@article{Catani:2001cc,
      author         = "Catani, S. and Krauss, F. and Webber, B. R. and Kuhn, R.",
      title          = "{QCD Matrix Elements + Parton Showers}",
      journal        = "JHEP",
      volume         = "11",
      year           = "2001",
      pages          = "063",
      doi            = "10.1088/1126-6708/2001/11/063",
      eprint         = "hep-ph/0109231",
      archivePrefix  = "arXiv",
      reportNumber   = "CERN-TH-2000-367, CAVENDISH-HEP-00-03",
      SLACcitation   = "%%CITATION = HEP-PH/0109231;%%"
}

@article{Cascioli:2011va,
      author         = "Cascioli, Fabio and Maierh{\"o}fer, Philipp and Pozzorini, Stefano",
      title          = "{Scattering Amplitudes with Open Loops}",
      journal        = "Phys. Rev. Lett.",
      volume         = "108",
      year           = "2012",
      pages          = "111601",
      doi            = "10.1103/PhysRevLett.108.111601",
      eprint         = "1111.5206",
      archivePrefix  = "arXiv",
      primaryClass   = "hep-ph",
      reportNumber   = "ZU-TH-23-11, LPN11-66",
      SLACcitation   = "%%CITATION = ARXIV:1111.5206;%%"
}

@article{Buccioni:2019sur,
    author = "Buccioni, Federico and Lang, Jean-Nicolas and Lindert, Jonas M. and Maierh{\"o}fer, Philipp and Pozzorini, Stefano and Zhang, Hantian and Zoller, Max F.",
    title = "{OpenLoops 2}",
    eprint = "1907.13071",
    archivePrefix = "arXiv",
    primaryClass = "hep-ph",
    reportNumber = "IPPP/19/62, FR-PHENO-2019-12, PSI-PR-19-15, ZU-TH 37/19",
    doi = "10.1140/epjc/s10052-019-7306-2",
    journal = "Eur. Phys. J. C",
    volume = "79",
    number = "10",
    pages = "866",
    year = "2019"
}

@article{Denner:2016kdg,
      author         = "Denner, Ansgar and Dittmaier, Stefan and Hofer, Lars",
      title          = "{\textsc{Collier}: A fortran-based complex one-loop library in
                        extended regularizations}",
      journal        = "Comput. Phys. Commun.",
      volume         = "212",
      year           = "2017",
      pages          = "220-238",
      doi            = "10.1016/j.cpc.2016.10.013",
      eprint         = "1604.06792",
      archivePrefix  = "arXiv",
      primaryClass   = "hep-ph",
      reportNumber   = "FR-PHENO-2016-003, ICCUB-16-016",
      SLACcitation   = "%%CITATION = ARXIV:1604.06792;%%"
}

@article{Ciccolini:2003jy,
      author         = "Ciccolini, M. L. and Dittmaier, S. and Kr{\"a}mer, M.",
      title          = "{Electroweak radiative corrections to associated \(WH\) and
                        \(ZH\) production at hadron colliders}",
      journal        = "Phys. Rev. D",
      volume         = "68",
      year           = "2003",
      pages          = "073003",
      doi            = "10.1103/PhysRevD.68.073003",
      eprint         = "hep-ph/0306234",
      archivePrefix  = "arXiv",
      primaryClass   = "hep-ph",
      reportNumber   = "EDINBURGH-2003-05, MPI-PHT-2003-24",
      SLACcitation   = "%%CITATION = HEP-PH/0306234;%%"
}

@article{Brein:2003wg,
      author         = "Brein, Oliver and Djouadi, Abdelhak and Harlander,
                        Robert",
      title          = "{NNLO QCD corrections to the Higgs-strahlung processes at
                        hadron colliders}",
      journal        = "Phys. Lett. B",
      volume         = "579",
      year           = "2004",
      pages          = "149-156",
      doi            = "10.1016/j.physletb.2003.10.112",
      eprint         = "hep-ph/0307206",
      archivePrefix  = "arXiv",
      XprimaryClass   = "hep-ph",
      reportNumber   = "MPP-2003-35, CERN-TH-2003-161, PM-03-16",
      SLACcitation   = "%%CITATION = HEP-PH/0307206;%%"
}

@article{Brein:2011vx,
      author         = "Brein, Oliver and Harlander, Robert V. and Wiesemann, Marius
                        and Zirke, Tom",
      title          = "{Top-Quark Mediated Effects in Hadronic Higgs-Strahlung}",
      journal        = "Eur. Phys. J. C",
      volume         = "72",
      year           = "2012",
      pages          = "1868",
      doi            = "10.1140/epjc/s10052-012-1868-6",
      eprint         = "1111.0761",
      archivePrefix  = "arXiv",
      primaryClass   = "hep-ph",
      reportNumber   = "CERN-PH-TH-2011-268, FR-PHENO-2011-016, WUB-11-15",
      SLACcitation   = "%%CITATION = ARXIV:1111.0761;%%"
}

@article{Altenkamp:2012sx,
      author         = "Altenkamp, Lukas and Dittmaier, Stefan and Harlander,
                        Robert V. and Rzehak, Heidi and Zirke, Tom J. E.",
      title          = "{Gluon-induced Higgs-strahlung at next-to-leading order
                        QCD}",
      journal        = "JHEP",
      volume         = "02",
      year           = "2013",
      pages          = "078",
      doi            = "10.1007/JHEP02(2013)078",
      eprint         = "1211.5015",
      archivePrefix  = "arXiv",
      primaryClass   = "hep-ph",
      reportNumber   = "CERN-PH-TH-2012-312, FR-PHENO-2012-023, WUB-12-21",
      SLACcitation   = "%%CITATION = ARXIV:1211.5015;%%"
}

@article{Denner:2014cla,
      author         = "Denner, Ansgar and Dittmaier, Stefan and Kallweit, Stefan
                        and M{\"u}ck, Alexander",
      title          = "{HAWK 2.0: A Monte Carlo program for Higgs production
                        in vector-boson fusion and Higgs strahlung at hadron
                        colliders}",
      journal        = "Comput. Phys. Commun.",
      volume         = "195",
      year           = "2015",
      pages          = "161-171",
      doi            = "10.1016/j.cpc.2015.04.021",
      eprint         = "1412.5390",
      archivePrefix  = "arXiv",
      primaryClass   = "hep-ph",
      reportNumber   = "FR-PHENO-2014-013, MITP-14-101, TTK-14-36",
      SLACcitation   = "%%CITATION = ARXIV:1412.5390;%%"
}

@article{Brein:2012ne,
      author         = "Brein, Oliver and Harlander, Robert V. and Zirke, Tom J.
                        E.",
      title          = "{vh@nnlo -- Higgs Strahlung at hadron colliders}",
      journal        = "Comput. Phys. Commun.",
      volume         = "184",
      year           = "2013",
      pages          = "998-1003",
      doi            = "10.1016/j.cpc.2012.11.002",
      eprint         = "1210.5347",
      archivePrefix  = "arXiv",
      primaryClass   = "hep-ph",
      SLACcitation   = "%%CITATION = ARXIV:1210.5347;%%"
}

@article{Harlander:2014wda,
      author         = "Harlander, Robert V. and Kulesza, Anna and Theeuwes,
                        Vincent and Zirke, Tom",
      title          = "{Soft gluon resummation for gluon-induced Higgs
                        Strahlung}",
      journal        = "JHEP",
      volume         = "11",
      year           = "2014",
      pages          = "082",
      doi            = "10.1007/JHEP11(2014)082",
      eprint         = "1410.0217",
      archivePrefix  = "arXiv",
      primaryClass   = "hep-ph",
      reportNumber   = "MS-TP-14-17, WUB-14-10, LPN14-112",
      SLACcitation   = "%%CITATION = ARXIV:1410.0217;%%"
}

@article{Demartin:2016axk,
    author = "Demartin, Federico and Maier, Benedikt and Maltoni, Fabio and Mawatari, Kentarou and Zaro, Marco",
    title = "{tWH associated production at the LHC}",
    eprint = "1607.05862",
    archivePrefix = "arXiv",
    primaryClass = "hep-ph",
    reportNumber = "MCNET-16-30, CP3-16-40, LPSC16158",
    doi = "10.1140/epjc/s10052-017-4601-7",
    journal = "Eur. Phys. J. C",
    volume = "77",
    number = "1",
    pages = "34",
    year = "2017"
}

@article{Agostinelli:2002hh,
      author         = "{GEANT4 Collaboration} and Agostinelli, S. and others",
      title          = "{\textsc{Geant4} -- a simulation toolkit}",
      journal        = "Nucl. Instrum. Meth. A",
      volume         = "506",
      year           = "2003",
      pages          = "250",
      doi            = "10.1016/S0168-9002(03)01368-8",
      reportNumber   = "SLAC-PUB-9350, FERMILAB-PUB-03-339",
      SLACcitation   = "%%CITATION = NUIMA,A506,250;%%"
}

@Article{Read:2002hq,
    Author = {Read, Alexander L.},
    Journal = {J. Phys. G},
    Pages = {2693},
    doi = "10.1088/0954-3899/28/10/313",
    Title = "{Presentation of search results: the \(CL_S\) technique}",
    Volume = {28},
    Year = {2002}
}

@Article{Cowan:2010js,
   Author = {Cowan, Glen and Cranmer, Kyle and Gross, Eilam and Vitells, Ofer},
   Title = {{Asymptotic formulae for likelihood-based tests of new physics}},
   Journal = {Eur. Phys. J. C},
   Volume = {71},
   Year = {2011},
   Pages = {1554},
   doi = "10.1140/epjc/s10052-011-1554-0",
   Eprint = "1007.1727",
   Archiveprefix = {arXiv},
   Primaryclass = {physics.data-an},
   Addendum = "Erratum: \href{http://dx.doi.org/10.1140/epjc/s10052-013-2501-z}{Eur. Phys. J. C {\textbf{73}} (2013) 2501}",
}

@Article{PERF-2007-01,
    author         = "{ATLAS Collaboration}",
    title          = "{The ATLAS Experiment at the CERN Large Hadron Collider}",
    journal        = "JINST",
    volume         = "3",
    year           = "2008",
    pages          = "S08003",
    doi            = "10.1088/1748-0221/3/08/S08003",
    primaryClass   = "hep-ex",
}

@Article{SOFT-2010-01,
    author         = "{ATLAS Collaboration}",
    title          = "{The ATLAS Simulation Infrastructure}",
    journal        = "Eur. Phys. J. C",
    volume         = "70",
    year           = "2010",
    pages          = "823",
    doi            = "10.1140/epjc/s10052-010-1429-9",
    eprint         = "1005.4568",
    archivePrefix  = "arXiv",
    primaryClass   = "physics.ins-det",
}

@Article{PERF-2014-03,
    author         = "{ATLAS Collaboration}",
    title          = "{Performance of pile-up mitigation techniques for jets in \(pp\) collisions at \(\sqrt{s} = 8\,\text{TeV}\) using the ATLAS detector}",
    journal        = "Eur. Phys. J. C",
    volume         = "76",
    year           = "2016",
    pages          = "581",
    doi            = "10.1140/epjc/s10052-016-4395-z",
    reportNumber   = "CERN-PH-EP-2015-206",
    eprint         = "1510.03823",
    archivePrefix  = "arXiv",
    primaryClass   = "hep-ex",
}

@Article{PERF-2015-09,
    author         = "{ATLAS Collaboration}",
    title          = "{Jet reconstruction and performance using particle flow with the ATLAS Detector}",
    journal        = "Eur. Phys. J. C",
    volume         = "77",
    year           = "2017",
    pages          = "466",
    doi            = "10.1140/epjc/s10052-017-5031-2",
    reportNumber   = "CERN-EP-2017-024",
    eprint         = "1703.10485",
    archivePrefix  = "arXiv",
    primaryClass   = "hep-ex",
}

@Article{EXOT-2016-04,
    author         = "{ATLAS Collaboration}",
    title          = "{Search for heavy Higgs bosons \(A/H\) decaying to a top quark pair in \(pp\) collisions at \(\sqrt{s} = 8\,\text{TeV}\) with the ATLAS detector}",
    journal        = "Phys. Rev. Lett.",
    volume         = "119",
    year           = "2017",
    pages          = "191803",
    doi            = "10.1103/PhysRevLett.119.191803",
    reportNumber   = "CERN-EP-2017-134",
    eprint         = "1707.06025",
    archivePrefix  = "arXiv",
    primaryClass   = "hep-ex",
}

@Article{EXOT-2016-16,
    author         = "{ATLAS Collaboration}",
    title          = "{Search for new phenomena in events with same-charge leptons and \(b\)-jets in \(pp\) collisions at \(\sqrt{s} = 13\,\text{TeV}\) with the ATLAS detector}",
    journal        = "JHEP",
    volume         = "12",
    year           = "2018",
    pages          = "039",
    doi            = "10.1007/JHEP12(2018)039",
    reportNumber   = "CERN-EP-2018-171",
    eprint         = "1807.11883",
    archivePrefix  = "arXiv",
    primaryClass   = "hep-ex",
}

@Article{HIGG-2016-32,
    author         = "{ATLAS Collaboration}",
    title          = "{Search for heavy neutral Higgs bosons produced in association with \(b\)-quarks and decaying into \(b\)-quarks at \(\sqrt{s} = 13\,\text{TeV}\) with the ATLAS detector}",
    journal        = "Phys. Rev. D",
    volume         = "102",
    year           = "2020",
    pages          = "032004",
    doi            = "10.1103/PhysRevD.102.032004",
    reportNumber   = "CERN-EP-2019-092",
    eprint         = "1907.02749",
    archivePrefix  = "arXiv",
    primaryClass   = "hep-ex",
}

@Article{PERF-2016-07,
    author         = "{ATLAS Collaboration}",
    title          = "{Performance of missing transverse momentum reconstruction with the ATLAS detector using proton--proton collisions at \(\sqrt{s} = 13\,\text{TeV}\)}",
    journal        = "Eur. Phys. J. C",
    volume         = "78",
    year           = "2018",
    pages          = "903",
    doi            = "10.1140/epjc/s10052-018-6288-9",
    reportNumber   = "CERN-EP-2017-274",
    eprint         = "1802.08168",
    archivePrefix  = "arXiv",
    primaryClass   = "hep-ex",
}

@Article{TOPQ-2016-11,
    author         = "{ATLAS Collaboration}",
    title          = "{Measurement of the \(t\bar{t}Z\) and \(t\bar{t}W\) cross sections in proton--proton collisions at \(\sqrt{s} = 13\,\text{TeV}\) with the ATLAS detector}",
    journal        = "Phys. Rev. D",
    volume         = "99",
    year           = "2019",
    pages          = "072009",
    doi            = "10.1103/PhysRevD.99.072009",
    reportNumber   = "CERN-EP-2018-331",
    eprint         = "1901.03584",
    archivePrefix  = "arXiv",
    primaryClass   = "hep-ex",
}

@Article{TOPQ-2016-14,
    author         = "{ATLAS Collaboration}",
    title          = "{Measurement of the production cross-section of a single top quark in association with a \(Z\) boson in proton--proton collisions at \(13\,\text{TeV}\) with the ATLAS detector}",
    journal        = "Phys. Lett. B",
    volume         = "780",
    year           = "2018",
    pages          = "557",
    doi            = "10.1016/j.physletb.2018.03.023",
    reportNumber   = "CERN-EP-2017-188",
    eprint         = "1710.03659",
    archivePrefix  = "arXiv",
    primaryClass   = "hep-ex",
}

@Article{TRIG-2016-01,
    author         = "{ATLAS Collaboration}",
    title          = "{Performance of the ATLAS trigger system in 2015}",
    journal        = "Eur. Phys. J. C",
    volume         = "77",
    year           = "2017",
    pages          = "317",
    doi            = "10.1140/epjc/s10052-017-4852-3",
    reportNumber   = "CERN-EP-2016-241",
    eprint         = "1611.09661",
    archivePrefix  = "arXiv",
    primaryClass   = "hep-ex",
}

@Article{TOPQ-2017-12,
    author         = "{ATLAS Collaboration}",
    title          = "{Measurements of inclusive and differential fiducial cross-sections of \(t\bar{t}\) production with additional heavy-flavour jets in proton--proton collisions at \(\sqrt{s} = 13\,\text{TeV}\) with the ATLAS detector}",
    journal        = "JHEP",
    volume         = "04",
    year           = "2019",
    pages          = "046",
    doi            = "10.1007/JHEP04(2019)046",
    reportNumber   = "CERN-EP-2018-276",
    eprint         = "1811.12113",
    archivePrefix  = "arXiv",
    primaryClass   = "hep-ex",
}

@Article{DAPR-2018-01,
    author         = "{ATLAS Collaboration}",
    title          = "{ATLAS data quality operations and performance for 2015--2018 data-taking}",
    journal        = "JINST",
    volume         = "15",
    year           = "2020",
    pages          = "P04003",
    doi            = "10.1088/1748-0221/15/04/P04003",
    reportNumber   = "CERN-EP-2019-207",
    eprint         = "1911.04632",
    archivePrefix  = "arXiv",
    primaryClass   = "physics.ins-det",
}

@Article{EGAM-2018-01,
    author         = "{ATLAS Collaboration}",
    title          = "{Electron and photon performance measurements with the ATLAS detector using the 2015--2017 LHC proton--proton collision data}",
    journal        = "JINST",
    volume         = "14",
    year           = "2019",
    pages          = "P12006",
    doi            = "10.1088/1748-0221/14/12/P12006",
    reportNumber   = "CERN-EP-2019-145",
    eprint         = "1908.00005",
    archivePrefix  = "arXiv",
    primaryClass   = "hep-ex",
}

@Article{FTAG-2018-01,
    author         = "{ATLAS Collaboration}",
    title          = "{ATLAS \(b\)-jet identification performance and efficiency measurement with \(t\bar{t}\) events in \(pp\) collisions at \(\sqrt{s} = 13\,\text{TeV}\)}",
    journal        = "Eur. Phys. J. C",
    volume         = "79",
    year           = "2019",
    pages          = "970",
    doi            = "10.1140/epjc/s10052-019-7450-8",
    reportNumber   = "CERN-EP-2019-132",
    eprint         = "1907.05120",
    archivePrefix  = "arXiv",
    primaryClass   = "hep-ex",
}

@Article{HDBS-2018-13,
    author         = "{ATLAS Collaboration}",
    title          = "{Search for a heavy Higgs boson decaying into a \(Z\) boson and another heavy Higgs boson in the \(\ell\ell bb\) and \(\ell\ell WW\) final states in \(pp\) collisions at \(\sqrt{s} = 13\,\text{TeV}\) with the ATLAS detector}",
    journal        = "Eur. Phys. J. C",
    volume         = "81",
    year           = "2020",
    pages          = "396",
    doi            = "10.1140/epjc/s10052-021-09117-5",
    reportNumber   = "CERN-EP-2020-191",
    eprint         = "2011.05639",
    archivePrefix  = "arXiv",
    primaryClass   = "hep-ex",
}

@Article{HDBS-2018-46,
    author         = "{ATLAS Collaboration}",
    title          = "{Search for Heavy Higgs Bosons Decaying into Two Tau Leptons with the ATLAS Detector Using \(pp\) Collisions at \(\sqrt{s} = 13\,\text{TeV}\)}",
    journal        = "Phys. Rev. Lett.",
    volume         = "125",
    year           = "2020",
    pages          = "051801",
    doi            = "10.1103/PhysRevLett.125.051801",
    reportNumber   = "CERN-EP-2020-014",
    eprint         = "2002.12223",
    archivePrefix  = "arXiv",
    primaryClass   = "hep-ex",
}

@Article{HIGG-2018-09,
    author         = "{ATLAS Collaboration}",
    title          = "{Search for heavy resonances decaying into a pair of \(Z\) bosons in the \(\ell^+\ell^-\ell'^+\ell'^-\) and \(\ell^+\ell^-\nu\bar{\nu}\) final states using \(139\,\text{fb}^{-1}\) of proton--proton collisions at \(\sqrt{s} = 13,\text{TeV}\) with the ATLAS detector}",
    journal        = "Eur. Phys. J. C",
    volume         = "81",
    year           = "2020",
    pages          = "332",
    doi            = "10.1140/epjc/s10052-021-09013-y",
    reportNumber   = "CERN-EP-2020-153",
    eprint         = "2009.14791",
    archivePrefix  = "arXiv",
    primaryClass   = "hep-ex",
}

@Article{HIGG-2018-27,
    author         = "{ATLAS Collaboration}",
    title          = "{Search for resonances decaying into photon pairs in \(139\,\text{fb}^{-1}\) of \(pp\) collisions at \(\sqrt{s} = 13\,\text{TeV}\) with the ATLAS detector}",
    journal        = "Phys. Lett. B",
    volume         = "822",
    year           = "2021",
    pages          = "136651",
    doi            = "10.1016/j.physletb.2021.136651",
    reportNumber   = "CERN-EP-2020-248",
    eprint         = "2102.13405",
    archivePrefix  = "arXiv",
    primaryClass   = "hep-ex",
}

@Article{HIGG-2018-57,
    author         = "{ATLAS Collaboration}",
    title          = "{Combined measurements of Higgs boson production and decay using up to \(80\,\text{fb}^{-1}\) of proton--proton collision data at \(\sqrt{s} = 13\,\text{TeV}\) collected with the ATLAS experiment}",
    journal        = "Phys. Rev. D",
    volume         = "101",
    year           = "2020",
    pages          = "012002",
    doi            = "10.1103/PhysRevD.101.012002",
    reportNumber   = "CERN-EP-2019-097",
    eprint         = "1909.02845",
    archivePrefix  = "arXiv",
    primaryClass   = "hep-ex",
}

@Article{JETM-2018-05,
    author         = "{ATLAS Collaboration}",
    title          = "{Jet energy scale and resolution measured in proton--proton collisions at \(\sqrt{s} = 13\,\text{TeV}\) with the ATLAS detector}",
    journal        = "Eur. Phys. J. C",
    volume         = "81",
    year           = "2020",
    pages          = "689",
    doi            = "10.1140/epjc/s10052-021-09402-3",
    reportNumber   = "CERN-EP-2020-083",
    eprint         = "2007.02645",
    archivePrefix  = "arXiv",
    primaryClass   = "hep-ex",
}

@Article{MUON-2018-03,
    author         = "{ATLAS Collaboration}",
    title          = "{Muon reconstruction and identification efficiency in ATLAS using the full Run~2 \(pp\) collision data set at \(\sqrt{s} = 13\,\text{TeV}\)}",
    journal        = "Eur. Phys. J. C",
    volume         = "81",
    year           = "2021",
    pages          = "578",
    doi            = "10.1140/epjc/s10052-021-09233-2",
    reportNumber   = "CERN-EP-2020-199",
    eprint         = "2012.00578",
    archivePrefix  = "arXiv",
    primaryClass   = "hep-ex",
}

@Article{PIX-2018-001,
    author         = "Abbott, B. and others",
    title          = "{Production and integration of the ATLAS Insertable B-Layer}",
    journal        = "JINST",
    volume         = "13",
    year           = "2018",
    pages          = "T05008",
    doi            = "10.1088/1748-0221/13/05/T05008",
    eprint         = "1803.00844",
    archivePrefix  = "arXiv",
    primaryClass   = "physics.ins-det",
}

@Article{TOPQ-2018-05,
    author         = "{ATLAS Collaboration}",
    title          = "{Evidence for \(t\bar{t}t\bar{t}\) production in the multilepton final state in proton--proton collisions at \(\sqrt{s} = 13\text {TeV}\) with the ATLAS detector}",
    journal        = "Eur. Phys. J. C",
    volume         = "80",
    year           = "2020",
    pages          = "1085",
    doi            = "10.1140/epjc/s10052-020-08509-3",
    reportNumber   = "CERN-EP-2020-111",
    eprint         = "2007.14858",
    archivePrefix  = "arXiv",
    primaryClass   = "hep-ex",
}

@Article{TRIG-2018-01,
    author         = "{ATLAS Collaboration}",
    title          = "{Performance of the ATLAS muon triggers in Run~2}",
    journal        = "JINST",
    volume         = "15",
    year           = "2020",
    pages          = "P09015",
    doi            = "10.1088/1748-0221/15/09/p09015",
    reportNumber   = "CERN-EP-2020-031",
    eprint         = "2004.13447",
    archivePrefix  = "arXiv",
    primaryClass   = "hep-ex",
}

@Article{TRIG-2018-05,
    author         = "{ATLAS Collaboration}",
    title          = "{Performance of electron and photon triggers in ATLAS during LHC Run~2}",
    journal        = "Eur. Phys. J. C",
    volume         = "80",
    year           = "2020",
    pages          = "47",
    doi            = "10.1140/epjc/s10052-019-7500-2",
    reportNumber   = "CERN-EP-2019-169",
    eprint         = "1909.00761",
    archivePrefix  = "arXiv",
    primaryClass   = "hep-ex",
}

@Article{TOPQ-2020-10,
    author         = "{ATLAS Collaboration}",
    title          = "{Measurement of the \(t\bar{t}t\bar{t}\) production cross section in \(pp\) collisions at \(\sqrt{s} = 13\,\text{TeV}\) with the ATLAS detector}",
    journal        = "JHEP",
    volume         = "11",
    year           = "2021",
    pages          = "118",
    doi            = "10.1007/JHEP11(2021)118",
    reportNumber   = "CERN-EP-2021-075",
    eprint         = "2106.11683",
    archivePrefix  = "arXiv",
    primaryClass   = "hep-ex",
}

@Booklet{ATL-SOFT-PUB-2021-001,
    author         = "{ATLAS Collaboration}",
    title          = "{The ATLAS Collaboration Software and Firmware}",
    howpublished   = "{ATL-SOFT-PUB-2021-001}",
    url            = "https://cds.cern.ch/record/2767187",
    year           = "2021",
}

@Booklet{ATL-SOFT-PUB-2021-003,
    author         = "{ATLAS Collaboration}",
    title          = "{ATLAS Computing Acknowledgements}",
    howpublished   = "{ATL-SOFT-PUB-2021-003}",
    url            = "https://cds.cern.ch/record/2776662",
}

@Report{ATLAS-TDR-19,
    author         = "{ATLAS Collaboration}",
    title          = "{ATLAS Insertable B-Layer: Technical Design Report}",
    type           = "ATLAS-TDR-19; CERN-LHCC-2010-013",
    year           = "2010",
    url            = "https://cds.cern.ch/record/1291633",
    related        = "ATLAS-TDR-19-addm",
    relatedstring  = "Addendum:",
}

@Article{CMS-HIG-17-027,
    author         = "{CMS Collaboration}",
    title          = "{Search for heavy Higgs bosons decaying to a top quark pair in proton--proton collisions at \(\sqrt{s} = 13\,\text{TeV}\)}",
    journal        = "JHEP",
    volume         = "04",
    year           = "2020",
    pages          = "171",
    doi            = "10.1007/JHEP04(2020)171",
    reportNumber   = "CERN-EP-2019-147",
    eprint         = "1908.01115",
    archivePrefix  = "arXiv",
    primaryClass   = "hep-ex",
}

@Article{CMS-HIG-17-031,
    author         = "{CMS Collaboration}",
    title          = "{Combined measurements of Higgs boson couplings in proton--proton collisions at \(\sqrt{s} = 13\,\text{TeV}\)}",
    journal        = "Eur. Phys. J. C",
    volume         = "79",
    year           = "2019",
    pages          = "421",
    doi            = "10.1140/epjc/s10052-019-6909-y",
    reportNumber   = "CERN-EP-2018-263",
    eprint         = "1809.10733",
    archivePrefix  = "arXiv",
    primaryClass   = "hep-ex",
}

@Article{CMS-HIG-17-033,
    author         = "{CMS Collaboration}",
    title          = "{Search for a heavy Higgs boson decaying to a pair of \(W\) bosons in proton--proton collisions at \(\sqrt{s} = 13\,\text{TeV}\)}",
    journal        = "JHEP",
    volume         = "03",
    year           = "2020",
    pages          = "034",
    doi            = "10.1007/JHEP03(2020)034",
    reportNumber   = "CERN-EP-2019-230",
    eprint         = "1912.01594",
    archivePrefix  = "arXiv",
    primaryClass   = "hep-ex",
}

@Article{CMS-TOP-17-019,
    author         = "{CMS Collaboration}",
    title          = "{Search for the production of four top quarks in the single-lepton and opposite-sign dilepton final states in proton--proton collisions at \(\sqrt{s} = 13\,\text{TeV}\)}",
    journal        = "JHEP",
    volume         = "11",
    year           = "2019",
    pages          = "082",
    doi            = "10.1007/JHEP11(2019)082",
    reportNumber   = "CERN-EP-2019-098",
    eprint         = "1906.02805",
    archivePrefix  = "arXiv",
    primaryClass   = "hep-ex",
}

@Article{CMS-HIG-18-005,
    author         = "{CMS Collaboration}",
    title          = "{Search for a heavy pseudoscalar boson decaying to a \(Z\) and a Higgs boson at \(\sqrt{s} = 13\,\text{TeV}\)}",
    journal        = "Eur. Phys. J. C",
    volume         = "79",
    year           = "2019",
    pages          = "564",
    doi            = "10.1140/epjc/s10052-019-7058-z",
    reportNumber   = "CERN-EP-2018-343",
    eprint         = "1903.00941",
    archivePrefix  = "arXiv",
    primaryClass   = "hep-ex",
}

@Article{CMS-HIG-18-010,
    author         = "{CMS Collaboration}",
    title          = "{Search for MSSM Higgs bosons decaying to \(\mu^+\mu^-\) in proton--proton collisions at \(\sqrt{s} = 13\,\text{TeV}\)}",
    journal        = "Phys. Lett. B",
    volume         = "798",
    year           = "2019",
    pages          = "134992",
    doi            = "10.1016/j.physletb.2019.134992",
    reportNumber   = "CERN-EP-2019-109",
    eprint         = "1907.03152",
    archivePrefix  = "arXiv",
    primaryClass   = "hep-ex",
}

@Article{CMS-HIG-18-012,
    author         = "{CMS Collaboration}",
    title          = "{Search for new neutral Higgs bosons through the \(H \to ZA \to \ell^{+}\ell^{-} b\bar{b}\) process in \(pp\) collisions at \(\sqrt{s} = 13\,\text{TeV}\)}",
    journal        = "JHEP",
    volume         = "03",
    year           = "2020",
    pages          = "055",
    doi            = "10.1007/JHEP03(2020)055",
    reportNumber   = "CERN-EP-2019-254",
    eprint         = "1911.03781",
    archivePrefix  = "arXiv",
    primaryClass   = "hep-ex",
}

@Article{CMS-HIG-18-023,
    author         = "{CMS Collaboration}",
    title          = "{Search for a heavy pseudoscalar Higgs boson decaying into a \(125\,\text{GeV}\) Higgs boson and a \(Z\) boson in final states with two tau and two light leptons at \(\sqrt{s} = 13\,\text{TeV}\)}",
    journal        = "JHEP",
    volume         = "03",
    year           = "2020",
    pages          = "065",
    doi            = "10.1007/JHEP03(2020)065",
    reportNumber   = "CERN-EP-2019-231",
    eprint         = "1910.11634",
    archivePrefix  = "arXiv",
    primaryClass   = "hep-ex",
}

@Article{CMS-TOP-18-003,
    author         = "{CMS Collaboration}",
    title          = "{Search for production of four top quarks in final states with same-sign or multiple leptons in proton--proton collisions at \(\sqrt{s} = 13\,\text{TeV}\)}",
    journal        = "Eur. Phys. J. C",
    volume         = "80",
    year           = "2020",
    pages          = "75",
    doi            = "10.1140/epjc/s10052-019-7593-7",
    reportNumber   = "CERN-EP-2019-163",
    eprint         = "1908.06463",
    archivePrefix  = "arXiv",
    primaryClass   = "hep-ex",
}

@Booklet{ATLAS-CONF-2014-018,
    author         = "{ATLAS Collaboration}",
    title          = "{Tagging and suppression of pileup jets with the ATLAS detector}",
    howpublished   = "{ATLAS-CONF-2014-018}",
    url            = "https://cds.cern.ch/record/1700870",
    year           = "2014",
}

@Booklet{ATLAS-CONF-2015-029,
    author         = "{ATLAS Collaboration}",
    title          = "{Selection of jets produced in \(13~\text{TeV}\) proton--proton collisions with the ATLAS detector}",
    howpublished   = "{ATLAS-CONF-2015-029}",
    url            = "https://cds.cern.ch/record/2037702",
    year           = "2015",
}

@Booklet{ATLAS-CONF-2019-021,
    author         = "{ATLAS Collaboration}",
    title          = "{Luminosity determination in \(pp\) collisions at \(\sqrt{s} = 13\,\text{TeV}\) using the ATLAS detector at the LHC}",
    howpublished   = "{ATLAS-CONF-2019-021}",
    url            = "https://cds.cern.ch/record/2677054",
    year           = "2019",
}

@Booklet{ATL-PHYS-PUB-2014-021,
    author         = "{ATLAS Collaboration}",
    title          = "{ATLAS Pythia~8 tunes to \(7~\text{TeV}\) data}",
    howpublished   = "{ATL-PHYS-PUB-2014-021}",
    url            = "https://cds.cern.ch/record/1966419",
    year           = "2014",
}

@Booklet{ATL-PHYS-PUB-2015-026,
    author         = "{ATLAS Collaboration}",
    title          = "{Vertex Reconstruction Performance of the ATLAS Detector at \(\sqrt{s} = 13~\text{TeV}\)}",
    howpublished   = "{ATL-PHYS-PUB-2015-026}",
    url            = "https://cds.cern.ch/record/2037717",
    year           = "2015",
}

@Booklet{ATL-PHYS-PUB-2016-017,
    author         = "{ATLAS Collaboration}",
    title          = "{The Pythia~8 A3 tune description of ATLAS minimum bias and inelastic measurements incorporating the Donnachie--Landshoff diffractive model}",
    howpublished   = "{ATL-PHYS-PUB-2016-017}",
    url            = "https://cds.cern.ch/record/2206965",
    year           = "2016",
}

@Booklet{ATL-PHYS-PUB-2016-020,
    author         = "{ATLAS Collaboration}",
    title          = "{Studies on top-quark Monte Carlo modelling for Top2016}",
    howpublished   = "{ATL-PHYS-PUB-2016-020}",
    url            = "https://cds.cern.ch/record/2216168",
    year           = "2016",
}

@Booklet{ATL-PHYS-PUB-2017-003,
    author         = "{ATLAS Collaboration}",
    title          = "{Identification of Jets Containing \(b\)-Hadrons with Recurrent Neural Networks at the ATLAS Experiment}",
    howpublished   = "{ATL-PHYS-PUB-2017-003}",
    url            = "https://cds.cern.ch/record/2255226",
    year           = "2017",
}

@Booklet{ATL-PHYS-PUB-2017-013,
    author         = "{ATLAS Collaboration}",
    title          = "{Optimisation and performance studies of the ATLAS \(b\)-tagging algorithms for the 2017-18 LHC run}",
    howpublished   = "{ATL-PHYS-PUB-2017-013}",
    url            = "https://cds.cern.ch/record/2273281",
    year           = "2017",
}

\clearpage
 
 
\begin{flushleft}
\hypersetup{urlcolor=black}
{\Large The ATLAS Collaboration}

\bigskip

\AtlasOrcid[0000-0002-6665-4934]{G.~Aad}$^\textrm{\scriptsize 101}$,
\AtlasOrcid[0000-0002-5888-2734]{B.~Abbott}$^\textrm{\scriptsize 119}$,
\AtlasOrcid[0000-0002-7248-3203]{D.C.~Abbott}$^\textrm{\scriptsize 102}$,
\AtlasOrcid[0000-0002-1002-1652]{K.~Abeling}$^\textrm{\scriptsize 55}$,
\AtlasOrcid[0000-0002-8496-9294]{S.H.~Abidi}$^\textrm{\scriptsize 29}$,
\AtlasOrcid[0000-0002-9987-2292]{A.~Aboulhorma}$^\textrm{\scriptsize 35e}$,
\AtlasOrcid[0000-0001-5329-6640]{H.~Abramowicz}$^\textrm{\scriptsize 150}$,
\AtlasOrcid[0000-0002-1599-2896]{H.~Abreu}$^\textrm{\scriptsize 149}$,
\AtlasOrcid[0000-0003-0403-3697]{Y.~Abulaiti}$^\textrm{\scriptsize 116}$,
\AtlasOrcid[0000-0003-0762-7204]{A.C.~Abusleme~Hoffman}$^\textrm{\scriptsize 136a}$
% ...
\bigskip
\\

% ...
$^{29}$Physics Department, Brookhaven National Laboratory, Upton NY; United States of America.\\
% ...
$^{35}$$^{(a)}$Facult\'e des Sciences Ain Chock, R\'eseau Universitaire de Physique des Hautes Energies - Universit\'e Hassan II, Casablanca;$^{(b)}$Facult\'{e} des Sciences, Universit\'{e} Ibn-Tofail, K\'{e}nitra;$^{(c)}$Facult\'e des Sciences Semlalia, Universit\'e Cadi Ayyad, LPHEA-Marrakech;$^{(d)}$LPMR, Facult\'e des Sciences, Universit\'e Mohamed Premier, Oujda;$^{(e)}$Facult\'e des sciences, Universit\'e Mohammed V, Rabat;$^{(f)}$Institute of Applied Physics, Mohammed VI Polytechnic University, Ben Guerir; Morocco.\\
% ...
$^{55}$II. Physikalisches Institut, Georg-August-Universit\"{a}t G\"ottingen, G\"ottingen; Germany.\\
% ...
$^{101}$CPPM, Aix-Marseille Universit\'e, CNRS/IN2P3, Marseille; France.\\
$^{102}$Department of Physics, University of Massachusetts, Amherst MA; United States of America.\\
% ...
$^{116}$Department of Physics, New York University, New York NY; United States of America.\\
% ...
$^{119}$Homer L. Dodge Department of Physics and Astronomy, University of Oklahoma, Norman OK; United States of America.\\
% ...
$^{136}$$^{(a)}$Departamento de F\'isica, Pontificia Universidad Cat\'olica de Chile, Santiago;$^{(b)}$Millennium Institute for Subatomic physics at high energy frontier (SAPHIR), Santiago;$^{(c)}$Instituto de Investigaci\'on Multidisciplinario en Ciencia y Tecnolog\'ia, y Departamento de F\'isica, Universidad de La Serena;$^{(d)}$Universidad Andres Bello, Department of Physics, Santiago;$^{(e)}$Instituto de Alta Investigaci\'on, Universidad de Tarapac\'a, Arica;$^{(f)}$Departamento de F\'isica, Universidad T\'ecnica Federico Santa Mar\'ia, Valpara\'iso; Chile.\\
% ...
$^{149}$Department of Physics, Technion, Israel Institute of Technology, Haifa; Israel.\\
$^{150}$Raymond and Beverly Sackler School of Physics and Astronomy, Tel Aviv University, Tel Aviv; Israel.\\
% ...
$^{a}$ Also Affiliated with an institute covered by a cooperation agreement with CERN.\\
$^{b}$ Also at Borough of Manhattan Community College, City University of New York, New York NY; United States of America.\\
$^{c}$ Also at Bruno Kessler Foundation, Trento; Italy.\\
$^{d}$ Also at Center for High Energy Physics, Peking University; China.\\
$^{e}$ Also at Centro Studi e Ricerche Enrico Fermi; Italy.\\
$^{f}$ Also at CERN, Geneva; Switzerland.\\
% ...

\end{flushleft}


\end{document}